\documentclass[12pt]{amsart}
\usepackage{setspace}
\usepackage[margin=1.25in]{geometry}
\usepackage[utf8]{inputenc} 
\usepackage[T1]{fontenc}    
\usepackage{hyperref}       
\usepackage{url}            
\usepackage{booktabs}       
\usepackage{amsfonts}       
\usepackage{nicefrac}       
\usepackage{microtype}      
\usepackage{xcolor}         
\usepackage[flushleft]{threeparttable}
\usepackage{natbib}
\usepackage{amsmath}
\usepackage{amsthm}
\usepackage{amssymb}
\usepackage{todonotes}
\usepackage{multirow}
\usepackage{comment}
\usepackage[justification=centering]{caption}
\usepackage{siunitx}
\usepackage{caption}
\usepackage{float}
\usepackage[foot]{amsaddr}

\usepackage[shortlabels]{enumitem} 

\usepackage[linesnumbered,ruled]{algorithm2e}
\SetKwInput{KwInput}{Input}                
\SetKwInput{KwOutput}{Output}              

\newtheorem{thm}{\protect\theoremname}[section]
\newtheorem{asm}{Assumption}[section]
\providecommand{\theoremname}{Theorem}
\newtheorem{lem}{Lemma}[section]

\theoremstyle{definition}

\newcommand{\changed}[1]{\textcolor{black}{#1}}
\newcommand{\changedSL}[1]{\textcolor{black}{#1}}

\title[Fast Inference for Quantile Regression]{Fast Inference for Quantile Regression \\ with Tens of Millions of Observations}\thanks{%
\changedSL{We would like to thank the guest associate editor, two anonymous referees, Roger Koenker, and seminar participants at CIREQ 2022, Albany, Wisconsin--Madison, Washington, UIUC, and SFU for helpful comments. Shin is grateful for the partial support from the Social Sciences and Humanities Research Council of Canada (SSHRC-20015659). This work was made possible by the facilities of Calcul Queb\'{e}c (\url{www.calculquebec.ca}) and Compute Canada (\url{www.computecanada.ca}). Seo acknowldges support from the Research Grant of the Center for Distributive Justice at the Institute of Economic Research, Seoul National University, and from the Ministry of Education of the Republic of Korea and the National Research Foundation of Korea (NRF-2018S1A5A2A01033487). }}

\author[Lee]{Sokbae  Lee}
\address{Department of Economics, Columbia University, New York, NY 10027, USA}
\email {sl3841@columbia.edu} 

\author[Liao]{Yuan  Liao}
\address{Department of Economics, Rutgers University, New Brunswick, NJ 08901, USA}
\email{yuan.liao@rutgers.edu}

\author[Seo]{Myung Hwan  Seo}
\address{Department of Economics, Seoul National University, Seoul, 08826, Korea}
\email{myunghseo@snu.ac.kr}

\author[Shin]{Youngki  Shin}
\address{Department of Economics, McMaster University, Hamilton, ON L8S 4L8, Canada}
\email{shiny11@mcmaster.ca}

\date{\today}

\begin{document}

\begin{abstract}
Big data analytics has opened new avenues in economic research, but the challenge of analyzing datasets with tens of millions of observations is substantial. Conventional econometric methods based on extreme estimators require large amounts of computing resources and memory, which are often not readily available.
In this paper, we focus on linear quantile regression applied to ``ultra-large'' datasets, such as U.S. decennial censuses. A fast inference framework is presented, utilizing stochastic subgradient descent (S-subGD) updates.
The inference procedure handles cross-sectional data sequentially: (i) updating the parameter estimate with each incoming ``new observation'', (ii) aggregating it as a \emph{Polyak-Ruppert} average, and (iii) computing a pivotal statistic for inference using only a solution path. The methodology draws from time-series regression to create an asymptotically pivotal statistic through random scaling.
Our proposed test statistic is calculated in a fully online fashion and critical values are calculated without resampling. We conduct extensive numerical studies to showcase the computational merits of our proposed inference.
For inference problems as large as $(n, d) \sim (10^7, 10^3)$, where $n$ is the sample size and $d$ is the number of regressors, our method generates new insights,  surpassing current inference methods in computation.
Our method specifically reveals trends in the gender gap in the U.S. college wage premium using millions of observations, while controlling over $10^3$ covariates to mitigate confounding effects.
\end{abstract}

\maketitle

\onehalfspacing

\newpage

\section{Introduction}\label{sec:intro}

Quantile regression (QR) has been increasingly popular in economics since the groundbreaking work of \citet{Koenker1978}. 
Many significant developments have been made to this important statistical methodology.  Meanwhile, economists are now able to analyze datasets that contain tens of millions of observations. For instance, one of the important economic applications of quantile regression is to study the wage structure. 
The 5\% sample in the 2000 U.S. Census contains more than 14 million observations and more than 4 million even after restricting the sample to a subpopulation of working \changedSL{White} adults. One motivation of using datasets with such   ``ultra-large\changed{''} sample sizes is to deal with econometric models with a large number of parameters within the framework of simple parametric models. Suppose that there are $d$ parameters to estimate with a sample of size $n$.
A parametric regression model can be as large as $d \sim 1,000$ because a large number of controls might be necessary. In the setting of ``ultra-large\changed{''} sample size with $n\sim 10^7$,  straightforward parametric estimators can be used with $d \sim 1,000$ under standard textbook  assumptions without demanding conditions such as sparsity assumptions.\footnote{\changedSL{A classical condition for a uniform normal approximation of an $M$-estimator is $ (d \log n)^{3/2}/n \rightarrow 0$ \citep{Portnoy:85}.}}

With datasets containing tens of millions of observations, however, one major challenge of  QR  is that it  would require huge computing powers and memory that are often not accessible using ordinary personal computers. 
While inference is  desirable in empirical studies to quantify statistical uncertainty, 
it often \changed{demands more computational resources} than point estimation.
Typically, asymptotic normal inference with QR estimators involves solving optimization problems and computing asymptotic covariance matrices. Computationally efficient implementation of both tasks is crucial with ultra-large datasets.

In this paper, we focus on this challenge, by studying the QR inference problem in the context of $n\sim 10^7$.  We propose a  \changed{fast statistical inference framework} to analyze cross-sectional data with millions of observations\changed{, a typical sample size for the census data}, embracing stochastic gradient descent techniques, one of the most active research areas in machine learning. This literature dates back to
\citet{Robbins-Monro-1951}. While we focus on the QR framework, whose inference has been well known to be a hard problem, our proposed method can be \changed{naturally} generalized to other econometric frameworks.

Empirically, we use the data from IPUMS USA \citep{IPUMS}, a common data source for the U.S. wage structure,  to study the trends in the gender gap in terms of the college wage premium. In our empirical application, we aim to control for work experience by flexibly interacting workers' age with gender and state fixed effects, which creates over one thousand regressors, thereby motivating the need of ultra-large datasets. 
We find that existing inference methods are not applicable due to time or memory constraints, while our inference method provides new insights into trends in the gender gap.  There has been a large literature on college wage premium and wage structure. See, for instance, 
\citet{chamberlain1994quantile,Buchinsky:1994,Buchinsky:1998,Gosling:2000,angrist2006quantile,CFM:2013} among others.
To the best of knowledge, our empirical illustration is the first to obtain confidence intervals based on quantile regression with more than 4 million observations and more than 1000 regressors.


\subsection{Standard inference for quantile regression}

Consider the setting of  a linear quantile regression model  for which
data $\{ Y_i \equiv (y_i,x_i) \in \mathbb{R}^{1+d} : i=1,\ldots,n\}$ are generated from 
\begin{align}\label{model}
y_i = x_i'\beta^* + \varepsilon_i, 
\end{align}
where
$\beta^*$ is a vector of unknown parameters
and the unobserved error $\varepsilon_i$ satisfies 
$P(\varepsilon_i \leq 0|x_i)= \tau$ for a fixed quantile $\tau \in(0,1)$.  
Note that $\beta^{*}$ is characterized by 
\[
\beta^{*}:=\arg\min_{\beta\in\mathbb{R}^{d}}Q\left(\beta\right),
\]
where
$
Q(\beta):= \mathbb E [q(\beta, Y_i)]$ with the check function
$
q(\beta, Y_i):= (y_i-x_i'\beta)(\tau-I\{y_i-x_i'\beta\leq 0\}). 
$
Here, $I(\cdot)$ is the usual indicator function. 
The standard approach for estimating $\beta^{*}$ is to use the following M-estimator
originally proposed by \citet{Koenker1978}:
\begin{align}\label{def:m-estimator}
\widehat{\beta}_n := \arg\min_{\beta\in\mathbb{R}^{d}} \; \frac{1}{n}  \sum_{i=1}^n q(\beta, Y_i).
\end{align}
See \citet{Koenker2005} and \citet{Koenker17} for a monograph and a review of recent
developments, respectively.

There have been two potential challenges for the inference of standard quantile regression. 
\changed{The first} is the computational problem. 
The optimization problem is typically   reformulated to a linear programming problem, and solved using interior-point algorithms \citep[e.g.,][]{Portnoy:Koenker:97}. 
\changed{The second} is associated with statistical inference. As was shown by
\citet{Koenker1978}, the asymptotic distribution of the M estimator is as follows:
\begin{align}\label{asymptotic-normality-H}
\sqrt{n}(\widehat\beta_n-\beta^*) \overset{d}{\to} N(0, \tau(1-\tau)H^{-1}\mathbb E[x_ix_i'] H^{-1}),\quad H= \mathbb E [f_{\varepsilon}(0|x_i)x_ix_i'],
\end{align}
where $f_{\varepsilon}(\cdot|x_i)$ is the conditional distribution of $\varepsilon_i$ given $x_i$, assuming data are independent and identically distributed (i.i.d.).
Hence, in the heteroskedastic setting, standard inference based on the ``plug-in'' method would require \changedSL{estimating matrix $H$, which involves the conditional density function $f_{\varepsilon}(\cdot|x_i)$}. An alternative inference would be based on bootstrap. 

One of the attempts to solving the computational/inference difficulties is to rely on the \emph{smoothing} idea, via either smoothing the estimating equation  \citep{Horowitz:1998}  or the convolution-type smoothing  \citep{FGH,TanWangZhou,HE2021}. Both require a choice of the smoothing bandwidth.    In particular, the   convolution-type smoothing (\texttt{conquer} as coined by \citet{HE2021}) has received recent attention in the literature because  the optimization problem is  convex, so it is  more scalable.  Figure 1 of \citet{HE2021}  shows that \texttt{conquer}  works well for point estimations when the sample size ranges from $10^3\sim 10^6$ with the number of regressors $d \approx n^{1/2}$. 


Meanwhile, in terms of inference,  both types of smoothed estimators are first-order asymptotically equivalent to the unsmoothed estimator $\widehat{\beta}_n$\changed{, and the scale of the numerical studies  of these solutions is less ambitious.} For instance,  the largest model considered for inference  in \citet{HE2021} is merely $(n, d) = (4000, 100)$ (see Figure 7 in their paper).
Therefore, there is a scalability gap between point estimation and inference in the literature.
In this paper, we aim to bridge this gap by proposing a method for fast inference. 

\subsection{The proposed fast S-subGD framework}

We focus on the inference problem at the scale up to 
\changed{$(n, d) \sim (10^7, 10^3)$, }
which we refer to as an ``ultra-large\changed{''} quantile regression problem.  
This is a possible scale with IPUMS USA, but very difficult to deal with using benchmark inference procedures for quantile regression. 
To tackle this large-scale problem, we estimate $\beta^*$ via stochastic (sub)gradient descent (S-subGD).  
Our   assumption on the asymptotic regime is that $d$ is fixed but $n$ grows to infinity. 

Suppose that we have i.i.d. data from a large cross-sectional survey:
$$Y_1,..., Y_n,\quad Y_i=(x_i, y_i),$$
where the ordering of these observations is randomized. 
Our method produces a sequence of 
\changed{estimators}, 
denoted by $\beta_i$, which is a solution path being updated as 
 $$
\beta_{i}=\beta_{i-1}-\gamma_{i}\nabla q\left(\beta_{i-1},Y_{i}\right).
$$ 
Here $Y_i$ is a 
\changed{new} 
observation in the randomized data sequence. Also,  $\nabla q\left(\beta_{i-1},Y_i\right)$ is a subgradient of the check function with respect to the current update, and   $\gamma_i$ is a predetermined learning rate.   Then\changed{,} we take the average of the sequence
$$
\bar\beta_n:=\frac{1}{n}\sum_{i=1}^n\beta_i,
$$
which is called as the \citet{Polyak1990}-\citet{ruppert1988efficient} average in the machine learning literature.
It will be shown in this paper that
$\sqrt{n}(\bar{\beta}_{n} - \beta^*)$ is asymptotically normal
with the asymptotic variance the same as that of the standard \citet{Koenker1978} estimator $\widehat{\beta}_n$. 
\changed{Therefore,}
it achieves the same first-order asymptotic efficiency. 

The main novelty of this paper comes from how to conduct inference based on 
$\bar{\beta}_{n}$. 
We use a sequential transformation of $\beta_i$'s \changed{in a suitable way for on-line updating, and construct asymptotically pivotal statistics.}
That is, we studentize $\sqrt{n}\left(\bar{\beta}_{n}-\beta^{*}\right)$ via a random scaling matrix $\widehat V_n$ whose exact form will be given later. 
The resulting statistic is not asymptotically normal but \emph{asymptotically pivotal} in the sense that its asymptotic distribution is free of any unknown nuisance parameters; thus, its critical values are easily available. 
Furthermore, the random scaling quantity $\widehat{V}_{n}$ does not require any additional input other than stochastic subgradient path $\beta_i$, and can be computed 
\changed{fast}. 
 
The main contribution of this paper is computational. We combine the idea of stochastic subgradient descent with random scaling and make large-scale inference for quantile regression much more practical. 
We provide a discussion of computational complexity and show that our algorithm is comparable to the existing ones in terms of point estimation and 
is superior to those in terms of inference (especially, subvector inference). See Section~\ref{sec:comp:complexity} for details.
In addition, we demonstrate the usefulness of our method via Monte Carlo experiments. Empirically, we apply it to studying the gender gap in college wage premiums using the IPUMS dataset. The proposed inference method of quantile regression to the ultra-big dataset reveals some interesting new features. First, it shows heterogeneous effects over different quantile levels. Second, we find that the female college wage premium is significantly higher \changed{than} that of male workers at the median, which is different from what have been concluded in the literature.  In fact, although the S-subGD inference tends to be conservative than the standard normal approximation, it reveals statistically significant results while controlling over $10^3$ covariates to mitigate confounding effects. With more availability of such a large dataset, the S-subGD method will make it possible to obtain convincing empirical evidence for other analyses. Another topic is to examine the relationship between seasonality and birth weight \citep[e.g.,][]{Buckles:Hungerman:2013,currie2013within,Chatterji:22} because the quantiles of birth weight are of interest and
birth data from vital statistics contain tens of millions of observations.\footnote{For example, the National Bureau of Economic Research has public use data at \url{https://www.nber.org/research/data/vital-statistics-natality-birth-data}.}





\subsection{The related literature}

Our estimator is motivated from the literature on \textit{online learning}, where data were collected in a streaming fashion, and as a new observation $Y_i$ arrives, the   estimator is updated to  $\beta_i$. Under the M-estimation framework with strongly convex, differentiable loss functions, \citet{polyak1992acceleration} showed that the    \citet{Polyak1990}-\citet{ruppert1988efficient} average is asymptotically normal. More recently, \cite{fang2018online}, \cite{chen2020statistical}, and \cite{zhu2021online} studied the statistical inference problem and proposed methods that require the implementation of bootstrap, consistent estimation of the asymptotic variance matrix,   or the use of ``batch-means'' approach, respectively. Besides the smoothness assumption that excludes quantile regression,
\changedSL{neither consistent estimation of the asymptotic variance matrix nor implementation of bootstrap is} computationally attractive for datasets as large as those with which this paper is concerned. 
\changedSL{The batch-means approach does not seem to provide adequate finite sample coverage even if $n$ is sufficiently large \citep[see, e.g.,][in the context of linear mean regression and logistic regression models]{lee2021fast}.}


The idea of random scaling is borrowed from 
the time-series literature on fixed bandwidth heteroskedasticity and autocorrelation robust (HAR) inference \citep[e.g.,][]{kiefer2000simple,sun2008optimal,sun2014fixed,lazarus2018har} and has been recently adopted in the context
of stochastic gradient descent (SGD) in machine-learning problems: online inference for linear mean regression \citep{lee2021fast},
federated learning \citep{li2021statistical},
and
Kiefer-Wolfowitz methods \citep{ChenLai2021}
among others.\footnote{\changedSL{\citet{li2021statistical} and \citet{ChenLai2021} applied the idea of random scaling after our previous work \citep{lee2021fast} appeared on arXiv.}}
The current paper shares the basic inference idea with our previous work \citep{lee2021fast}; however, quantile regression is sufficiently different from mean regression and it requires further theoretical development that is not covered in the existing work.  
In particular, our previous work is based on \citet{polyak1992acceleration}, which limits its analysis to 
differentiable and strongly convex objective functions; however, 
the check function $q(\beta, Y_i)$ for quantile regression is non-differentiable and is not strongly convex. 
We overcome these difficulties by using the results given in  \citet{gadat2022optimal}.
One could have adopted convolution-type smoothing for S-subGD as the corresponding objective function is differentiable and locally strongly convex. We do not go down this alternative route in this paper as the existence of a sequence of smoothing bandwidths converging to zero complicates theoretical analysis and it would be difficult to optimally choose the sequence of smoothing bandwidths along with the sequence of learning rates. 
Previously, 
\citet{Pesme:Flammarion:20} studied online robust regression in the context of a Gaussian linear model with an adversarial noise; in fact, their algorithm corresponds to our median regression case (see equation (2) in their paper) but they did not study quantile regression models.  
\citet{gadat2022optimal} considered, as their example, the recursive quantile estimator, which corresponds to the quantile regression estimator with an intercept being only the regressor.
Neither of these two papers delved into the issue of inference.

 In the econometric literature, \cite{forneron2021estimation} and \citet{forneron2022estimation} developed SGD-based resampling schemes that  deliver both point estimates and  standard errors within the same optimization framework. 
 The major difference is that their stochastic path is a Newton-Raphson type which requires computing the Hessian matrix or its approximation. Hence, their framework is not applicable to quantile regression because the check function is not twice differentiable. In other words, for quantile regression inferences with over millions of observations,  estimating the inverse Hessian matrix \changedSL{$H^{-1}$ in \eqref{asymptotic-normality-H}} is  a task that we particularly would like to avoid.  

As an alternative to full sample estimation, one may consider sketching \citep[e.g., see][for a review from an econometric perspective]{Lee:Ng:2020}. For example, 
\citet{Portnoy:Koenker:97} developed a preprocessing algorithm that can be viewed as a sketching method 
and
\citet{Yang:QR:ICML} proposed a fast randomized algorithm for large-scale quantile regression and solved the problem of size $n \sim 10^{10}$ and $d = 12$ by randomly creating a subsample of about $n = 10^5$. However, the theoretical analysis carried out in \citet{Yang:QR:ICML} is limited to approximations of the optimal value of the check function and does not cover the issue of inference.
As an alternative sketching method, one may employ \citet{wang2021optimal}  and construct a random subsample using data-dependent, nonuniform weights to improve the efficiency of the estimator. 
In mean regression models, the precision of a sketched estimator depends on the subsample size and is typically worse than that of the full sample estimator \citep[e.g.,][]{Lee:Ng:2020,Lee:Ng:2022}.  
We expect a similar phenomenon for quantile regression models. 
 

\subsection{Notation}

Let $a'$ and $A'$, respectively, denote the transpose of vector $a$ and matrix $A$. 
Let $\| a \|$ denote the Euclidean norm of vector $a$
and $\| A \|$ the Frobenius norm of matrix $A$. 
Also, let $\ell^{\infty}\left[0,1\right]$ denote the  set of bounded continuous functions on $[0,1].$
Let $I(A)$ be the usual indicator function, that is $I(A) = 1$ if $A$ is true and 0 otherwise.
For a symmetric, positive definite matrix $S$, let $\lambda_{\min}(S)$ denote its smallest eigenvalue.

\section{A Fast Algorithm for Quantile Inference}\label{sec:algorithm}

\subsection{Stochastic subgradient descent (S-subGD)}

In this section, we \changed{propose our}  inference algorithm.
Suppose that we have cross-sectional data $\{Y_1,..., Y_n \}$, where $Y_i=(x_i, y_i)$. 

First, we randomize the   ordering of the observed data:
$$
Y_1=(x_1, y_1),..., Y_n=(x_n, y_n).
$$
We start with an 
\changed{initial}
estimator, denoted by $\beta_0$, using a method which we shall discuss later. Then produce \changedSL{an S-subGD solution path} which updates   according to the rule:
\begin{align}\label{eq:SGD1}
\beta_{i}=\beta_{i-1}-\gamma_{i}\nabla q\left(\beta_{i-1},Y_{i}\right),
\end{align}
where the updating subgradient depends on a \textit{single ``next'' observation} $Y_i$: 
\begin{align}\label{eq:SS}
 \nabla q(\beta, Y_i) := x_i [I\{y_i \leq x_i'\beta\} - \tau].
\end{align}
This is the  subgradient
of $q\left(\beta,Y_i\right)$ with respect to $\beta$. Here $\beta_{i-1}$ is the ``current'' estimate, which is updated to $\beta_{i}$ when the next observation $Y_i$ comes into play. 
Also $\gamma_i$ is a pre-determined step size, which  is also called a learning rate and is assumed to have the form
$\gamma_{i}:=\gamma_0 i^{-a} $ for some constants $\gamma_0 > 0$ and $a \in (1/2,1)$. Then, we take the average of the sequence
$
\bar\beta_n:=\frac{1}{n}\sum_{i=1}^n\beta_i
$ as the final estimator. 

Computing this estimator does not require storing the historical data; thus, it is very fast and memory-efficient even when $n\sim 10^7$. In fact, as we shall see below, the computational gain is much more substantial for inference.  




\subsection{The pivotal statistic}

It is most common to make inference based on a consistent estimator of the asymptotic variance for $\sqrt{n}(\bar\beta_n-\beta^*)$, which  is 
\changed{computationally demanding.} For instance, one 
\changed{needs}
to estimate 
\changedSL{the inverse Hessian matrix $H^{-1}$ in \eqref{asymptotic-normality-H}}.
\changed{Instead of pursuing asymptotic normal inference using a consistent estimator of the asymptotic variance, we employ asymptotic mixed-normal inference based on a suitable random scaling method through the partial sum process of the solution path as done in the \emph{fixed-b} standardization.}
Motivated by the fact that the solution path $\beta_i$ is sequentially updated,  we apply the random scaling approach by defining:
\begin{align}\label{def:random-scaling}
\widehat{V}_{n} := \frac{1}{n}\sum_{s=1}^{n} 
\left\{ \frac{1}{\sqrt{n}} \sum_{i=1}^{s} \left( \beta_{i}-\bar{\beta}_{n} \right) \right \} 
\left\{ \frac{1}{\sqrt{n}} \sum_{i=1}^{s} \left( \beta_{i}-\bar{\beta}_{n} \right) \right \}'.
\end{align}
Inference will be conducted based on the standardization using $\widehat V_n.$  Computing $\widehat V_n$ can be 
\changed{efficient} even if $n\sim 10^7$ or larger, since it can be sequentially computed as detailed later. 
Furthermore, its subvector inference involves only the associated elements, substantially reducing the computational burden.

Once $\bar{\beta}_{n}$ and $\widehat{V}_{n}$ are obtained, the inference is straightforward. For example, for the $j$ th component of $\bar \beta_n$, let $\widehat V_{n,jj}$ denote the $(j,j)$ th diagonal entry of $\widehat V_n$. The t-statistic is then
\begin{align}\label{t-stat}
\frac{\sqrt{n}\left(\bar{\beta}_{n,j}-\beta_{j}^{*}\right)}{\sqrt{\widehat{V}_{n,jj}}}, 
\end{align}
whose asymptotic distribution is mixed normal and symmetric around zero\changed{. We will} formally derive it in the next section. 
The mixed normal asymptotic distribution for the t-statistic in \eqref{t-stat}  is the same as the distribution of the statistics observed in the estimation of the cointegration vector by \citet{johansen1991estimation} and \citet{Abadir:Paruolo:97} and in the heteroskedasticity and autocorrelation robust inference in \citet{kiefer2000simple}. 
They are different statistics but \changed{converge to} the identical distribution\changed{, a functional} of the standard Wiener process as shown by \citet{Abadir:Paruolo:02}.
We  can use the t-statistic  to construct 
 the $(1-\alpha)$ asymptotic confidence interval for the $j$-th element $\beta_{j}^{*}$ of $\beta^{*}$ by 
\[
\left[
\bar{\beta}_{n,j} - \textrm{cv} (1- \alpha/2) \sqrt{\frac{\widehat{V}_{n,jj}}{n}}, \;
\bar{\beta}_{n,j} + \textrm{cv} (1- \alpha/2) \sqrt{\frac{\widehat{V}_{n,jj}}{n}} \;
\right],
\]
where the critical value $\textrm{cv} (1- \alpha/2)$ is tabulated in \citet[Table I]{Abadir:Paruolo:97}. For easy reference, the critical values are reproduced in Table~\ref{tab:cv}.
For instance, the critical value for $\alpha = 0.05$ is 6.747.
Critical values for testing linear restrictions $H_{0}: R\beta^{*} = c$
can be found in \citet[Table II]{kiefer2000simple}.

\begin{table}[htbp]
\caption{Asymptotic critical values of the t-statistic}
\begin{center}
\begin{tabular}{lcccc}
\hline
Probability  & 90\% & 95\% & 97.5\% & 99\% \\
Critical Value & 3.875 & 5.323 & 6.747 & 8.613 \\
\hline
\end{tabular}
\end{center}
\label{tab:cv}
\begin{minipage}{1\textwidth} 
{Note. 
The table gives one-sided asymptotic critical values that satisfy 
$\mathrm{Pr}( \hat{t} \leq c ) = p$ asymptotically, where
$p \in \{0.9, 0.95, 0.975, 0.99\}$.
Source: \citet[Table I]{Abadir:Paruolo:97}. 
  \par}
\end{minipage}
\end{table}%

\subsection{Practical details for memory efficiency}

While computing $(\widehat V_n,\bar\beta_n)$ is straightforward for datasets of medium sample sizes, it is not practical when $n\sim 10^7$ or even larger size, as storing the entire solution path 
\changed{$\beta_i$ }
can be infeasible for memory of usual personal computing devices. 


We can construct a solution path for  $(\widehat V_n,\bar\beta_n)$ along the way of updating $\beta_i$, so all relevant quantities of the proposed inference can be obtained sequentially. Specifically, let $(\bar\beta_{i-1},\beta_{i-1})$ be the current update when we use up to $i-1$ observations. We then update using the new observation $Y_i$ by:
\begin{eqnarray}
\beta_i&=& \beta_{i-1} - \gamma_i \nabla q(\beta_{i-1}, Y_i), \label{eq.7beta}\\
\bar\beta_i&=& \bar\beta_{i-1}\frac{i-1}{i} + \beta_i\frac{1}{i}, \label{eqa7}\\
\widehat V_i&=& i^{-2}\left(A_i-\bar\beta_i b_i' -b_i\bar\beta_i'+\bar\beta_i\bar\beta_i'\sum_{s=1}^is^2\right),\label{eqa8}
\end{eqnarray}
where the intermediate quantities $A_i$ and $b_i$ are also updated \changedSL{sequentially}:
$$
A_i=A_{i-1} +i^2\bar\beta_i\bar\beta_i',\quad b_i= b_{i-1} + i^2\bar\beta_i.
$$
Therefore computing $(\bar\beta_n, \widehat V_n)$ can be cast sequentially  until $i=n$, and does not require saving the entire solution path of $\beta_i$. 

 A good initial value could help achieve the computational stability in practice, though it does not matter in the theoretical results. 
We recommend applying \texttt{conquer} in \citet{HE2021} to the subsample (e.g. 5\% or 10\%) to estimate the initial value, which we adopted in our numerical experiments.
Specifically, we start with a smooth quantile regression proposed in \citet{FGH} and \citet{HE2021}:
$$
\beta_0=\arg\min\sum_{i\in \mathcal S}q_{n}(\beta, Y_i),
$$
where $q_n$ is a convolution-type-smoothed checked function, and $\mathcal S$ is a randomized subsample. 
The \texttt{conquer} algorithm
is very fast to compute a point estimate of $\beta_0$ but it is much less scalable with respect to inference (see the Monte Carlo results later in the paper).
For the intermediate quantities $A_i$ and $b_i$, we set $A_0 = 0$ and $b_0 = 0$.

Another approach is the burn-in method that applies the S-subGD to some initial observations, and throw them away for the main inference procedure. 
In our experiments, initial values estimated by \texttt{conquer} outperform those by the burn-in method because \texttt{conquer} 
\changedSL{tends} to produce high-quality initial solutions.




\subsection{Choice of the learning rate}\label{sec:rule-of-thumb}

We introduce a practical guideline for determining the value of $\gamma_i = \gamma_0 i^{-a}$. 
Let $\widehat{\sigma}$ denote the sample standard deviation of $y_i$ using subsample $\mathcal{S}$. 
Then, we suggest to fix $a = 0.501$ and select the initial learning rate $\gamma_0$
via
\begin{align}\label{gamma0:def}
   \gamma_0 = \frac{1}{\widehat{\sigma}} \frac{\phi( \Phi^{-1}(\tau)) }{\sqrt{\tau(1-\tau)}},
\end{align}
where $\phi(\cdot)$ and $\Phi(\cdot)$ are standard normal density and distribution functions, respectively. For example, if $\tau = 0.5$, then $\gamma_0 \approx 0.798/\widehat{\sigma}$;
if $\tau = 0.1$ or $0.9$, then $\gamma_0 \approx 0.585/\widehat{\sigma}$. The rule-of-thumb selection of $\gamma_0$ is heuristically motivated from the asymptotic distribution of the sample quantile when the regression errors are normally distributed.

\subsection{Subvector inference} In most empirical studies, while many covariates are being controlled in the model, of interest are often just one or two key independent variables that are policy-related. In such cases, we are particularly interested in  making inference for subvectors, say $\beta^*_1$, of the original vector $\beta^*$. Subvector inference is not often direct target of interest in the quantile regression literature, partially because inference regarding the full vector $\beta^*$ and then converting to the subvector is often scalable up to the medium sample size. But this is no longer the case for the ultra-large sample size that is being considered here. 

It is straightforward to tailor our updating algorithm to focusing on subvectors. Continue denoting  $\beta_i$ as the $i$-th update of the full vector. While the full vector $\beta_i$ still needs to be computed in the S-subGD, both the Polyak-Ruppert average and the random scale can  be updated only up to the scale of the subvector. Specifically, let $\bar\beta_{i, sub}$ denote the subvector of $\bar\beta_i$, corresponding to the subvector of interest. Also, let $\widehat V_{i,sub}$ denote the sub-matrix of $\widehat V_i$; both are updated according to the following rule: 
\begin{eqnarray}
\beta_i&=& \text{updated the same as (\ref{eq.7beta})}, \\
\bar\beta_{i,sub}&=& \bar\beta_{i-1, sub}\frac{i-1}{i} + \beta_{i,sub}\frac{1}{i}, \label{eqa7-sub}\\
\widehat V_{i,sub}&=& i^{-2}\left(A_{i,sub}-\bar\beta_{i,sub} b_{i,sub}' -b_{i,sub}\bar\beta_{i,sub}'+\bar\beta_{i,sub}\bar\beta_{i,sub}'\sum_{s=1}^is^2\right), \\
A_{i,sub}&=&A_{i-1,sub} +i^2\bar\beta_{i,sub}\bar\beta_{i,sub}',\quad b_{i,sub}= b_{i-1,sub} + i^2\bar\beta_{i,sub}.\label{end123}
\end{eqnarray}
In most cases, steps (\ref{eqa7-sub})-(\ref{end123}) only involve small dimensional objects.


\subsection{Computational complexity}\label{sec:comp:complexity}

\citet{Portnoy:Koenker:97} demonstrated that the computational complexity of the interior point algorithm for quantile regression can be on the order of $\mathcal{O} (n d^3 \log^2 n)$ (see discussions after Theorem 5.1 of \citet{Portnoy:Koenker:97}).
They further proposed an initial phase of preprocessing that effectively reduces computational complexity to $\mathcal{O} (n^{2/3} d^3 \log^2 n) + \mathcal{O} (nd)$ \cite[see][p. 290]{Portnoy:Koenker:97}. The basic idea behind preprocessing is that using a preliminary estimate of $\beta^*$, one can remove observations that are above and below the quantile regression line with high probability and add two pseudo-observations called ``globs'', resulting in the reduction of the sample size from $n$ to $\mathcal{O} (n^{2/3} )$.\footnote{It is an interesting future research topic to examine whether and how preprocessing can be used in our inference procedure.}
These results indicate that the interior point algorithm works well with a large $n$ but not with a large $d$. In addition, \citet{Chen:07:JCGS} developed a smoothing algorithm that has the computational complexity of  $\mathcal{O} (n d^{2.5})$, which reduces the dependence on $d$ from $d^3$ to $d^{2.5}$. 
In \citet{HE2021}, \texttt{conquer} is computed using gradient descent (GD), which has the computational complexity of $\mathcal{O}(n d k)$, where $k$ is the number of iterations in GD. 

Table~\ref{tb_comp_cost} summarizes the computational cost of estimation and inference.  Importantly, the computational gain of our stochastic subgradient method is \textit{not} in terms of point estimation, but of inference.  Specifically, our S-subGD method has the computational complexity of $\mathcal{O}(n d)$. If the GD algorithm of \texttt{conquer} converges fast in a finite number of iterations, which seems to be the case empirically, we expect that our algorithm will be comparable---will not be superior---to \texttt{conquer} in terms of point estimation. 
However, note that GD method can suffer from memory constraints when $n d$ is very large.

\begin{table}[htbp]
\centering
\caption{Computational Complexity for Different Methods} 
\label{tb_comp_cost}
\begin{tabular}{lr}
  \hline
\multicolumn{2}{r}{Computational Cost for Estimation}  \\ 
  \hline \\
  \vspace{0.5pt}
 QR (interior point)  &  $\widetilde{\mathcal{O}}( n d^3)$ \\
 QR (interior point with preprocessing)  &  $\widetilde{\mathcal{O}}( n^{2/3} d^3) + \mathcal{O}( n d)$ \\ 
 CONQUER (GD) &  $\mathcal{O}(n d k)$ \\
 S-subGD & $\mathcal{O}(n d)$ \\ 
   \hline     
 \multicolumn{2}{r}{Additional Computational Cost for Inference} \\ 
  \hline \\
  \vspace{0.5pt}
 QR (plugin)  &  $\mathcal{O}( n d^2)$ \\
 QR (bootstrap with preprocessing)  &    $\widetilde{\mathcal{O}}( b n^{2/3} d^3) + \mathcal{O}( b n d)$ \\
 CONQUER (plugin) &  $\mathcal{O}( n d^2)$ \\
 CONQUER (bootstrap) &  $\mathcal{O}(b n d k)$ \\
 S-subGD (online bootstrap) & $\mathcal{O}(b n d)$  \\ 
 S-subGD (random scaling) &  $\mathcal{O}(n d^2)$ \\ 
 S-subGD (subvector random scaling) & $\mathcal{O}(n  s^2)$ \\ 
 S-subGD (diagonal random scaling) & $\mathcal{O}(n  d)$ \\
   \hline
\end{tabular}
\vspace{0.5pt}
\parbox{5in}{
Notes. The symbol $\widetilde{\mathcal{O}}(\cdot)$ does not explicitly show logarithmic factors. 
Here, 
$n$ is the sample size;
$d$ is the dimension of regressors; 
$k$ is the number of iterations in CONQUER;
$b$ is the number of bootstrap draws;
$s$ is the dimension of the subvector of interest.
}
\end{table}


However, the comparative advantage is more pronounced in statistical inference, which incurs additional computational costs. Our inference, based on random scaling, entails an additional cost of order $\mathcal{O} (n d^2)$. Nevertheless, it reduces to $\mathcal{O} (n d)$ when the object of inference is a subvector of $\beta^*$, with a dimension of $s = \mathcal{O} (d^{1/2})$. This reduction occurs because we only need to update a submatrix of the random scaling matrix, as discussed in the previous section. Similar gains are achieved when focusing on the inference of each element separately, as is common in applied practice. This is again because we only need to update the diagonal elements, referred to as diagonal random scaling in Table~\ref{tb_comp_cost}. We emphasize that achieving the same order of computational cost for subvector inference is generally not possible with plug-in inference. This is due to the fact that the plug-in estimator of the sandwich formula $(H^{-1}\Sigma H^{-1})$ of asymptotic variance in \eqref{asymptotic-normality-H} cannot be computed using only diagonal elements or a submatrix.

As for the computational cost of competing methods for QR inference, first,  the bootstrap inference will be time consuming as the computational cost will be a multiple of bootstrap draws. For instance, bootstrap inference with \texttt{conquer} has  the computational complexity of $\mathcal{O}(b n d k)$, where $b$ is the number of bootstrap draws. Second, it requires access to the entire sample to conduct standard plug-in inference based on asymptotic normality. Thus, the plug-in inference has the additional computational complexity of $\mathcal{O} (n d^2)$ even with the \texttt{conquer} algorithm. This will be faster than bootstrap inference if $k > d$,  but cannot be reduced to $\mathcal{O} (n s^2)$ or $\mathcal{O} (n d)$ even for subvector inference.

In short, our algorithm can be as competitive as \texttt{conquer} in terms of computational cost, while offering the advantages of lower memory requirements and faster computations for subvector inference.

\section{Asymptotic Theory}\label{sec:theory}

In this section, we   present asymptotic theory that underpins our inference method.
We consider the following QR model:
$$
y_{i}= x_i'\beta^*_\tau + \varepsilon_i,
$$
where $P(\varepsilon_i\leq 0|x_i)= \tau$ for a fixed quantile $\tau \in(0,1)$. Denote by $\beta^*=\beta_{\tau}^*$ for simplicity.

\begin{asm}[Conditions for Quantile Regression]\label{asm:qr}
Suppose:
\begin{enumerate}[(i)]	

\item  The data  $\{ Y_i \equiv (y_i, x_i) \}_{i=1}^n$ are independent and identically distributed.

\item  The conditional density $f_{\varepsilon}(\cdot|x_i)$  of $\varepsilon_i$ given $x_i$ 
and its partial derivative
$\frac{d}{d\varepsilon}f_{\varepsilon}(\cdot|x_i)$ exist, and furthermore, $\sup_b\mathbb E [ \|x_i\|^3A(b, x_i) ] <C$ for some constant $C < \infty$, where  $$A(b, x_i) :=   \left|\frac{d}{d\varepsilon}f_{\varepsilon}(x_i'b|x_i)\right| + f_{\varepsilon}(x_i'b|x_i).$$ 

\item 
 There exist positive constants $\epsilon$ and $c_0$ such that 
  	$$
  \inf_{  | \beta-\beta^* |<\epsilon }\lambda_{\min} \left(  \mathbb E [x_ix_i'f_{\varepsilon}(x_i'(\beta-\beta^*)|x_i)] \right) > c_0.
  	$$

\item $\mathbb E[ (\|x_i\|^6+1)\exp(\|x_i\|^2) ]<C$ for some constant $C < \infty$.

\item $\gamma_{i}=\gamma_0 i^{-a} $ for some $1/2 < a <1$.





\end{enumerate}
 \end{asm}


Conditions (ii)  imposes some moment conditions on the conditional density, and is satisfied if 
both $f_{\varepsilon}(\cdot|x_i)$ and $\frac{d}{d\varepsilon}f_{\varepsilon}(\cdot|x_i)$ are uniformly bounded along with a moment condition on $x_i$. Both conditions (ii) and (iii)  are imposed here to establish the consistency of $\bar\beta_n$. 

Condition (iii) can be viewed as a local  identifiability condition, which ensures that the population loss function is locally strongly convex, and is reasonable as $\beta \mapsto q(\beta, Y_i)$ is convex.

Condition (iv) requires a tail condition on the regressors. It is stronger than the usual moment conditions for standard QR inference, but is required by both  consistency and the asymptotic inference for S-subGD. 
In the literature, it is common to adopt stronger conditions on the regressors than on the error depending on the context.
For example, \cite{zheng2018} and \cite{Chen:Lee:23} assumed the uniform boundedness of each of the covariates; the former paper emphasized that 
a linear quantile regression model with compact support for the regressors is most sensible to avoid the problem of quantile crossing.

One of the key technical steps in  the proof of such   one-observation-at-a-time updating, is to  show that the nonlinear  updating path is \textit{linearizable}. Specifically, define $\Delta_i= \beta_i- \beta^*$, then the S-subGD updating   can be rewritten using the estimation path:
\[
\Delta_{i}=\Delta_{i-1}-\gamma_{i}\nabla Q\left(\beta_{i-1}\right)+\gamma_{i}\xi_{i},
\]
where $\xi_i= \nabla Q(\beta_{i-1}) - \nabla q(\beta_{i-1}, Y_i)$ is  a martingale difference sequence.  We then show that the  path of $\Delta_{i}$ can be approximated by a linear path: 
\[
\Delta_{i}^{1}:=\Delta_{i-1}^{1}-\gamma_{i}H\Delta_{i-1}^{1}+\gamma_{i}\xi_{i}\quad\text{and}\quad\Delta_{0}^{1}=\Delta_{0},
\]
where $H=\nabla^2 Q(\beta^*)$ is the population Hessian matrix. Such an \textit{linearization}, as one of the technical  steps in the proof,  transfers the study of nonlinear SGD/S-subGD problems into linear problems.

Condition (v) gives a requirement for the learning rate, which is standard for \textit{online learning}.  \cite{polyak1992acceleration} showed that $a<1$ is needed for the learning rate to decrease   sufficiently slow so that  the asymptotic normality of the  Polyak-Ruppert averaging   estimator holds. Meanwhile, for nonlinear models as in the QR model, it should not degenerate too  slowly  so $a>1/2$ is also needed.

 We extend \citet{lee2021fast} (who assumed the sample loss function to be smooth and globally convex) to the quantile regression model, and establish a functional central limit theorem (FCLT) for the aggregated solution path of the S-subGD.   Under Assumption~\ref{asm:qr}, we  establish the following  FCLT:
\begin{equation}\label{eq4}
\frac{1}{\sqrt{n}}\sum_{i=1}^{\left[nr\right]}\left(\beta_{i}-\beta^{*}\right)\Rightarrow \Upsilon^{1/2}W\left(r\right),\quad r\in\left[0,1\right],
\end{equation}
where $\Rightarrow$ stands for the weak convergence in $\ell^{\infty}\left[0,1\right]$,
$W\left(r\right)$ stands for a vector of the independent standard
Wiener processes on $\left[0,1\right]$,
and
$ \Upsilon := H^{-1}SH^{-1}$, with 
 $
 S := \mathbb E [x_i x_i'] \tau(1-\tau)
 $ and $H := \mathbb E [x_i x_i' f_{\varepsilon}(0|x_i)]$. 
 
The FCLT in \eqref{eq4} states that the partial sum of the sequentially updated estimates $\beta_i$ converges weakly to a rescaled Wiener process, with the scaling matrix equal to a square root of the asymptotic variance of the usual quantile regression estimator $\widehat{\beta}_n$. 
Building on the FCLT, we propose
a large-scale inference procedure.

For any $\ell \leq d$ linear restrictions
\[
H_{0}: R\beta^{*} = c,
\]
where $R$ is  an $(\ell \times d)$-dimensional known matrix of  rank $\ell$
and
$c$ is an $\ell$-dimensional known vector,
the conventional Wald test based
on $\widehat{V}_{n}$ is asymptotically pivotal. 
We formally state the main theoretical \changed{result} in the following theorem.

\begin{thm}[Main Theorem]
\label{thm:Wald:qr} Suppose that
$H_{0}: R\beta^{*} = c$ holds with 
$\mathrm{rank}(R)=\ell$. Under Assumption~\ref{asm:qr},
the FCLT in (\ref{eq4}) holds and 
\begin{eqnarray*}
&&n\left(R\bar{\beta}_{n}-c\right)'\left(R\widehat{V}_{n}R'\right)^{-1}\left(R\bar{\beta}_{n}-c\right)\overset{d}{\to}W\left(1\right)'\left(\int_{0}^{1}\bar{W}(r)\bar{W}(r)'dr\right)^{-1}W\left(1\right),
\end{eqnarray*}
where $W$ is an $\ell$-dimensional vector of the standard Wiener
processes and $\bar{W}\left(r\right):=W\left(r\right)-rW\left(1\right)$. 
\end{thm}

As an important special case of Theorem \ref{thm:Wald:qr}, the t-statistic defined in \eqref{t-stat} converges in distribution to the following pivotal limiting distribution:
for each $j = 1,\ldots,d$, 
\begin{align}\label{t-stat-limit}
\frac{\sqrt{n}\left(\bar{\beta}_{n,j}-\beta_{j}^{*}\right)}{\sqrt{\widehat{V}_{n,jj}}}
\overset{d}{\to}
W_1\left(1\right)  \left[ \int_{0}^{1} \left\{ W_1\left(r\right)-rW_1\left(1\right) \right\}^2 dr\right]^{-1/2}, 
\end{align}
where 
 $W_1$ is a one-dimensional standard Wiener process. 

\subsection{Testing for homegeneity between two quantiles} 
Next, consider two quantiles $\tau_1, \tau_2\in(0,1)$, and let $\beta_{\tau_1}^*$ and $\beta_{\tau_2}^*$ denote the corresponding coefficients at these two quantiles.  We are interested in testing the heterogeneity at the two quantile levels:
$$
H_0: \beta_{\tau_1, sub}^*=\beta_{\tau_2, sub}^*. 
$$
which is to test whether  particular subvectors of the  coefficients at the two quantile levels are the same. For instance, the subvector could be the slope of a particular regressor of interest, and we are interested in testing whether its quantile effect stays the same  when the quantile level changes from  $\tau_1$ to  $\tau_2$.

Both subvector coefficients can be iteratively estimated using  S-subGD, whose stochastic paths $(\beta_i^1, \bar\beta_{i,sub}^1)$ and $(\beta_i^2, \bar\beta_{i,sub}^2)$  are constructed following  the Algorithm (\ref{eqa7-sub})-(\ref{end123}). In particular,  the stacked random scaling matrix $\bar V_{i,sub}$ corresponding to the two subvectors $(\bar\beta_{i,sub}^1, \bar\beta_{i,sub}^2)$ is updated as:
\begin{eqnarray*}
\theta_i&:=&\begin{pmatrix}
\bar\beta_{i,sub}^1\\
\bar\beta_{i,sub}^2
\end{pmatrix}, \cr
A_{i,sub}&=&A_{i-1,sub} +i^2\bar\theta_{i,sub}\bar\theta_{i,sub}',\quad b_{i,sub}= b_{i-1,sub} + i^2\bar\theta_{i,sub}, \cr
\bar V_i&=&   i^{-2}\left(A_i-\bar\theta_{i,sub} b_{i,sub}' -b_{i,sub}\bar\theta_{i,sub}'+\bar\theta_{i,sub}\bar\theta_{i,sub}'\sum_{s=1}^is^2\right).
\end{eqnarray*}
In this case, we let $G:=(I, -I)$, and form a test statistic:
$$
 n(\bar\beta_{n,sub}^1-\bar\beta_{n,sub}^2)' (G\bar V_{n,sub}G')^{-1}(\bar\beta_{n,sub}^1-\bar\beta_{n,sub}^2) .
$$
We reject the null if the test statistic is large. 
The following theorem establishes the asymptotic null distribution. 
\begin{thm} 
\label{thm:d} Suppose that
$H_{0}:\beta^*_{\tau_1,sub} = \beta^*_{\tau_2,sub} $ holds. Under Assumption~\ref{asm:qr}, 
\begin{eqnarray*}
&&  n(\bar\beta_{n,sub}^1-\bar\beta_{n,sub}^2)' (G\bar V_{n,sub}G')^{-1}(\bar\beta_{n,sub}^1-\bar\beta_{n,sub}^2) \overset{d}{\to}W\left(1\right)'\left(\int_{0}^{1}\bar{W}(r)\bar{W}(r)'dr\right)^{-1}W\left(1\right),
\end{eqnarray*}
where $W$ is an $2\times \dim(\beta^*_{\tau_1,sub})$-dimensional vector of the standard Wiener
processes and $\bar{W}\left(r\right):=W\left(r\right)-rW\left(1\right)$. 
\end{thm}

\section{Monte Carlo Experiments}\label{sec:MC}

In this section we investigate the performance of the S-subGD random scaling method via Monte Carlo experiments. 
The main question of these numerical experiments is whether the proposed method is feasible for a large scale model. 

The simulation is based on the following data generating process:
\begin{align*}
y_{i} =  x_i'\changed{\beta^*} + \varepsilon_i~~\mbox{for}~~i=1,\ldots,n,
\end{align*}
where $x_i$ is a $(d+1)$-dimensional covariate vector whose first element is 1 and the remaining $d$ elements are generated from  $\mathcal{N} (0,I_d)$. The error term $\varepsilon_i$ is generated from $\mathcal N(0,1)$, and the parameter value $\beta^*$ is set to be $(1,\ldots,1)$. 
Without loss of generality, we focus on the second element, say $\beta^*_{(2)}$, of $\beta^*$. 
That is, we focus on subvector inference: the confidence interval for $\beta^*_{(2)}$ as well as testing the null hypothesis of $\beta^*_{(2)} = 1$.
To accommodate a large scale model, the sample size varies in $n \in \{10^5, 10^6, 10^7\}$, and the dimension of $x$ varies in $d\in \{10, 20, 40, 80, 160, 320\}$ in the first set of experiments. Later, we extend the dimension of $x$ into $d\in \{500, 750, 1000, 1500\}$.
The initial value $\beta_0$ is estimated by the convolution-type smoothed quantile regression (\texttt{conquer}) in \citet{HE2021} and we do not burn in any observations. 
The learning rate is set to be $\gamma=\gamma_0 t^{-a}$ with $\gamma_0=1$ and $a=0.501$.
The simulation results are summarized from $1000$ replications of each design.

We compare the performance of the proposed method with four additional alternatives:

\begin{enumerate}
\item[(i)] S-subGD: the proposed S-subGD random scaling method.

\item[(ii)] QR: the classical quantile regression method. 
Given the scale of models, we apply the Frisch-Newton algorithm after preprocessing and the Huber sandwich estimate under the local linearity assumption of the conditional density function by selecting `pfn' and `nid' options in R package \texttt{quantreg} (CRAN version 5.88). 

\item[(iii)]  CONQUER-plugin: estimates the parameter using  the conquer method, and estimates the asymptotic variance by plugging in the parameter estimates; implemented using the R package \texttt{conquer} (CRAN version 1.3.0).

\item[(iv)]  CONQUER-bootstrap:  estimates the parameter using the conquer method, and applies the multiplier bootstrap method (see \citet{HE2021} for details); implemented using the R package \texttt{conquer}. We set the number of bootstrap samples as 1000.

\item[(v)] SGD-bootstrap: estimates using the S-subGD method and conduct the inference by an online bootstrap method (see \citet{fang2018online} for details). We set the number of bootstrap samples as 1000.

\end{enumerate}

\begin{figure}[!htbp]
    \caption{Computation time} \label{fig:time}
    \centering
    \includegraphics[width=\textwidth]{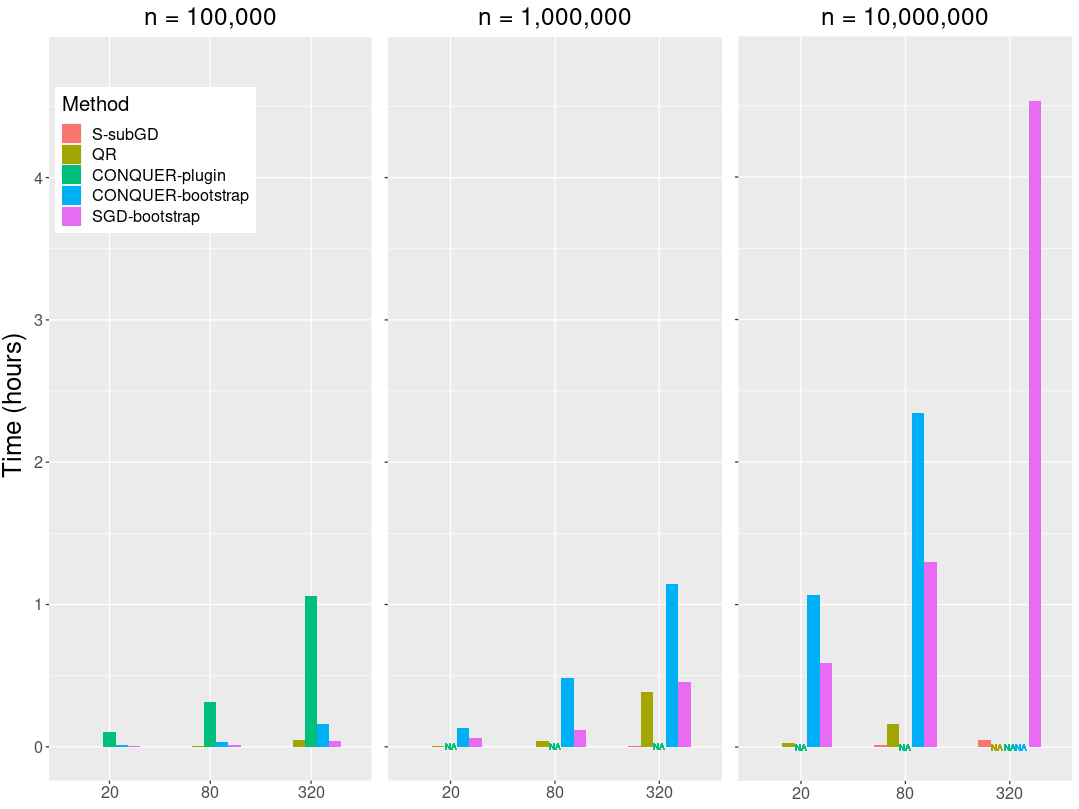}
    \includegraphics[width=\textwidth]{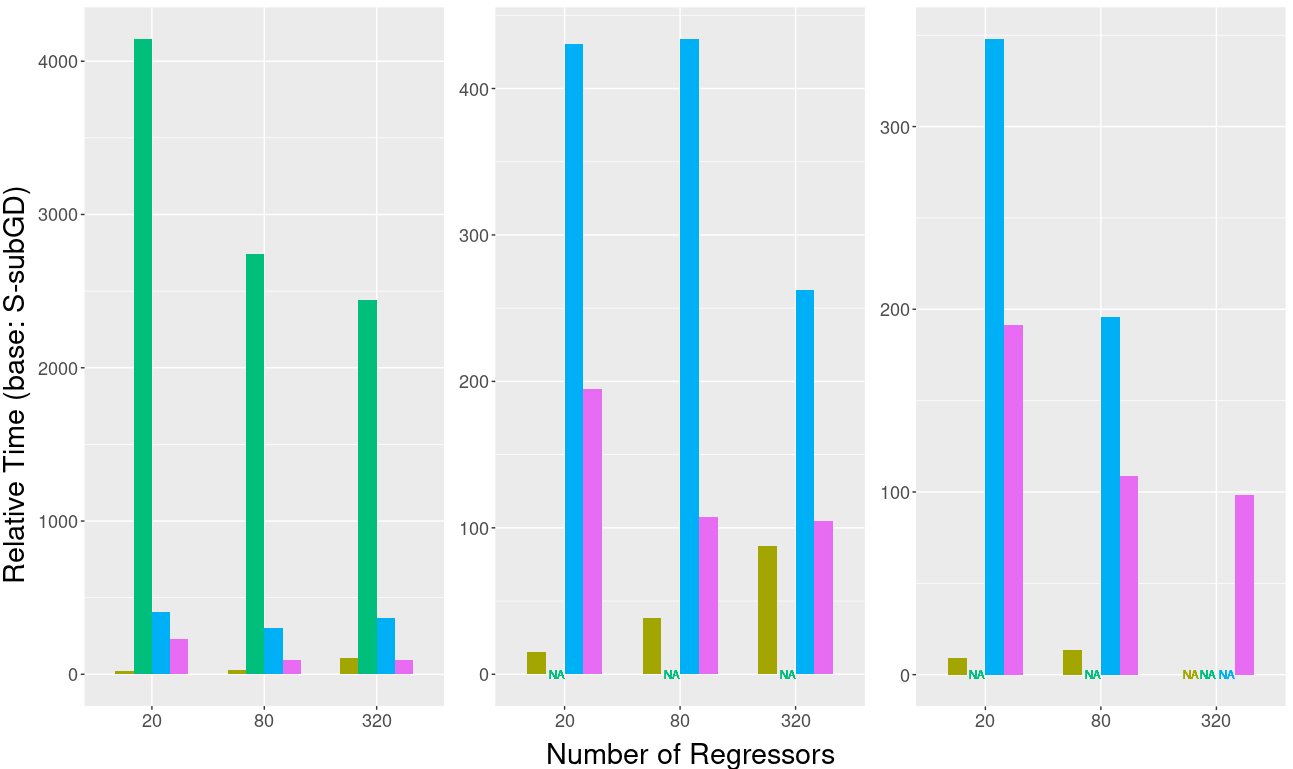}
\end{figure}

\begin{figure}[!htbp]
    \caption{Coverage and CI Length} \label{fig:ci}
    \centering
    \vskip10pt
    \includegraphics[width=\textwidth]{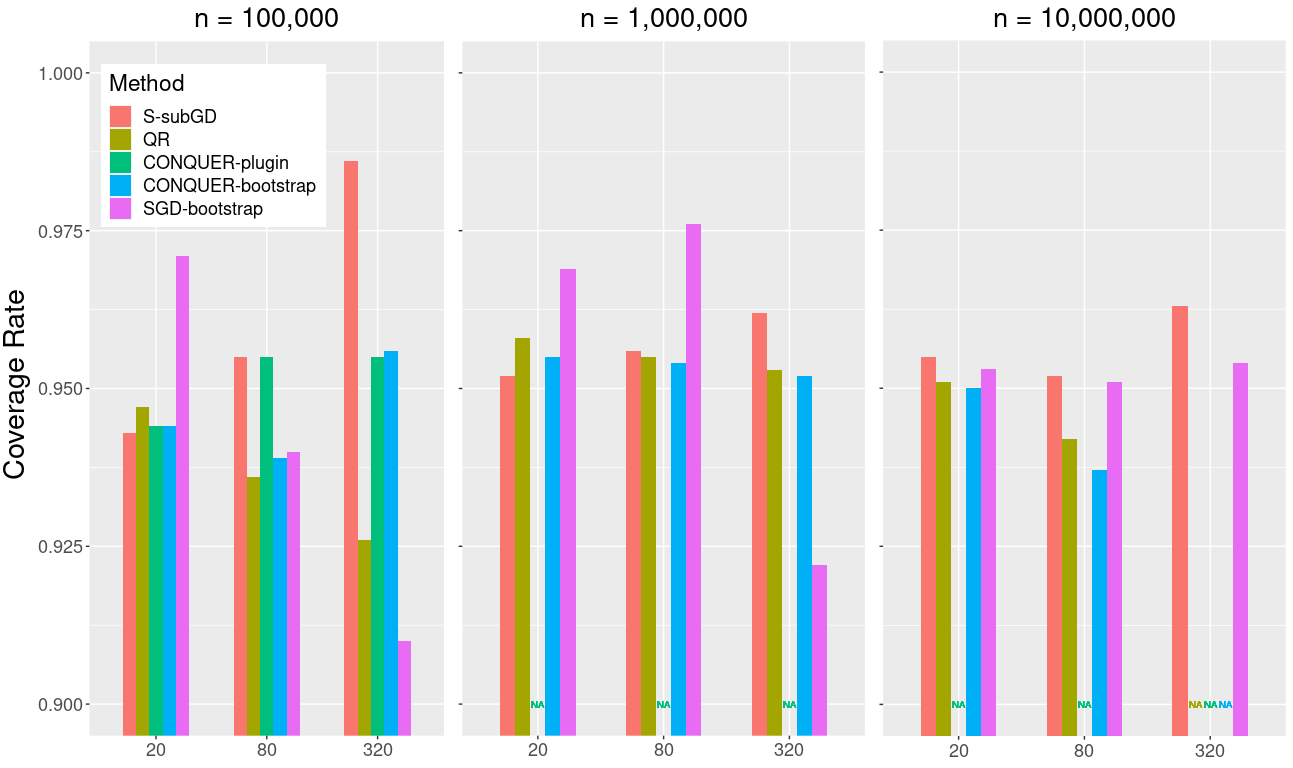}
    \includegraphics[width=\textwidth]{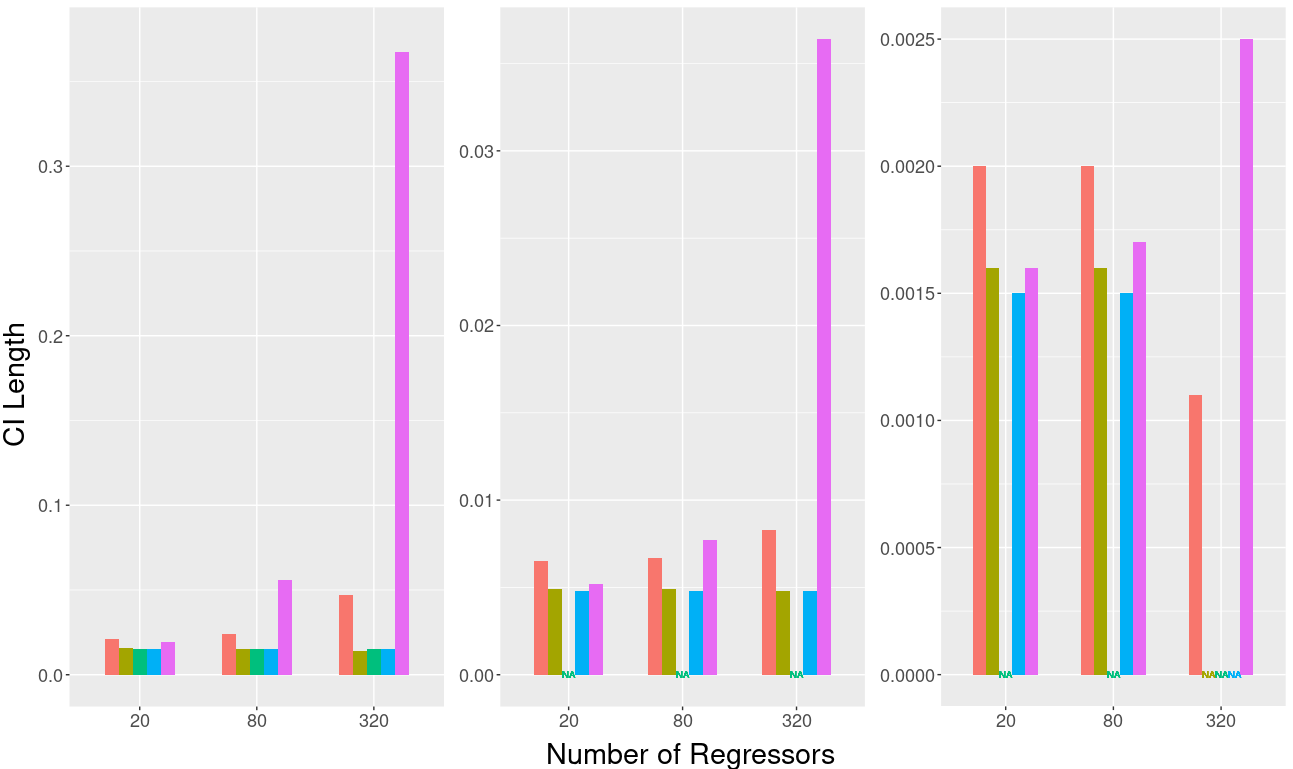}
\end{figure}

We use the following performance measures: the computation time, the coverage rate, and the length of the 95\% confidence interval.
Note that the nominal coverage probability is 0.95.
The computation time is measured by 10 replications on a separate desktop computer, equipped with an AMD Ryzen Threadripper 3970X CPU (3.7GHz) and a 192GB RAM. 
Other measures are computed by 1000 replications on the cluster system composed of several types of CPUs. 
We set the time budget as 10 hours for a single estimation and the RAM budget as 192GB.

Figures \ref{fig:time}--\ref{fig:ci} summarize the simulation result. To save space, we report the results of $d=20, 80,$ and $320$. We provide tables for all the simulation designs in the appendix. 
First, we observe that S-subGD is easily scalable to a large simulation design. On the contrary, some methods cannot complete the computation within the time/memory budget. For instance, CONQUER-plugin does not work for any $d$ when $n=10^6$ or $n=10^7$ because of the memory budget. QR shows a similar issue when $n=10^7$ and $d=320$. CONQUER-bootstrap cannot meet the 10-hour time budget for a single replication when $n=10^7$ and $d=320$. They are all denoted as `NA' in the figures.

Second, S-subGD outperforms the alternative in terms of computational efficiency. The bottom panel of Figure \ref{fig:time} shows the relative computation time, and 
\changedSL{the relative speed improvement is by a factor of 10 or 100.}
In case of CONQUER-plugin, it is slower than S-subGD in the scale of 1000's. If we convert them into the actual computation time,  S-subGD takes 16 seconds for the design of $n=10^6$ and $d=320$. However, QR, CONQUER-bootstrap, and SGD-bootstrap take 1374 seconds (22 minutes 54 seconds), 4113 seconds (1 hour 8 minutes 33 seconds), and 1643 seconds (27 minutes 23 seconds) on average of 10 replications.  

Third, all methods shows satisfactory results in terms of the coverage rate and the length of the confidence interval. The top panel of Figure \ref{fig:ci} reports that coverage rates are all around 0.95 although S-subGD and SGD-bootstrap are slightly under-reject and over-reject when $d=320$. The average lengths of the confidence interval are reported on the bottom panel. The performance of S-subGD is comparable to QR, CONQUER-plugin, and CONQUER-bootstrap. SGD-bootstrap shows much larger CI lengths, especially when $d=320$.

\begin{table}[htb]
\centering
\caption{Performance of S-subGD: $n=10^7$} 
\label{tb_SGD_rs}
\begin{tabular}{c S[table-format=3.2]S[table-format=1.3]S[table-format=1.4]}
  \hline
{$d$} & {Time (sec.)} & {Coverage Rate} & {CI Length} \\ 
  \hline
   10 & 4.80 & 0.965 & 0.0020 \\ 
   20 & 6.42 & 0.955 & 0.0020 \\ 
   40 & 11.02 & 0.954 & 0.0020 \\ 
   80 & 21.40 & 0.952 & 0.0020 \\ 
  160 & 29.87 & 0.953 & 0.0021 \\ 
  320 & 59.73 & 0.963 & 0.0011 \\ 
  500 & 154.12 & 0.945 & 0.0172 \\
  750 & 208.02 & 0.934 & 0.0359 \\
 1000 & 302.78 & 0.923 & 0.0461 \\ 
 1500 & 505.01 & 0.931 & 0.0645 \\ 
   \hline
\end{tabular}
\end{table}

In the second set of experiments, we stretch the computational burden by increasing the size of $d$ up to $d=1500$ when $n=10^7$. Because of the challenging model scale, we use the cluster system for these exercises and check only the performance of S-subGD focusing on inference of a single parameter. Table \ref{tb_SGD_rs} reports the simulation result along with different sizes of $d$. S-subGD still completes the computation within a reasonable time range. The average computation time is under 505 seconds (8 minutes 25 seconds) even for the most challenging design. The coverage rate and CI length becomes slightly worse than those of smaller $d$'s but they still fall in a satisfactory range. 

In sum, S-subGD performs well when we conduct a large scale inference problem in quantile regression. It is scalable to the range where the existing alternatives cannot complete the computation within a time/memory budget. In addition to computational efficiency, the coverage rate and CI lengths are also satisfactory. Thus, S-subGD provides a solution to those who hesitate to apply quantile regression with a large scale data for the computational burden.

\section{Inference for the College Wage Premium}\label{sec:empirical}

The study of the gender gap in the college wage premium has been a long-standing important question in labor economics.   The literature has pointed out a stylized fact that the higher college wage premium for women  as the major cause for attracting more women to attend and graduate from colleges than men (e.g., \cite{goldin2006homecoming, chiappori2009investment}). Meanwhile, \citet{hubbard2011phantom} pointed out  a bias issue associated with the ``topcoded'' wage data, where the use of the sample from Current Population Survey (CPS) often censors the wage data at a maximum value. As  a remedy for it, \citet{hubbard2011phantom} proposed to use  quantile regression that is robust to censoring. Analyzing the CPS data during 1970-2008, he found no gender difference in the college wage premium later years once the topcoded bias has been accounted for.

We revisit this problem by estimating and comparing the college wage premium between women and men.   While the data also contains the topcoded issue, the quantile regression is less sensitive to the values of upper-tail wages.  By applying the new S-subGD  inference to the ultra-big dataset we are using,  we aim at achieving the following goals in the empirical study: (1) \changedSL{to} identify (if any) the  heterogeneous effects across quantiles; (2) \changedSL{to} understand the trends in the college wage premium respectively for female and male; and (3) \changedSL{to} understand the gender difference in the college wage premium.

We use the data from IPUMS USA \citep{IPUMS} that contains several millions of workers. The main motivation of using  ultra-large datasets for wage regressions is to deal with high-dimensional regressors within a simple parametric quantile regression model. As has been pointed out in the literature,  work experience is 
 an important factor in wage regressions 
 but is difficult to be measured precisely. \changedSL{As career paths may be different across states and gender, one possible means of controlling for the experience is to flexibly interact workers' age, gender, and} state fixed effects, which would create over one thousand regressors\changed{. Thus,} a very large sample size would be desirable to obtain precise estimates. 
But the ultra-large dataset makes most existing inference procedures for quantile regression fail to work. 

\subsection{The data}

We use the samples over six different years (1980, 1990, 2000-2015) from IPUMS USA at \url{https://usa.ipums.org/usa/}. 
In the years from 1980 to 2000, we use the 5\% State sample which is a 1-in-20 national random sample of the population.
In the remaining years, we use the American Community Survey (ACS) each year. The sampling ratio varies from 1-in-261 to 1-in-232 in 2001-2004, but it is set to a 1-in-100 national random sample of the population after 2005. 
To balance the sample size, we bunch the sample every 5 year after 2001. 
We also provide the estimation results using separate but smaller samples in 2005, 2010, and 2015 in the appendix, which is similar to those reported in this section. 

We restrict our sample to $White$, $18 \le Age \le 65$,  and $Wage \ge \$62$, which is a half of minimum wage earnings in 1980 ($\$3.10 \times 40 \mbox{hours} \times 1/2$). 
$Wage$ denotes the implied weekly wage that is computed by dividing yearly earnings by weeks worked last year. 
Since 2006, weeks worked last year are available only as an interval value and we use the midpoint of an interval.
We only consider full-time workers who worked more than 30 hours per week.
Then, we compute the \textit{real} wage using the personal consumption expenditures price index (PCEPI) normalized in 1980. 
The data cleaning leaves us with 3.6-4.7 million observations besides 2001-2005, where we have around 2.5 million observations.
Table \ref{tb:summary_stat} reports some summary statistics on the key variables, where $Educ$ denotes an education dummy for some college or above. 
The table confirms that female college education has increased substantially over the years and female workers received more college education than male workers since 1990.

\begin{table}[ht]
\caption{Summary Statistics} \label{tb:summary_stat}
\centering
\begin{tabular}{crcccc}
  \hline
Year & Sample Size & $\mathbb E(Female$) & $\mathbb E(Educ$) & $\mathbb E(Educ$|$Male$) & $\mathbb E(Educ$|$Female$) \\ 
  \hline
  1980      & 3,659,684 & 0.390 & 0.433 & 0.444 & 0.416 \\ 
  1990      & 4,192,119 & 0.425 & 0.543 & 0.537 & 0.550 \\ 
  2000      & 4,479,724 & 0.439 & 0.600 & 0.578 & 0.629 \\ 
  2001-2005 & 2,493,787 & 0.447 & 0.642 & 0.619 & 0.670 \\ 
  2006-2010 & 4,708,119 & 0.447 & 0.663 & 0.631 & 0.701 \\ 
  2011-2015 & 4,542,874 & 0.447 & 0.686 & 0.646 & 0.735 \\ 
   \hline
\end{tabular}
\end{table}

\subsection{The quantile regression model}

We use the following baseline model:
\begin{footnotesize}
\begin{align*}
    \log(Wage_i) = \beta_0 + \beta_1 Female_i + \beta_2 Educ_i + \beta_3 Female_i\cdot Educ_i + \theta_1'X_i + \theta_2' (X_i\cdot Female_i) + \varepsilon_i,
\end{align*}
\end{footnotesize}where $Wage_i$ is a real weekly wage in 1980 terms, $Female_i$ is a female dummy, $Educ_i$ is an education dummy for some college or above, and $X_i$ is a vector of additional control variables. 
For control variable $X_i$, we use 12 age group dummies with a four-year interval, 51 states dummies (including D.C.), and their interactions. Note that $(X_i\cdot Female_i)$ implies that there exist up to 3-way interactions. 
The model contains 1226 covariates in total. 
We also add 4 additional year dummies for the 5-year combined samples after 2001. 
We estimate the model using the proposed S-subGD method over 9 different quantiles: $\tau=0.1,0.2,\ldots,0.9$. 
We obtain the starting value of S-subGD inference by applying \texttt{conquer} to the 10\% subsample.\footnote{Recall that $X_i$ contains a large number of dummy variables in this empirical application. If the subsample size is too small, \texttt{conquer} may face a singularity issue and cannot compute the initial value. Alternatively, one may use a randomazied algorithm of \citet{Yang:QR:ICML}.
} 
For the learning rate $\gamma_0 t^{-a}$, we set $a=0.501$ and $\gamma_0$ by the rule-of-thumb approach described in Section \ref{sec:rule-of-thumb}. As an sensitivity analysis, we also vary $a$ over 17 equi-spaced points in $[0.501,0.661]$. The results are quite robust to the choice of $a$ as reported in the online appendix. 
We use the Compute Canada cluster system equipped with the AMD Milan CPUs whose clock speed is 2.65 GHz and with 650 GB of RAMs. 
For comparison, we also try to estimate the model using the standard QR method with the `quantreg' R package, but it fails to compute the result in the given cluster environments.

\subsection{The results}

\begin{figure}[!htbp]
    \centering
    \caption{College Wage Premium: 
    Combining 5-Year Data 
    }
    \label{fig:premium_design_5yr_full}
    \hskip15pt
    \begin{tabular}{c c c}
        \includegraphics[scale=0.23]{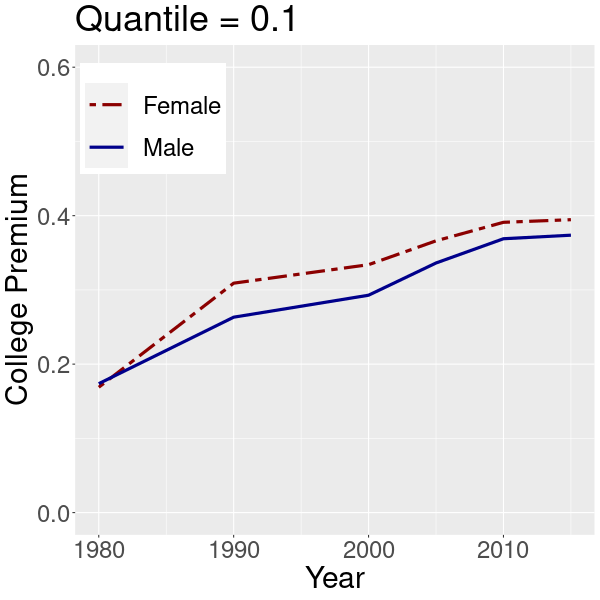} & \includegraphics[scale=0.23]{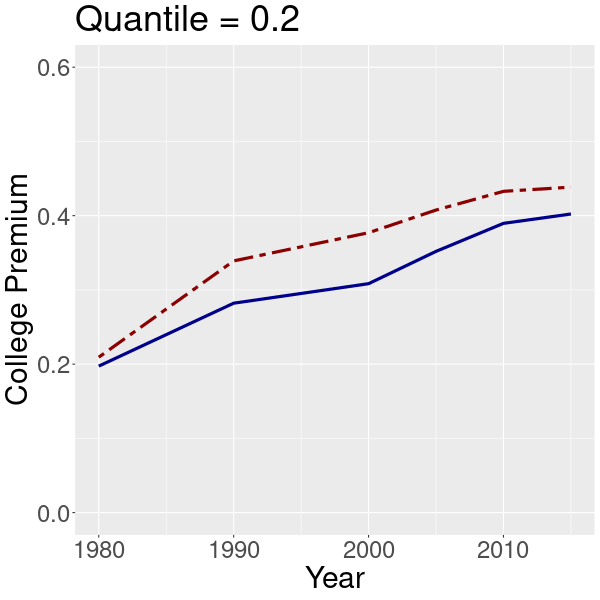} & \includegraphics[scale=0.23]{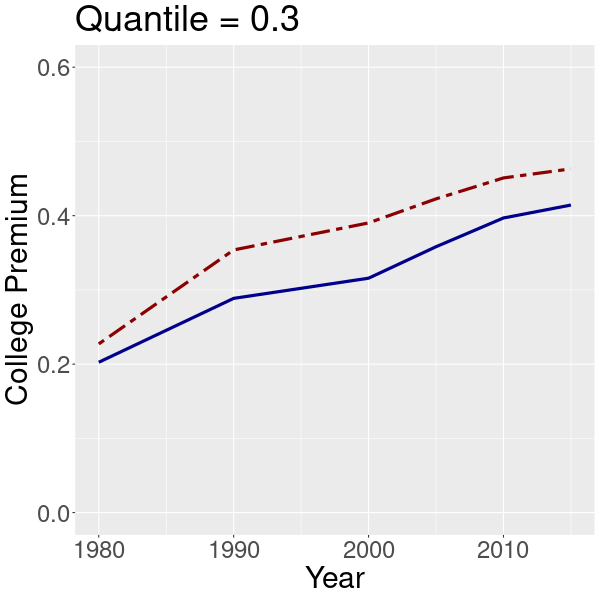} \\
        \includegraphics[scale=0.23]{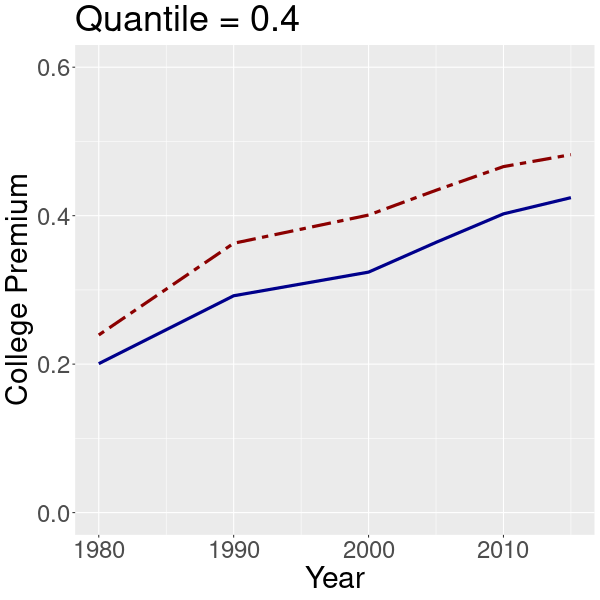} & \includegraphics[scale=0.23]{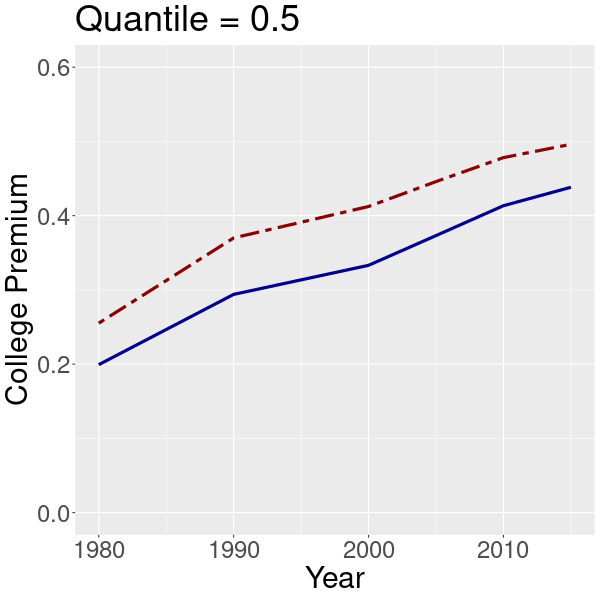} & \includegraphics[scale=0.23]{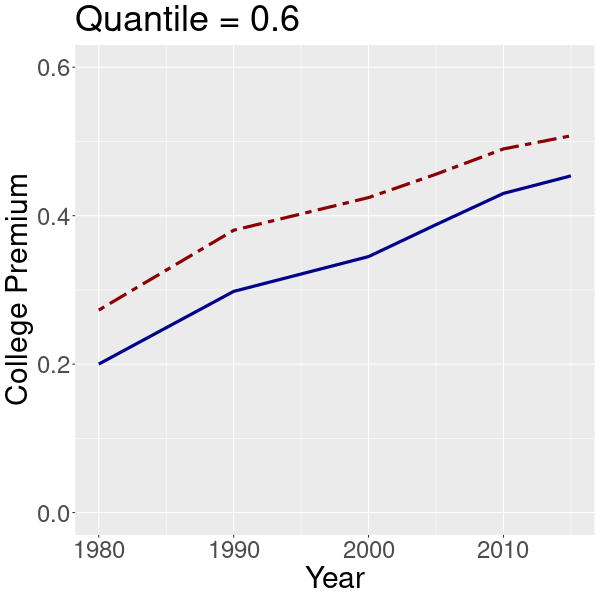} \\
        \includegraphics[scale=0.23]{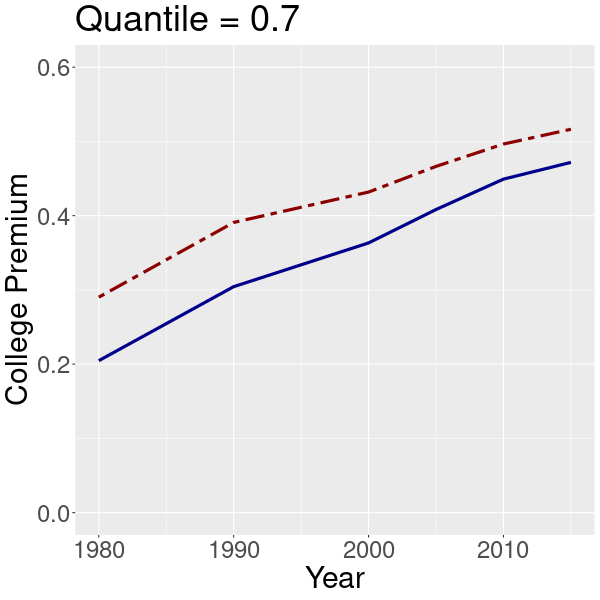} & \includegraphics[scale=0.23]{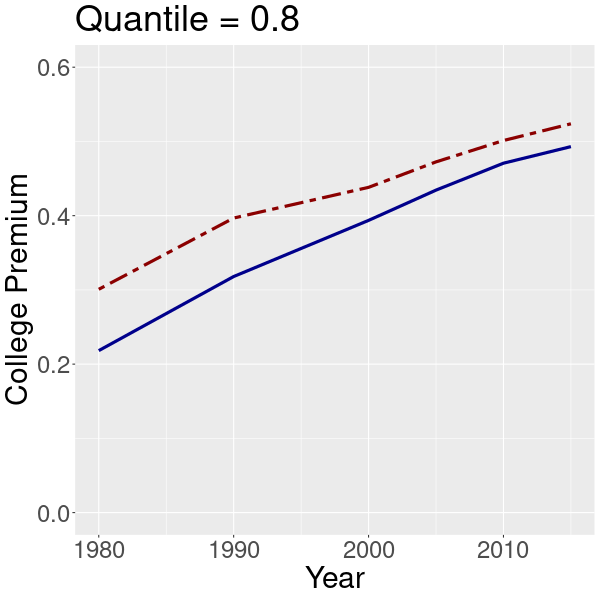} & \includegraphics[scale=0.23]{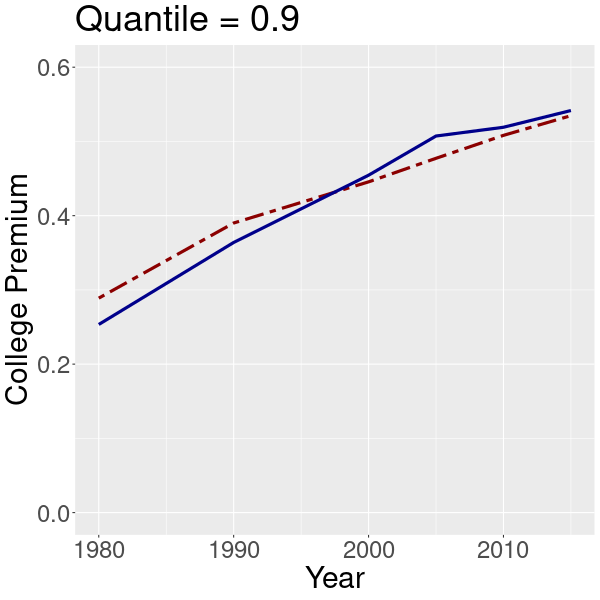} 
    \end{tabular}
\end{figure}

\begin{figure}
    \centering
    \caption{Difference of College Premium: Combining 5-Year Data} 
    \label{fig:diff_5yr_full}
    \hskip15pt
    \begin{tabular}{c c c}
        \includegraphics[scale=0.23]{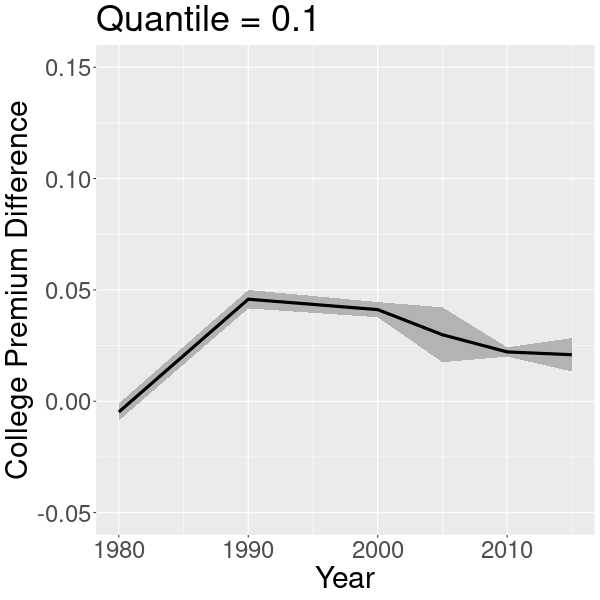} & \includegraphics[scale=0.23]{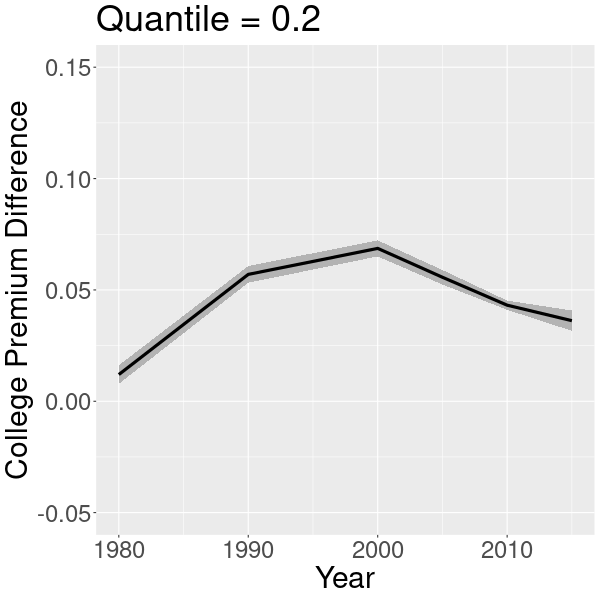} & \includegraphics[scale=0.23]{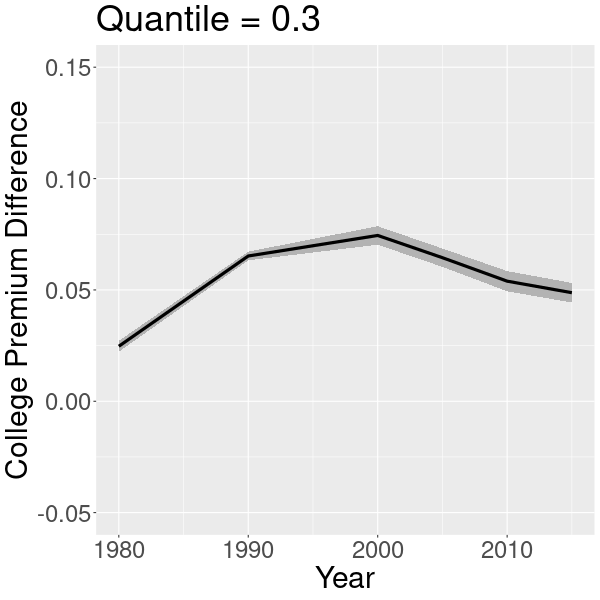} \\
        \includegraphics[scale=0.23]{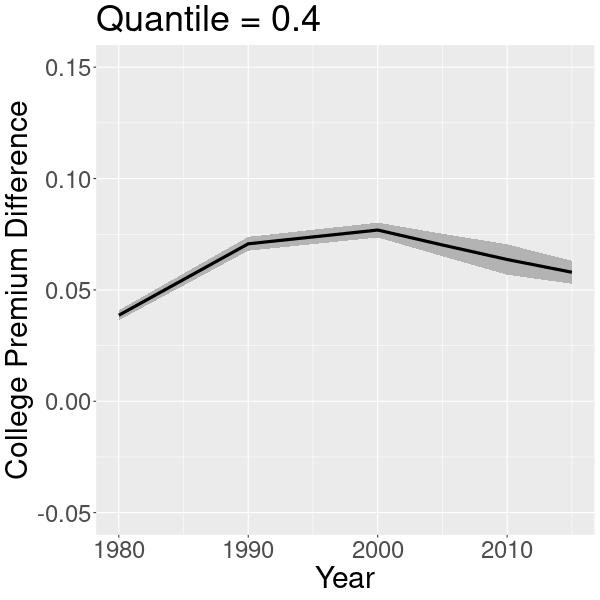} & \includegraphics[scale=0.23]{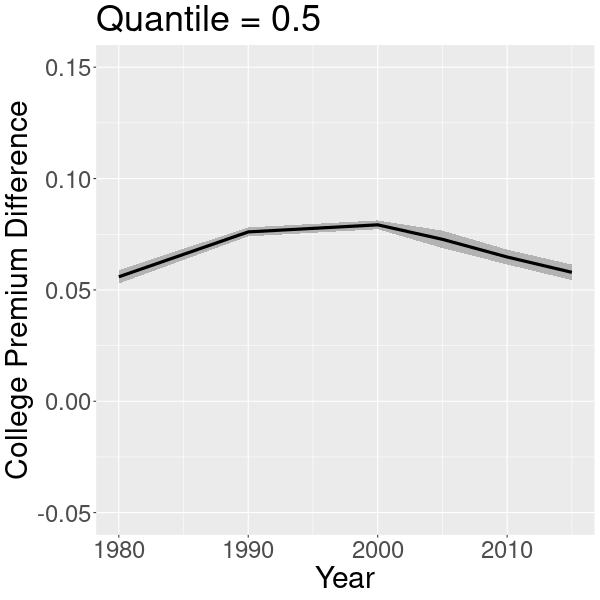} & \includegraphics[scale=0.23]{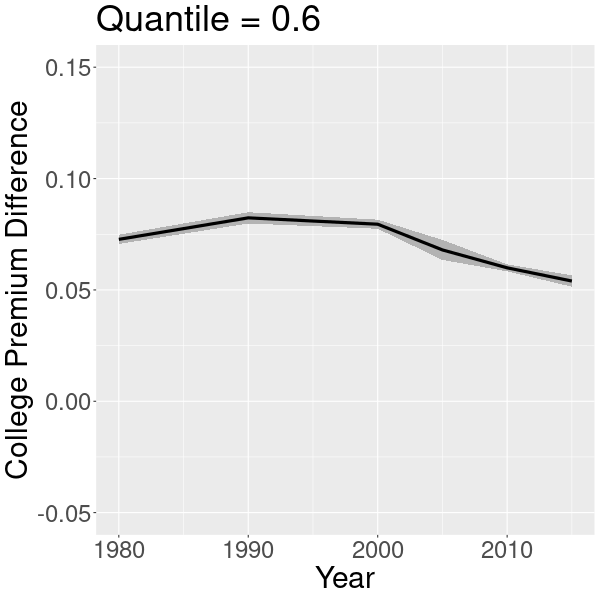} \\
        \includegraphics[scale=0.23]{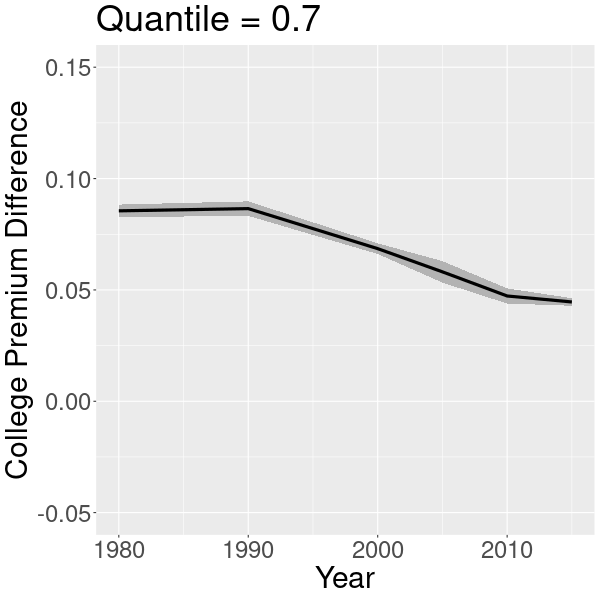} & \includegraphics[scale=0.23]{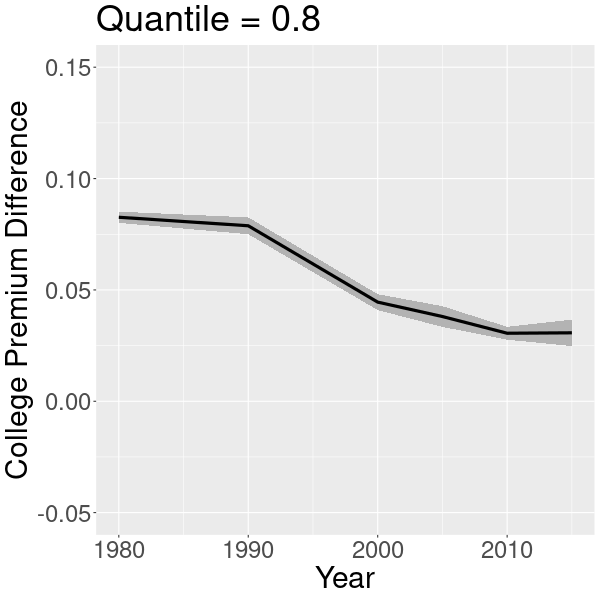} & \includegraphics[scale=0.23]{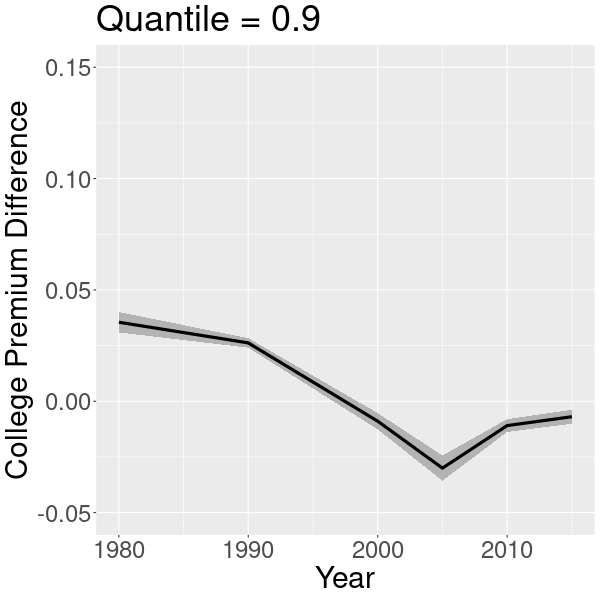} 
    \end{tabular}
\end{figure}

Figures \ref{fig:premium_design_5yr_full}--\ref{fig:diff_5yr_full} and Table \ref{tb:median_5yr_full} summarize the estimation results. 
We also provide the full estimation results in the appendix.

Our results share several interesting features about the estimated college wage premium.   First, some degrees of heterogeneity are found over different quantiles.  At the upper tail quantile, $\tau=0.9$, male college premium is slightly higher than the female one since 2000, while female college premiums are higher over all years in other quantiles. This shows an interesting feature about high-income individuals that cannot be viewed by \changed{mean regression}, though sensible economic interpretations might be subject to debates. Also, note that the 95\% confidence intervals are slightly wider for tail quantiles ($\tau=0.1$ and 0.9, respectively).

Second, college premiums increase over time for both male and female. This result coincides with the overall findings in the literature. It is interesting that college premiums get flatter since 2010 for lower quantiles ($\tau \le 0.5$).

Third, when we focus on the median ($\tau=0.5$) as reported in Table \ref{tb:median_5yr_full}, we observe that the female-male college premium difference shows an inverse ``U-shape'' pattern. 
In addition, the difference is always significantly positive, which is different from the result in \citet{hubbard2011phantom}.
He found the gender difference to be insignificant starting from around the year 2000 from the CPS data. 
We also note that the computation time is within a reasonable range meanwhile it is not feasible to estimate the model with the alternative approaches.



\begin{table}[t]
\centering
\caption{College Wage Premium: $\tau=0.5$}\label{tb:median_5yr_full}
\begin{tabular}{ccccc}
  \hline
Year & Female & Male & Difference & Time (min.) \\ 
  \hline
   \underline{$\tau = 0.5$}\\
   1980 & 0.2552 & 0.1993 & 0.0559 & 7.6 \\ 
   & [0.2521,0.2583] & [0.1979,0.2007] & [0.0529,0.0589] &  \\ 
  1990 & 0.3700 & 0.2940 & 0.0761 & 6.7 \\ 
   & [0.3681,0.3720] & [0.2934,0.2946] & [0.0740,0.0781] &  \\ 
  2000 & 0.4124 & 0.3331 & 0.0792 & 7.1 \\ 
   & [0.4101,0.4147] & [0.3322,0.3341] & [0.0773,0.0812] &  \\ 
  2001-2005 & 0.4459 & 0.3732 & 0.0727 & 3.1 \\ 
   & [0.4435,0.4483] & [0.3708,0.3756] & [0.0688,0.0766] &  \\ 
  2006-2010 & 0.4782 & 0.4135 & 0.0648 & 7.6 \\ 
   & [0.4765,0.4799] & [0.4112,0.4157] & [0.0614,0.0681] &  \\ 
  2011-2015 & 0.4962 & 0.4383 & 0.0579 & 7.3 \\ 
   & [0.4940,0.4984] & [0.4337,0.4429] & [0.0544,0.0614] &  \\ 
   \hline
\end{tabular}
    \flushleft{\footnotesize Notes. The male college premium is from $\hat{\beta}_2$ and the female college premium is from $\hat{\beta}_2+\hat{\beta}_3$. Thus, the college premium difference between male and female workers is from $\hat{\beta}_3$. }
\end{table}





%


\section{Discussions}

We have proposed an inference method for large-scale quantile regression to analyze datasets whose sizes are of order $(n, p) \sim (10^7, 10^3)$. 
\changed{Based on the stochastic subgradient descent updates, our method runs fast and constructs asymptotically pivotal statistics via random scaling.}

As an important extension, one can embed variable selections in the S-subGD framework.    At the $i$ th update, define the t-statistic 
$ 
t_{i,j} = \bar\beta_{i,j}/ \sqrt{V_{i,jj}}.
$   for $j=1,...,d. $
Then, variable $j$ is selected if 
$$
|t_{i,j}|>\lambda
$$
for a tuning parameter $\lambda$. For instance, we can set $\lambda= 6.747$ which is the critical value for the 95\% confidence level of the asymptotic distribution. For each variable $j$, we  obtain a cumulative selection  path along the iteration:
$$
SP(j,i)= \frac{1}{i}\sum_{k=1}^i1\{|t_{k,j}|>\lambda\}.
$$
The  $SP$ path   provides rich  information about variable selections, which is particularly attractive  when the signal-noise ratio is relatively weak. We would like to leave the theoretical justification of the selection path   for future studies.

Besides the variable selection, there are a few more extensions worth pursuing in future research. 
First, we may build on \citet{chernozhukov2022fast} to develop fast inference for the quantile regression process.
Second, while we focus on the regular quantile regression models where the quantile is assumed to be bounded away from both zero and one, our framework is also potentially applicable to extreme quantile regression models \citep[see, e.g.,][for a review]{chernozhukov2016extremal}. The  ability of handling ultra-large datasets is also appealing for extreme quantile regression models  because the resulting sample size is considerably larger at the extreme quantiles. 
A formal treatment for these cases  is out of the scope of this paper, and we leave it for future studies. 
Third, \citet{Shan:Yang:09} considered, among other things, an online setting for which time series data arrive sequentially and several quantile estimators are combined with weights being updated sequentially.
It is an interesting topic to study the properties of our proposed method with time series data.
Finally, \citet{Liu-et-al:ICML:2023} developed online inference on population quantiles with privacy concerns. It is another research topic to adapt our inference method for privacy-focused applications.


\appendix

\section{Proofs}

In the appendix, we provide the proof of Theorem~\ref{thm:Wald:qr}.
Especially, we prove Theorem~\ref{thm:Wald} which is a more general version than
Theorem~\ref{thm:Wald:qr}.
Theorem~\ref{thm:Wald} includes quantile regression as a special case but can be of independent interest. 
To do so, we first present high-level regularity conditions, which we will verify for quantile regression
in Theorem~\ref{thm:qr}.

\subsection{General Theory for M-estimation that allows possibly non-smooth and locally-strongly-convex loss}

We formalize a general inferential theory for M-estimations that allows non-smooth (like quantile regression) and non-globally-strongly-convex loss functions (but locally strongly convex), in the online fashion. 

Recall that  $\beta^{*}$ is characterized by 
\[
\beta^{*}:=\arg\min_{\beta\in\mathbb{R}^{d}}Q\left(\beta\right),
\]
where 
$
Q(\beta):= \mathbb E [q(\beta, Y_i)]$   is the population loss function. In this subsection, it  can be the loss function for general M-estimations, not just the quantile regression.  The updating rule is 
 $$
\beta_{i}=\beta_{i-1}-\gamma_{i}\nabla q\left(\beta_{i-1},Y_{i}\right),
$$ 
where $\nabla q $ belongs to the subgradient of $q$. 
Next, define
$$\xi_{i}\left(\beta\right):=\nabla Q\left(\beta\right)-\nabla q\left(\beta,Y_{i}\right),
$$
and 
$\xi_{i}:=\xi_i(\beta_{i-1}).
$

We impose the following conditions. 
\begin{asm}[IID Data]\label{asm:iid}
Data $\{ Y_i: i=1,\ldots,n \}$ are i.i.d. 
\end{asm}

Assumption~\ref{asm:iid} restricts our attention to i.i.d. data. 
Assumption~\ref{asm:iid} implies among other things that the sequence $\{ \xi_{i} \}_{i\geq1}$
    is a martingale-difference sequence (mds) defined on a probability space
    $(\Omega,\mathcal{{F}},\mathcal{{F}}_{i},P)$. That is, $\mathbb E(\xi_{i}|\mathcal{{F}}_{i-1})=0$  almost surely.

\begin{asm}[Population Criterion]\label{asm:obj}
The real-valued function $Q\left(\beta\right):=\mathbb{E} \left[ q\left(\beta,Y_i \right) \right]$ is twice continuously differentiable, with:

(i) $\sup_{\beta\in\Theta}\|\nabla^2 Q(\beta)\|<\infty.$

(ii) There is $K_1>0$ and  $\varepsilon>0$ that for all $\| \beta-\beta^* \| \leq\varepsilon$,
$$
\|\nabla^2Q(\beta) - \nabla^2Q(\beta^*)\|\leq K_1\|\beta-\beta^*\|.
$$

(iii) In addition, $\nabla ^2 Q(\beta)$ is positive semi-definite for all $\beta$. 
\end{asm}

Assumption~\ref{asm:obj} imposes mild regularity conditions on the population criterion function $Q(\beta)$. It is similar to Assumption 3.3 of  \citet{polyak1992acceleration}, but with an important extension that allows for non-differentiable loss functions as in quantile regression.

\begin{asm}[Learning Rate]\label{asm:learning-rate}
$\gamma_{i}=\gamma_0 i^{-a} $ for some $1/2 < a <1$.

\end{asm}

 Assumption~\ref{asm:learning-rate} is 
the standard condition on the learning rate.

\begin{asm}[local strong convexity]\label{aslcox} There are constants $\epsilon>0$ and $c_0>0$ that satisfy 
$$
  \inf_{  \| \beta-\beta^* \|<\epsilon }\lambda_{\min} \left(  \nabla^2 Q(\beta) \right) > c_0.
  	$$
\end{asm}

 The key difference between  Assumption~\ref{aslcox}  and those of \citet{polyak1992acceleration}  is that in our setting, the strong convexity is imposed only locally on a neighborhood of the true value.  In contrast, \citet{polyak1992acceleration} imposed the global strong convexity. They also imposed additional global conditions, which are relaxed here. 
This relaxation is important  for the quantile regression model, which may not be globally strongly convex.

\begin{asm}[Sufficient Conditions for Consistency]\label{as:suff:cons}  
The sequence $\left\{ \xi_{i} \right\}_{i\geq1}$ 
satisfy the following additional conditions.
 \begin{enumerate}[(i)]
  
    \item $\mathbb E\left[\left\|\xi_{i}\right\|^{6}\exp\left(\left(1+\left\|\xi_{i}\right\|^{4}\right)^{1/2}\right) \bigg| \mathcal{F}_{i-1}\right]<\infty\quad\text{a.s.}$
    
    \item
    $S_{i}\left(\beta\right):=\mathbb E\left[\xi_{i}\left(\beta\right)\xi_{i}\left(\beta\right)'|\mathcal{F}_{i-1}\right]$
    is uniformly Lipschitz continuous, that is, 
    \[
    \left\Vert S_{i}\left(\beta_{1}\right)-S_{i}\left(\beta_{2}\right)\right\Vert \leq L\|\beta_{1}-\beta_{2}\| \;
   \text{ a.s. for some $L < \infty$}.
   \]

    \item    In addition, for some $p\geq\left(1-a\right)^{-1}$,
$\mathbb{E}\left\Vert \xi_{i}\right\Vert ^{2p}$
is bounded. Here, $a$ is defined in Assumption \ref{asm:learning-rate}.
\end{enumerate}
\end{asm}


Conditions (i) and (ii) in Assumption~\ref{as:suff:cons} are imposed to apply Theorem 2  of \citet{gadat2022optimal}.
In addition, under Assumption~\ref{as:suff:cons}~(i)-(ii) combined with the local strong convexity in Assumption~\ref{aslcox},    
a global Kurdyka-Łojasiewicz  condition  in \citet{gadat2022optimal} is satisfied. From here, we can establish the consistency.  
Condition (iii) of  Assumption \ref{as:suff:cons} is  an additional moment condition to strengthen the results for 
 the FCLT.

\begin{asm}[Stochastic Disturbances]\label{as2fda}

\begin{enumerate}[(i)]

       \item  For $K_{2}>0$ and  for all $i\geq1$, $$\|\nabla Q(\beta_{i-1})\|^{2}+\mathbb E( \|\nabla q(\beta_{i-1},Y_i)\|^2|\mathcal F_{i-1}) \leq K_{2}(1+\|\beta_{i-1}\|^{2}) \;\;  a.s.$$
  
      \item Denote $\nabla q_i^*:= \nabla q\left(\beta^*,Y_{i}\right)$. Then

   \begin{enumerate}[(a)]

    \item There is  a  symmetric and positive
    definite matrix $S$, such that    as $i\rightarrow\infty$, 
    $$\text{Var}\left( \nabla q_i^*  |\mathcal{{F}}_{i-1} \right) = S_i(\beta^*)\xrightarrow{P} S,$$
    where $S_i(\cdot)$ is as defined in Assumption \ref{as:suff:cons}.
 
    \item As $C\to\infty$, 
    $$\sup\limits _{i}\mathbb{E} (\|\nabla q_i^*\|^{2}I(\| \nabla q_i^*\|>C|\mathcal{\mathcal{{F}}}_{i-1})\xrightarrow{P}0.$$

    \end{enumerate}

    \item There is a function $h(\cdot)$ such that 
    $h(x)\rightarrow0$ as $\|x\|\rightarrow0$, which satisfies:
    for all $i$ large enough, almost surely, $$\mathbb E(\|  \nabla q\left(\beta_{i-1},Y_{i}\right) -\nabla q\left(\beta^*,Y_{i}\right)\|^2|\mathcal F_{i-1})\leq h(\beta_{i-1}-\beta^*). $$
    \end{enumerate}
\end{asm}

We now establish the general theorem under these assumptions.

\begin{thm}[General Theorem]
\label{thm:Wald}  Under Assumptions
\ref{asm:iid}-\ref{as2fda},  
\begin{equation} 
\frac{1}{\sqrt{n}}\sum_{i=1}^{\left[nr\right]}\left(\beta_{i}-\beta^{*}\right)\Rightarrow \Upsilon^{1/2}W_d\left(r\right),\quad r\in\left[0,1\right],
\end{equation}
where $W_d$ is a $d$-dimensional vector of the standard Wiener
processes; $\Rightarrow$ stands for the weak convergence in $\ell^{\infty}\left[0,1\right]$, 
and
$ \Upsilon := H^{-1}SH^{-1}$, with 
 $
 S $ being the probability limit of  $\text{Var}\left(  \nabla q\left(\beta^*,Y_{i}\right)  |\mathcal{{F}}_{i-1} \right)$ and $H:= \nabla^2 Q(\beta^*)$.

In addition, under 
$H_{0}: R\beta^{*} = c$  for an $\ell\times d$ matrix $R$ with 
$\mathrm{rank}(R)=\ell$,  \begin{eqnarray*}
&&n\left(R\bar{\beta}_{n}-c\right)'\left(R\widehat{V}_{n}R'\right)^{-1}\left(R\bar{\beta}_{n}-c\right)\overset{d}{\to}W_\ell\left(1\right)'\left(\int_{0}^{1}\bar{W}_\ell(r)\bar{W}_\ell(r)'dr\right)^{-1}W_\ell\left(1\right),
\end{eqnarray*}
where $\bar{W}_\ell\left(r\right):=W_\ell\left(r\right)-rW_\ell\left(1\right)$.

\end{thm}

\subsection{Proof of Theorem \ref{thm:Wald}}

\begin{proof}[Proof of Theorem \ref{thm:Wald}] 

We extend the proof of \citet{polyak1992acceleration} and \citep{lee2021fast} in two aspects: allow non-globally-strong-convex  and possibly nonsmooth loss functions. For the former, we first establish the consistency. Then  we can focus on a local neighborhood of the true value.  It is  under Assumption \ref{aslcox} that the loss function is locally strongly convex. 
The formal proof is divided in the following steps.
Some of them are delegated to Lemma \ref{leafdafa} - \ref{lema.3} in the online supplement.

\textbf{step 1: consistency.}  Lemma \ref{leafdafa} shows that $$
\bar\beta_n\to^P \beta^*.
$$

\textbf{step 2: local-linearization.} 
Rewrite (\ref{eq:SGD1})
as 
\begin{equation}
\beta_{i}=\beta_{i-1}-\gamma_{i}\nabla Q\left(\beta_{i-1}\right)+\gamma_{i}\xi_{i}\label{eq:SGD2}.
\end{equation}
Let $\Delta_{i}:=\beta_{i}-\beta^{*}$
and $\bar{\Delta}_{i}:=\bar{\beta}_{i}-\beta^{*}$ to denote the errors
in the $i$-th iterate and that in the average estimate at $i$, respectively.
Then, subtracting $\beta^{*}$ from both sides of (\ref{eq:SGD2})
yields that
\[
\Delta_{i}=\Delta_{i-1}-\gamma_{i}\nabla Q\left(\beta_{i-1}\right)+\gamma_{i}\xi_{i}.
\]
Now consider the following linear process $\Delta_i^1$, defined as:
\[
\Delta_{i}^{1}:=\Delta_{i-1}^{1}-\gamma_{i}H\Delta_{i-1}^{1}+\gamma_{i}\xi_{i}\quad\text{and}\quad\Delta_{0}^{1}=\Delta_{0},
\]
where  $H=\nabla^2 Q(\beta^*)$. Furthermore, for $r\in\left[0,1\right]$, introduce a partial sum process
\[
\bar{\Delta}_{i}\left(r\right):=i^{-1}\sum_{j=1}^{\left[ir\right]}\Delta_{j},\quad \bar{\Delta}_{i}^1\left(r\right):=i^{-1}\sum_{j=1}^{\left[ir\right]}\Delta_{j}^1.
\]
This step establishes  a uniform approximation of the partial sum process  $\Delta_i^r(r)$ to   $\bar{\Delta}_{i}^{1}\left(r\right)$. To this end, we adopt  Part 4 in the proof of \citet{polyak1992acceleration}'s Theorem 2.
To adopt their proof,  we need to  establish that the local convexity of the loss function can ensure that the nonlinear updating process $\Delta_i$ can be approximated locally (as $\beta_i\to^P\beta^*$) by the linear process $\Delta_i^1$. Indeed, as Lemma \ref{ass:PJ's Assumption 3} shows, in the neighbourhood of $\beta^*$, this is the case. In fact, Lemma \ref{ass:PJ's Assumption 3} verifies that the required conditions, Assumptions 3.1 and 3.2 of \citet{polyak1992acceleration}, hold locally.  In addition,  carefully examining their proof of Part 4, it is only required that the population loss function be twice differentiable, as the linear process $\Delta_i^1$ only depends on the Hessian of the population loss function. 

Therefore, with  Lemma \ref{ass:PJ's Assumption 3}, the proof of Part 4 in \citet{polyak1992acceleration} goes through, which establishes: 
$$
\sqrt{i}\sup_{r}\left\|\bar{\Delta}_{i}\left(r\right)-\bar{\Delta}_{i}^{1}\left(r\right)\right\|=o_{p}\left(1\right).
$$
 Hence it suffices to analyze the sequence $\bar{\Delta}_{i}^{1}$ in place of $\bar{\Delta}_{i}$, and  establish the weak convergence of $\bar{\Delta}_{i}^{1}$. 
 
\textbf{step 3: weak convergence of $\sqrt{i}\bar{\Delta}_{i}^{1}\left(r\right)$.}   Following the decomposition in (A10) in \citet{polyak1992acceleration}, write
\[
\sqrt{i}\bar{\Delta}_{i}^{1}\left(r\right)=I^{\left(1\right)}\left(r\right)+I^{\left(2\right)}\left(r\right)+I^{\left(3\right)}\left(r\right),
\]
where 
\begin{align*}
I^{\left(1\right)}\left(r\right)  :=\frac{1}{\gamma_{0}\sqrt{i}}\alpha_{\left[ir\right]}\Delta_{0}, 
I^{\left(2\right)}\left(r\right)  :=\frac{1}{\sqrt{i}}\sum_{j=1}^{\left[ir\right]}H^{-1}\xi_{j}, \text{ and }
I^{\left(3\right)}\left(r\right)  :=\frac{1}{\sqrt{i}}\sum_{j=1}^{\left[ir\right]}w_{j}^{\left[ir\right]}\xi_{j},
\end{align*}
where $\alpha_{i}\leq K$ and $\left\{ w_{j}^{\left[ir\right]}\right\} $
is a bounded sequence such that $i^{-1}\sum_{j=1}^{i}\left\Vert w_{j}^{i}\right\Vert \to0$.
Then, $\sup_{r}\left\Vert I^{\left(1\right)}\left(r\right)\right\Vert =o_{p}\left(1\right)$. 
Suppose for now that $\mathbb E\sup_r \| I^{(3)}\|^p=o(1)$ for some $p\geq 1$. Bounding $I^{(3)}$ requires some involved arguments, as $w_j^{\left[ir\right]} \xi_j$ is not mds, even though $\xi_j$ is (note that $w_j^{\left[ir\right]} \xi_j$ is indexed by $r$). 
The next step of the proof develops new techniques to bound this term. 

We now apply the FCLT for mds, see
e.g. Theorem 4.2 in \citet{Hall-Heyde}. To   verify  its sufficient conditions, we  prove
  Lemma \ref{lema.3}. Then   Lemma \ref{lema.3} is sufficient to
  verify the   conditions in  \citet{Hall-Heyde} using the
  same argument of Part 1 of the proof in \citet{polyak1992acceleration}. Hence,  $I^{(2)}$ converges weakly to a rescaled Wiener process $\Upsilon^{1/2}W\left(r\right)$. 
Hence 
$$
\sqrt{i}\bar{\Delta}_{i}^{1}\left(r\right)\Rightarrow \Upsilon^{1/2}W\left(r\right).
$$

\textbf{step 4: uniform convergence of $I^{(3)}(r)$.}   Let $S_{i}=\sum_{j=1}^{i}w_{j}^{i}\xi_{j}$.
Let $p>\left(1-a\right)^{-1}$ and note that
\[
\mathbb{E}\sup_{r}\left\Vert I^{\left(3\right)}\left(r\right)\right\Vert ^{2p}\leq t^{-p}\mathbb{E}\sup_{r}\left\Vert S_{\left[ir\right]}\right\Vert ^{2p}\leq t^{-p}\sum_{m=1}^{i}\mathbb{E}\left\Vert S_{m}\right\Vert ^{2p}.
\]
Note that for each $m$, $S_m$ is the sum of mds. Hence, due to Burkholder's inequality \citep[e.g., Theorem 2.10 in][]{Hall-Heyde},
\begin{align*}
\mathbb{E}\left\Vert S_{m}\right\Vert ^{2p} & \leq 
C_p \mathbb{E} \left|\sum_{j=1}^{m}\left\Vert w_{j}^{m}\right\Vert {}^{2}\left\Vert \xi_{j}\right\Vert ^{2}\right|^{p} \\
&= C_p \sum_{j_{1},...,j_{p}=1}^{m}\left\Vert w_{j_{1}}^{m}\right\Vert ^{2}\cdots\left\Vert w_{j_{p}}^{m}\right\Vert ^{2}\mathbb{E}\left\Vert \xi_{j_{1}}\right\Vert ^{2}\cdots\left\Vert \xi_{j_{p}}\right\Vert ^{2},
\end{align*}
where the universal constant $C_p$ depends only on $p$.
Note that $\mathbb{E}\left\Vert \xi_{j_{1}}\right\Vert ^{2}\cdots\left\Vert \xi_{j_{p}}\right\Vert ^{2}$
is bounded since $\mathbb{E}\left\Vert \xi_{j}\right\Vert ^{2p}$
is bounded. Also, the boundedness of $\left\Vert w_{j}^{m}\right\Vert $ yields $\sum_{j=1}^{m}\left\Vert w_{j}^{m}\right\Vert ^{b}=O\left(\sum_{j=1}^{m}\left\Vert w_{j}^{m}\right\Vert \right)$
for any $b$.
By  Lemma 2 in \citet{Zhu:Dong}, $\sum_{j=1}^{m}\left\Vert w_{j}^{m}\right\Vert =o\left(m^{a}\right)$.
These facts yield that 
\begin{equation}
\mathbb{E}\left\Vert S_{m}\right\Vert ^{2p}=O\left(\left(\sum_{j=1}^{m}\left\Vert w_{j}^{m}\right\Vert \right)^{p}\right)=o(m^{ap}),\label{eq:boundloos}
\end{equation}
which holds uniformly for $m,k$. It in turn implies the desired result that 
\[
\mathbb{E}\sup_{r}\left\Vert I^{\left(3\right)}\left(r\right)\right\Vert ^{2p}\leq t^{-p}\sum_{m=1}^{i}o\left(m^{ap}\right)=o\left(t^{1+ap-p}\right)=o\left(1\right).
\]

\textbf{step 5: FCLT.}  By the previous three steps:
\begin{equation}\label{eq20}
\frac{1}{\sqrt{n}}\sum_{i=1}^{\left[nr\right]}\left(\beta_{i}-\beta^{*}\right)=\sqrt{n}\bar\Delta_n(r)=\sqrt{n} \bar\Delta_n^1(r) +o_P(1)= \frac{1}{\sqrt{n}}\sum_{j=1}^{\left[nr\right]}H^{-1}\xi_{j}+o_P(1).
\end{equation}
We have $ \frac{1}{\sqrt{n}}\sum_{j=1}^{\left[nr\right]}H^{-1}\xi_{j}  \Rightarrow \Upsilon^{1/2}W\left(r\right)$.
 This establishes the FCLT in (\ref{eq4}). 

\textbf{step 6: random scaling.}  Now let $C_n(r):=R\frac{1}{\sqrt{n}}\sum_{i=1}^{\left[nr\right]}\left(\beta_{i}-\beta^{*}\right)$. Also let $\Lambda= (R\Upsilon R')^{1/2}$. It is well-defined and invertible when $l\leq d$. By the FCLT in the previous step,  for some vector of independent standard Wiener process $W^*(r)$,
$$
C_n(r)\Rightarrow \Lambda W^*(r).
$$
In addition,
$R\widehat V_nR'= \frac{1}{n}\sum_{s=1}^n[C_n(\frac{s}{n})-\frac{s}{n}C_n(1)][C_n(\frac{s}{n})-\frac{s}{n}C_n(1)]'$. 
Here the sum is also an integral over $ r $ as $ C_n(r) $ is a partial sum process, 
and  $R(\bar\beta_n-\beta^*)=\frac{1}{\sqrt{n}}C_n(1)$. Hence
 $
n\left(R\bar{\beta}_{n}-c\right)'\left(R\widehat{V}_{n}R'\right)^{-1}\left(R\bar{\beta}_{n}-c\right) $ is a continuous functional of $C_n(\cdot)$.  Then the continuous mapping theorem proves the theorem. 
\end{proof}

\subsection{Proof of Theorem~\ref{thm:Wald:qr}}

\begin{proof}[Proof of Theorem~\ref{thm:Wald:qr}]
This follows directly from Theorem~\ref{thm:Wald} because all the regularity conditions in 
Theorem~\ref{thm:Wald} are verified in Theorem~\ref{thm:qr}.
\end{proof}

We verify the high-level assumptions with more primitive conditions for the quantile regression  model.

\begin{thm}[Verifying High-Level Conditions for Quantile regression]\label{thm:qr}
Suppose that Assumption~\ref{asm:qr} holds. Then, 
Assumptions
\ref{asm:iid}-\ref{as2fda}  
 hold with  $S = \mathbb E [x_i x_i'] \tau(1-\tau)$ and $H= \mathbb E [x_i x_i' f_{\varepsilon}(0|x_i)]$.
 \end{thm}

\begin{proof}[Proof of Theorem~\ref{thm:qr}]

  \textit{Verify Assumption \ref{asm:iid}.}   It is  directly imposed.



\bigskip
\noindent

 \textit{Verify Assumption \ref{asm:obj}.}    It is straightforward to verify the twice differentiability, where
  $$
 \nabla Q(\beta) = \mathbb Ex_i[P(\varepsilon_i\leq x_i'(\beta-\beta^*)|x_i)- \tau].
 $$ and 
$$\nabla^2Q(\beta)=G(\beta)= \mathbb E [x_ix_i'f_{\varepsilon}(x_i'(\beta-\beta^*)|x_i)]. $$

(i)  $\sup_{\beta\in\Theta}\|\nabla^2 Q(\beta)\| \leq \sup_{b\Theta}\|\mathbb E [x_ix_i'f_{\varepsilon}(x_i'b)|x_i)]  \|<\infty. $

(ii) For all $\| \beta-\beta^* \| \leq\varepsilon$,
$$
\|\nabla^2Q(\beta) - \nabla^2Q(\beta^*)\|\leq  \|\nabla^3 Q(\widetilde\beta)\| \|\beta-\beta^*\|
$$for some $\widetilde\beta$. 
Here 
 $\nabla^3 Q(\beta)$ is a $\dim(\beta)\times \dim(\beta)^2$ matrix, whose $j$ th row is given by $[\text{vec}\nabla^2 (\partial_j Q(\beta)) ]'$. We note that $$\|\nabla^3 Q(\widetilde\beta)\|
  \leq\sup_{| \beta-\beta^* |<\epsilon}  \mathbb E\|x_i\|^3 \left|\frac{d}{d\varepsilon}f_{\varepsilon}(x_i'(\beta-\beta^*)|x_i)\right|<K_1.
  $$	
The last inequality follows from Assumption 
  \ref{asm:qr}(ii). 

  (iii) It is clear that $  \mathbb E [x_ix_i'f_{\varepsilon}(x_i'(\beta-\beta^*)|x_i)]$   is globally semipositive definite.

   \bigskip
\noindent

  \textit{Verify Assumption \ref{asm:learning-rate}.}   It is  directly imposed.

   \bigskip
\noindent

\textit{Verify Assumption \ref{aslcox}.}  This is imposed by Assumption \ref{asm:qr} (i) that the eigenvalues of $\mathbb E [x_ix_i'f_{\varepsilon}(x_i'(\beta-\beta^*)|x_i)]$ are locally bounded away from zero.

   \bigskip
\noindent

\textit{Verify Assumption \ref{as:suff:cons}.}
For (i), we have for any $r>0,$
  $\|\xi_{i} \|^r\leq C(\|x_i\|^r+\mathbb E \|x_i\|^r).
  $ Hence  $$\mathbb E\left[\left\|\xi_{i}\right\|^{6}\exp\left(\left(1+\left\|\xi_{i}\right\|^{4}\right)^{1/2}\right)|\mathcal{F}_{i-1}\right] 
  \leq C\mathbb E[\|x_i\|^6+1)\exp(\|x_i\|^2|\mathcal F_{i-1} ]<C.
  $$

  For (ii),  it is a straightforward application of the Cauchy-Schwarz and triangular inequalities. For instance, 
$
 \mathbb E \|x_i\|^2|I\{\varepsilon_i<\beta_1\}-I\{\varepsilon_i<\beta_2\}|
 \leq \sup_{b}\mathbb E \|x_i\|^2f_{\varepsilon}(x_i'b)\|\beta_1-\beta_2\|.
$
 We omit the details.
For (iii), note, for all $\beta$, $
\mathbb E\|\xi_i(\beta)\|^{2p}
\leq 2\times 4^p\mathbb E\|x_i\|^{2p} <\infty.
$

\textit{Verify Assumption \ref{as2fda}.}
  (i) 
Write
  		 \begin{eqnarray*}
  	&&	 \|\nabla Q\left(\beta_{i-1}\right)\|^2  \leq  4(\mathbb E\|x_i\| )^2\leq K\cr
  	&&\mathbb E(\|\nabla q(\beta_{i-1}, Y_i)|^{2}\|\mathcal{{F}}_{i-1})\leq	4\mathbb E(\|x_i\|^2|\mathcal F_{i-1}) \leq 4\mathbb E \|x_i\|^2\leq K.
  		 \end{eqnarray*}
  		 This implies 
$$
 \|\nabla Q\left(\beta_{i-1}\right)\|^2+ \mathbb E(\|\nabla q(\beta_{i-1}, Y_i)\|^{2}|\mathcal{{F}}_{i-1}) \leq 2K\leq 2K (1+\|\beta_{i-1}\|^{2}) \;\; a.s.
$$

(ii)  Here $\nabla q_i^*=\xi_i(\beta^*)=x_i[I\{\varepsilon_i<0\}-\tau] .$ Clearly, $\mathbb{E} (\xi_{i}(\beta^*)|\mathcal{{F}}_{i-1})=0$ and  
 (a) $
 \text{Var}(\xi_{i}(\beta^*)|\mathcal F_{i-1})=\mathbb E [x_i x_i'] \tau(1-\tau)=S.
 $
 (b) Also,  as $C\to \infty, $
 \begin{eqnarray*}
 &&\sup_{i}\mathbb{E} (\|\xi_{i}(\beta^*)\|^{2}I(\|\xi_{i}(\beta^*)|>C\|\mathcal{\mathcal{{F}}}_{i-1})\leq \sup\limits _{i}(\mathbb E \|\xi_{i}(\beta^*)\|^{3}|\mathcal F_{i-1})^{2/3} P(\|\xi_{i}(\beta^*)\|>C|\mathcal F_{i-1})^{1/3} \cr
&\leq& 4(\mathbb E\|x_i\|^3)^{2/3}\mathbb E (\|\xi_{i}(\beta^*)\||\mathcal F_{i-1}) ^{1/3}C^{-1/3}\xrightarrow{P}0. 
 \end{eqnarray*}

 (iii)  Recall $\nabla q(\beta, Y_i) := x_i [I\{y_i \leq x_i'\beta\} - \tau].$  Write $$A_i = x_i [ I\{\varepsilon_i< x_i'(\beta_{i-1}-\beta^*)\}-I\{\varepsilon_i<0\}].$$
Hence  for $h(x)=2\sup_{b}\mathbb E\|x_i\|^3  f_{\varepsilon}(x_i'b|x_i) \|x\|$, 
$$\mathbb E(\|  \nabla q\left(\beta_{i-1},Y_{i}\right) -\nabla q\left(\beta^*,Y_{i}\right)\|^2|\mathcal F_{i-1})\leq
\mathbb E(\|A_i\|^2|\mathcal F_{i-1})
\leq h(\beta_{i-1}-\beta^*).
$$
\end{proof}

\subsection{Proof of Theorem~\ref{thm:d}}

  \begin{proof}[Proof of Theorem \ref{thm:d}] 

 By (\ref{eq20}), uniformly in $r$, and $d=1,2,$
$$
i^{-1/2} \sum_{j=1}^{[ir]} [\beta_{j,sub}^d- \beta^*_{\tau_d,sub} ]=\frac{1}{\sqrt{i}}\sum_{j=1}^{\left[ir\right]}(H_{d}^{-1})_{sub}\xi_{j,d}
+o_{P}(1)
$$
where $(H_{d}^{-1})_{sub}$ denotes the rows of $H_d= \mathbb E[x_ix_i'f_{\varepsilon,d}(0|x_i)]$ corresponding to the subvector $\beta_{\tau_d,sub}^*$;
 $f_{\varepsilon,d}(|x)$ denotes the conditional density of $\varepsilon_{i,d}$ for $d=1,2$,
$$\xi_{i,d}\left(\beta\right):
=\mathbb Ex_i[P(\varepsilon_{i,d}\leq x_i'(\beta-\beta_{\tau_d}^*)|x_i)- \tau_d]-x_i[I\{\varepsilon_{i,d}\leq x_i'(\beta-\beta_{\tau_d}^*)\}-\tau_d].
$$
and $\xi_{i,d}:
=\xi_{i,d}(\beta_{i-1}^d)
$.  
Then we have 
$$
\sqrt{i}\bar\Delta_i(r):= i^{-1/2} \sum_{j=1}^{[ir]}\begin{pmatrix}
 \beta_{j,sub}^1- \beta^*_{\tau_1,sub} \\
  \beta_{j,sub}^2- \beta^*_{\tau_2,sub} 
\end{pmatrix}
=\frac{1}{\sqrt{i}}\sum_{j=1}^{\left[ir\right]}\bar Hu_{j}
+o_{P}(1),\quad \bar H=\begin{pmatrix}
  (H_1^{-1})_{sub} &0\\
 0& (H_2^{-1})_{sub}
\end{pmatrix}
$$
$$
u_j:= u_j(\beta^1_{j-1},\beta^2_{j-1}),\quad 
u_j(\beta^1, \beta^2):= \begin{pmatrix}
  \xi_{j,1}(\beta^1)\\
\xi_{j,2}(\beta^2)
\end{pmatrix}
$$
Then 
$$
u_i(\beta^*_{\tau_1}, \beta^*_{\tau_2})=    - \begin{pmatrix}
   [I\{\varepsilon_{i,1}\leq 0\}-\tau_1] \\
 [I\{\varepsilon_{i,2}\leq 0\}-\tau_2]
\end{pmatrix}\otimes x_i.
$$
Hence,   $
 \text{Var}(u_i(\beta^*_{\tau_1}, \beta^*_{\tau_2})|\mathcal F_{i-1}) \to^P  A\otimes\mathbb E x_ix_i',
 $ 
where 
$$
  A:= \begin{pmatrix}
  \tau_1(1-\tau_1) & \tau_1\wedge\tau_2 -\tau_1\tau_2 \\
\tau_1\wedge\tau_2 -\tau_1\tau_2 &  \tau_2(1-\tau_2).
\end{pmatrix} . 
$$

 Note that $\bar Hu_{j}$ is an mds. We apply the mds CLT to  the partial sum $\frac{1}{\sqrt{i}}\sum_{j=1}^{\left[ir\right]}\bar Hu_{j}$, which converges weakly: 
$$
\sqrt{i}\bar\Delta_i(r)\Rightarrow \bar\Upsilon^{1/2}W(r),\quad r\in[0,1]
$$
 where $W(r)$ stands for a vector of independent standard Wiener process on $[0,1]$, and $\bar\Upsilon:= \bar H  (A\otimes\mathbb E x_ix_i')\bar H $.  
Now let $C_n(r):= G \sqrt{n}\bar\Delta_n(r)$, with $G=(I, -I)$. Also let 
$\Lambda= (G\bar \Upsilon G')^{1/2}$. 
Then   for some vector of independent standard Wiener process $W^*(r)$,
$
C_n(r)\Rightarrow \Lambda W^*(r).
$
In addition, 
\begin{eqnarray*}
G\bar V_nG'= &=& \frac{1}{n}\sum_{s=1}^n[C_n(\frac{s}{n})- \frac{s}{n}C_n(1)][C_n(\frac{s}{n})- \frac{s}{n}C_n(1)]'\cr
&=&\int_0^1[C_n(r)-rC_n(1)][C_n(r)-rC_n(1)]'dr,
\end{eqnarray*}
the last equality holds because $C_n(r)$ is a partial sum.
Also, under the null that $\beta^*_{\tau_1,sub}= \beta^*_{\tau_2,sub}$, 
$$
\bar\beta_{n,sub}^1-\bar\beta_{n,sub}^2= G\begin{pmatrix}
\bar\beta_{n,sub}^1- \beta^*_{\tau_1,sub}\\
\bar\beta_{n,sub}^2 -\beta^*_{\tau_2,sub}
\end{pmatrix}=\frac{1}{\sqrt{n}}C_n(1). 
$$
Hence  the test statistic $n(\bar\beta_{n,sub}^1-\bar\beta_{n,sub}^2)' (G\bar V_{n,sub}G')^{-1}(\bar\beta_{n,sub}^1-\bar\beta_{n,sub}^2)$ equals
$$
  C_n(1)'\left(\int_0^1[C_n(r)-rC_n(1)][C_n(r)-rC_n(1)]'dr\right)^{-1} C_n(1)
$$
  is a continuous functional of $C_n(\cdot)$.  By the continuous mapping theorem, the test statistic converges weakly to the desired limit.
  \end{proof}


\bibliographystyle{chicago}
\bibliography{LLSS_bib}

 \clearpage

\renewcommand{\thepage}{S-\arabic{page}}
\setcounter{page}{1}
\setcounter{section}{0}

\section*{Online Supplements for ``Fast Inference for Quantile Regression with Tens of Millions of Observations'' (Not for Publication)}

\section*{Sokbae Lee, Yuan Liao, Myung Hwan Seo, and Youngk Shin.}

This part of the appendix is only for online supplements. It contains additional Lemma's with proofs and additional tables from Monte Carlo experiments and empirical application.

\section{Additional Lemma's}

\begin{lem}[$L^2$-Consistency]\label{leafdafa}  
	Let Assumptions
	\ref{asm:iid}- \ref{as:suff:cons} hold. 
	Then, as $n \rightarrow \infty$, $\mathbb E \| \bar \beta_n - \beta^* \|^2=o(1)$.
\end{lem}

\begin{proof}[Proof of Lemma~\ref{leafdafa}]
	The desired conclusion follows immediately from Theorem 2  of \citet{gadat2022optimal} if we verify the conditions imposed in  their Theorem 2. In particular, we provide sufficient conditions for a Kurdyka--Łojasiewicz inequality (with $r=0$ using the notation in  \citet{gadat2022optimal}):
	\begin{align*}
		\lim\inf_{\left|h\right|\to\infty} \left\|\nabla Q\left(h\right)\right\|>0 
		\ \  \text{ and } \ \
		\lim\sup_{\left|h\right|\to\infty} \left\|\nabla Q\left(h\right)\right\|< + \infty.
	\end{align*}
	As the second condition is trivially satisfied, it suffices to show that  
	under Assumption~\ref{as:suff:cons}(i)-(ii), $\lim\inf_{\left|h\right|\to\infty}\|\nabla Q(h)\|>0$.
	We write $G(\beta):= \nabla^2 Q(\beta)$. Then the mean-value theorem with the integral remainder yields 
	$$
	\nabla Q(\beta) =\int_{0}^{| \beta-\beta^* |}G\left(\frac{\left(\beta-\beta^{*}\right)}{\left\|\beta-\beta^{*}\right\|}s+\beta^{*}\right)\frac{\left(\beta-\beta^{*}\right)}{\left\|\beta-\beta^{*}\right\|}ds.
	$$ Let $A(\beta,s):= G\left(\frac{\left(\beta-\beta^{*}\right)}{\left\|\beta-\beta^{*}\right\|}s+\beta^{*}\right)$ and $K(\beta):=\int_0^{\| \beta-\beta^* \|} A(\beta,s)ds$.
	Then $K(\beta)$ is  positive semi-definite. For any $\left\| \beta-\beta^{*} \right\|>\epsilon$, write, with $b:=(\beta-\beta^*)$, 
	\begin{align*}
		\| \nabla Q(\beta)\|^2 & = | b |^{-2}b^T K(\beta)^2 b
		\geq \lambda^2_{\min}( K(\beta)) \\
		&\geq \lambda^2_{\min}( \int_0^{  \epsilon} A(\beta,s)ds) \geq\inf_{s<\epsilon, | h |=1 }\lambda^2_{\min}( G(hs+\beta^*)) \epsilon  > \epsilon c_0^2.
	\end{align*}
	Note that $\inf_{s<\epsilon, | h |=1 }\lambda_{\min}( G(hs+\beta^*)) = \inf_{  | \beta-\beta^* |<\epsilon }\lambda_{\min}( G( \beta)) $.
	Then, with Assumptions~\ref{as:suff:cons},
	it follows from Theorem 2  of \citet{gadat2022optimal} 
	that $\mathbb E \| \bar \beta_n - \beta^* \|^2=o(1)$.
\end{proof}

\begin{lem} 
	[Local around the True Minimizer]\label{ass:PJ's Assumption 3}
	Let $H= \nabla^2 Q(\beta^*)$ and  $\Psi(x)=Q(x+\beta^*)- Q(\beta^*)$. Suppose Assumptions \ref{asm:obj},  \ref{aslcox}, \ref{as2fda} hold. 
	There exists  $\varepsilon>0$ that for all $\| \beta-\beta^* \| \leq\varepsilon$,  the following  results hold. 
	\begin{enumerate} [(i)]
		\item $\Psi(0)=0$.
		\item $\Psi(\beta-\beta^*)\geq\alpha \| \beta-\beta^* \|^{2}$ for some $\alpha>0$. 
		\item 
		$|\nabla \Psi(x)-\nabla \Psi(y)|\leq L\|x-y\|$ for some $ L>0$.
		
		\item $\nabla \Psi(\beta-\beta^{*})' \nabla Q(\beta)>0$
		for $\beta\neq\beta^{*}$. 
		
		\item  $\nabla \Psi(\beta-\beta^{*})' \nabla Q(\beta)\geq\lambda \Psi(\beta-\beta^*)$ for some $\lambda   >0$.
		
		\item
		$\|\nabla Q(\beta)-H(\beta-\beta^{*})\|\leq K_{1}\|\beta-\beta^{*}\|^{1+\eta}$
		for some $K_{1}<\infty$ and $0<\eta\leq1$.
	\end{enumerate}   
\end{lem}

\begin{proof} Let $L=\sup_{\beta\in\Theta}\|\nabla^2 Q(\beta)\|$ and   $\nabla^2 Q(\beta)= G(\beta)$.

	(i) is naturally satisfied.
	
	(ii) Note $\nabla Q(\beta^*)=0$.   By the second-order Taylor expansion,  $\Psi(x)=0.5x'\nabla^2 Q(\widetilde \beta) x$, for some $\widetilde\beta$, and $\nabla^2 Q(\widetilde\beta)= G(\widetilde\beta)$ whose minimum eigenvalue is locally lower bounded by Assumption \ref{aslcox}.   So $\Psi(x)\geq \alpha\|x\|^2$, where $x= 0.5c_0$ for the constant $c_0$ in  Assumption \ref{aslcox}.
	
	(iii)     For any $\beta_1, \beta_2, $ there exists $\widetilde\beta$,  
	\begin{eqnarray*}
		\|\nabla \Psi(\beta_1)-\nabla \Psi(\beta_2)\| &=& \|\nabla^2Q(\widetilde\beta)(\beta_1-\beta_2)\|
		\leq L\|\beta_1-\beta_2\|.
	\end{eqnarray*}
	
	(iv) 
	We have  $\nabla \Psi(\beta-\beta^*)= \nabla Q(\beta)= G(\widetilde\beta)(\beta-\beta^*)$ for some $\widetilde\beta$ on the segment joining $\beta$ and $\beta^*$. So  $\|\widetilde\beta-\beta^*\|\leq\|\beta-\beta^*\|$.
	Hence uniformly in $\|\beta-\beta\|<\epsilon$, and $\beta\neq\beta^*$,  and $c_0>0$ in  Assumption \ref{aslcox}, 
	$$
	\nabla \Psi(\beta-\beta^{*})'\nabla Q(\beta)=\|\nabla Q(\beta)\|^2=(\beta-\beta^*)'G(\widetilde\beta)(\beta-\beta^*)
	\geq c_0 \| \beta-\beta^* \|^2>0.
	$$

	(v) For some $\bar\beta$,
	$$
	\Psi(\beta-\beta^*)=Q(\beta)-Q(\beta^*)= \frac{1}{2}  (\beta-\beta^*)' G(\bar\beta)(\beta-\beta^*),
	$$
	which is upper bounded by $0.5  L | \beta-\beta^* |^2$.
	Hence 
	$
	\nabla \Psi(\beta-\beta^{*})'\nabla Q(\beta)\geq\lambda \Psi(\beta-\beta^*)$ holds for $\lambda\leq 2c_0/L.$
	
	(vi)  The Taylor expansion yields, for some $\widetilde\beta$ so that $\|\widetilde\beta-\beta^*\|\leq \|\beta-\beta^*\| $, we have  
	$\nabla Q(\beta)= G(\widetilde\beta) (\beta-\beta^*)$. Hence by Assumption \ref{as2fda}
	$$
	\|\nabla Q(\beta)- H(\beta-\beta^*)\|\leq \| G(\widetilde\beta) - H \| \|\beta-\beta^*\|
	\leq K_1 \| \beta-\beta^*\|^2.
	$$
	
\end{proof}

\begin{lem}\label{lema.3}

	The sequence $\left\{ \xi_{i} \right\}_{i\geq1}$ satisfies the following conditions. 
	\begin{enumerate}[(i)]

		\item  For some $K_{3} < \infty$ and  for all $i\geq1$, $$\mathbb E(\|\xi_{i}\|^{2}|\mathcal{{F}}_{i-1})+|\nabla Q(\beta_{i-1})|^{2}\leq K_{3}(1+\|\beta_{i-1}\|^{2}) \;\; a.s.$$
		
		\item The following decomposition holds:
		$\xi_{i}=\xi_{i}(\beta^*)+\zeta_{i}(\beta_{i-1}).$ 
		They satisfy:

		\begin{enumerate}[(a)]

			\item $\mathbb{E} \left[ \xi_{i}(\beta^*)\xi_{i}(\beta^*)'|\mathcal{{F}}_{i-1} \right] \xrightarrow{P}S$
			as $i\rightarrow\infty$ and $S>0$ ($S$ is symmetric and positive
			definite), 
			\item
			$\sup\limits _{i}\mathbb{E} (\|\xi_{i}(\beta^*)\|^{2}I(\|\xi_{i}(\beta^*)\|>C|\mathcal{\mathcal{{F}}}_{i-1})\xrightarrow{P}0$
			as $C\rightarrow\infty$, 
			\item
			for all $i$ large enough, $\mathbb{E} \left[ \|\zeta_{i}(\beta_{i-1})\|^{2}\|\mathcal{{F}}_{i-1} \right] \leq\delta(\beta_{i-1}-\beta^*)$
			a.s. with $\delta(x)\rightarrow0$ as $\|x\|\rightarrow0$.
		\end{enumerate}
	\end{enumerate}

\end{lem}

\begin{proof}
	(i) We have 
	$
	\xi_i= \nabla Q\left(\beta_{i-1}\right)- \nabla q\left(\beta_{i-1},Y_{i}\right).
	$
	Hence  Assumption \ref{as2fda}(i) implies 
	$$\mathbb E(\|\xi_{i}\|^{2}|\mathcal{{F}}_{i-1})+\|\nabla Q(\beta_{i-1})\|^{2}\leq 3 \|\nabla Q(\beta_{i-1})\|^{2}+2\mathbb E( \|\nabla q(\beta_{i-1},Y_i)\|^2|\mathcal F_{i-1}) \leq 3K_{2}(1+\|\beta_{i-1}\|^{2}) \;\; a.s.
	$$
	
	(ii) We have 
	$
	\xi_{i}(\beta^*)= -\nabla q\left(\beta^*,Y_{i}\right)
	$
	and
	$$
	\zeta_{i}(\beta_{i-1})= \nabla Q\left(\beta_{i-1}\right)-\left[\nabla q\left(\beta_{i-1},Y_{i}\right) -\nabla q\left(\beta^*,Y_{i}\right)\right].
	$$
	The desired results (a)(b) then follows from  Assumption \ref{as2fda}(ii).  As for (c),
	\begin{eqnarray*}
		\mathbb{E} \left[ \|\zeta_{i}(\beta_{i-1})\|^{2}|\mathcal{{F}}_{i-1} \right]  &\leq& a_1+ a_2 \cr 
		a_1&=&   2 \| \nabla Q\left(\beta_{i-1}\right)-  \nabla Q\left(\beta^* \right)\|^2 
		\leq 2\sup_b\|\nabla^2 Q(b)\|\|\beta_{i-1}-\beta^*\|^2
		\cr 
		a_2&=&2\mathbb E(\|  \nabla q\left(\beta_{i-1},Y_{i}\right) -\nabla q\left(\beta^*,Y_{i}\right)\|^2|\mathcal F_{i-1})\leq 2h(\beta_{i-1}-\beta^*)
	\end{eqnarray*}
	where the function $h()$ for term $a_2$ exists, from   Assumption \ref{as2fda}(iii).  Hence we can define 
	$$\delta(x)
	=2\sup_b\|\nabla^2 Q(b)\|\|x\|^2 +2 h(x)
	$$
	and $\delta(x)\to0$  as $\|x\|\to0$ because $\sup_b\|\nabla^2 Q(b)\|<\infty$ and $h(x)\to0$.
	
\end{proof}

%

\section{Additional Tables}

In this section we provide additional results from Monte Carlo simulation studies and the empirical application. Tables \ref{tb_d_10}--\ref{tb_d_320}
report performance measures for all simulation designs. Recall that S-subGD stands for the SGD random scaling method, QR for the standard quantile regression method, CONQUER-plugin for the conquer method with the plug-in asymptotic variance, CONQUER-bootstrap for the conquer method with bootstrapping, SGD-bootstrap for the SGD bootstrap method. In addition, we report the performance of S-subGD-all, which applies the S-subGD random scaling method for the full variance-covariance matrix. In contrast, S-subGD  \changedSL{updates only the asymptotic variance for $\beta_1$, which is the parameter of interest}.

\clearpage

\begin{table}[ht]
	\centering
	\caption{Simulation Results: $d=10$. Reported time is the average computational time over 10 replications. Coverage Rate an CI Length are based on 1000 replications.} 
	\label{tb_d_10}
	\begin{tabular}{lS[table-format=4.2]S[table-format=4.2]S[table-format=1.3]S[table-format=1.4]}
		\hline
		{Method} & {Time (sec.)} & {Relative Time} & {Coverage Rate} & {CI Length} \\ 
		\hline
		\underline{$n=10^5$}\\
		S-subGD & 0.06 & 1.00 & 0.945 & 0.0204 \\ 
		S-subGD-all & 0.10 & 1.67 & 0.945 & 0.0204 \\ 
		QR & 0.91 & 15.17 & 0.957 & 0.0155 \\ 
		CONQUER-plugin & 154.04 & 2567.33 & 0.959 & 0.0152 \\ 
		CONQUER-bootstrap & 16.72 & 278.67 & 0.948 & 0.0151 \\ 
		SGD-bootstrap & 14.59 & 243.17 & 0.966 & 0.0171 \\ 
		\\
		\underline{$n=10^6$}\\
		S-subGD & 0.58 & 1.00 & 0.939 & 0.0065 \\ 
		S-subGD-all & 1.26 & 2.17 & 0.939 & 0.0065 \\ 
		QR & 8.52 & 14.69 & 0.939 & 0.0049 \\ 
		CONQUER-plugin &{NA}  &{NA}  &    {NA} &       {NA} \\ 
		CONQUER-bootstrap & 254.15 & 438.19 & 0.937 & 0.0048 \\ 
		SGD-bootstrap & 139.84 & 241.10 & 0.944 & 0.0050 \\ 
		\\
		\underline{$n=10^7$}\\
		S-subGD & 5.87 & 1.00 & 0.965 & 0.0020 \\ 
		S-subGD-all & 12.26 & 2.09 & 0.965 & 0.0020 \\ 
		QR & 78.43 & 13.36 & 0.963 & 0.0016 \\ 
		CONQUER-plugin &{NA}  &{NA}  &    {NA} &       {NA} \\ 
		CONQUER-bootstrap & 3329.80 & 567.26 & 0.965 & 0.0015 \\ 
		SGD-bootstrap & 1413.33 & 240.77 & 0.959 & 0.0016 \\ 
		\hline
	\end{tabular}
\end{table}
\begin{table}[ht]
	\centering
	\caption{Simulation Results: $d=20$} 
	\label{tb_d_20}
	\begin{tabular}{lS[table-format=4.2]S[table-format=4.2]S[table-format=1.3]S[table-format=1.4]}
		\hline
		{Method} & {Time (sec.)} & {Relative Time} & {Coverage Rate} & {CI Length} \\ 
		\hline
		\underline{$n=10^5$}\\
		S-subGD & 0.09 & 1.00 & 0.943 & 0.0208 \\ 
		S-subGD-all & 0.28 & 3.11 & 0.943 & 0.0208 \\ 
		QR & 1.73 & 19.22 & 0.947 & 0.0155 \\ 
		CONQUER-plugin & 373.00 & 4144.44 & 0.944 & 0.0152 \\ 
		CONQUER-bootstrap & 36.74 & 408.22 & 0.944 & 0.0151 \\ 
		SGD-bootstrap & 20.94 & 232.67 & 0.971 & 0.0193 \\ 
		\\
		\underline{$n=10^6$}\\
		S-subGD & 1.10 & 1.00 & 0.952 & 0.0065 \\ 
		S-subGD-all & 3.06 & 2.78 & 0.952 & 0.0065 \\ 
		QR & 16.96 & 15.42 & 0.958 & 0.0049 \\ 
		CONQUER-plugin &{NA}  &{NA}  &    {NA} &       {NA} \\ 
		CONQUER-bootstrap & 473.33 & 430.30 & 0.955 & 0.0048 \\ 
		SGD-bootstrap & 214.13 & 194.66 & 0.969 & 0.0052 \\ 
		\\
		\underline{$n=10^7$}\\
		S-subGD & 11.05 & 1.00 & 0.955 & 0.0020 \\ 
		S-subGD-all & 30.10 & 2.72 & 0.955 & 0.0020 \\ 
		QR & 100.17 & 9.07 & 0.951 & 0.0016 \\ 
		CONQUER-plugin &{NA}  &{NA}  &    {NA} &       {NA} \\ 
		CONQUER-bootstrap & 3844.13 & 347.89 & 0.950 & 0.0015 \\ 
		SGD-bootstrap & 2112.46 & 191.17 & 0.953 & 0.0016 \\ 
		\hline
	\end{tabular}
	
\end{table}
\begin{table}[ht]
	\centering
	\caption{Simulation Results: $d=40$} 
	\label{tb_d_40}
	\begin{tabular}{lS[table-format=4.2]S[table-format=4.2]S[table-format=1.3]S[table-format=1.4]}
		\hline
		{Method} & {Time (sec.)} & {Relative Time} & {Coverage Rate} & {CI Length} \\ 
		\hline
		\underline{$n=10^5$}\\
		S-subGD & 0.15 & 1.00 & 0.946 & 0.0215 \\ 
		S-subGD-all & 0.72 & 4.80 & 0.946 & 0.0215 \\ 
		QR & 3.65 & 24.33 & 0.947 & 0.0154 \\ 
		CONQUER-plugin & 668.34 & 4455.60 & 0.960 & 0.0153 \\ 
		CONQUER-bootstrap & 63.01 & 420.07 & 0.950 & 0.0151 \\ 
		SGD-bootstrap & 27.15 & 181.00 & 0.955 & 0.0277 \\ 
		\\
		\underline{$n=10^6$}\\
		S-subGD & 1.92 & 1.00 & 0.953 & 0.0066 \\ 
		S-subGD-all & 7.82 & 4.07 & 0.953 & 0.0066 \\ 
		QR & 46.23 & 24.08 & 0.944 & 0.0049 \\ 
		CONQUER-plugin &{NA}  &{NA}  &    {NA} &       {NA} \\ 
		CONQUER-bootstrap & 869.24 & 452.73 & 0.936 & 0.0048 \\ 
		SGD-bootstrap & 275.98 & 143.74 & 0.958 & 0.0056 \\ 
		\\
		\underline{$n=10^7$}\\
		S-subGD & 21.86 & 1.00 & 0.954 & 0.0020 \\ 
		S-subGD-all & 76.17 & 3.48 & 0.954 & 0.0020 \\ 
		QR & 221.46 & 10.13 & 0.955 & 0.0016 \\ 
		CONQUER-plugin &{NA}  &{NA}  &    {NA} &       {NA} \\ 
		CONQUER-bootstrap & 5769.39 & 263.92 & 0.952 & 0.0015 \\ 
		SGD-bootstrap & 2854.69 & 130.59 & 0.963 & 0.0016 \\ 
		\hline
	\end{tabular}
\end{table}
\begin{table}[ht]
	\caption{Simulation Results: $d=80$} 
	\label{tb_d_80}
	\centering
	\begin{tabular}{lS[table-format=4.2]S[table-format=4.2]S[table-format=1.3]S[table-format=1.4]}
		\hline
		{Method} & {Time (sec.)} & {Relative Time} & {Coverage Rate} & {CI Length} \\  
		\hline
		\underline{$n=10^5$}\\
		S-subGD & 0.41 & 1.00 & 0.955 & 0.0240 \\ 
		S-subGD-all & 2.33 & 5.68 & 0.955 & 0.0240 \\ 
		QR & 12.15 & 29.63 & 0.936 & 0.0153 \\ 
		CONQUER-plugin & 1124.94 & 2743.76 & 0.955 & 0.0152 \\ 
		CONQUER-bootstrap & 123.20 & 300.49 & 0.939 & 0.0151 \\ 
		SGD-bootstrap & 38.42 & 93.71 & 0.940 & 0.0560 \\ 
		\\
		\underline{$n=10^6$}\\
		S-subGD & 4.01 & 1.00 & 0.956 & 0.0067 \\ 
		S-subGD-all & 24.47 & 6.10 & 0.956 & 0.0067 \\ 
		QR & 154.75 & 38.59 & 0.955 & 0.0049 \\ 
		CONQUER-plugin &{NA}  &{NA}  &    {NA} &       {NA} \\ 
		CONQUER-bootstrap & 1739.38 & 433.76 & 0.954 & 0.0048 \\ 
		SGD-bootstrap & 429.33 & 107.06 & 0.976 & 0.0077 \\ 
		\\
		\underline{$n=10^7$}\\
		S-subGD & 43.12 & 1.00 & 0.952 & 0.0020 \\ 
		S-subGD-all & 239.50 & 5.55 & 0.952 & 0.0020 \\ 
		QR & 571.91 & 13.26 & 0.942 & 0.0016 \\ 
		CONQUER-plugin &{NA}  &{NA}  &    {NA} &       {NA} \\ 
		CONQUER-bootstrap & 8439.91 & 195.73 & 0.937 & 0.0015 \\ 
		SGD-bootstrap & 4680.84 & 108.55 & 0.951 & 0.0017 \\ 
		\hline
	\end{tabular}
\end{table}
\begin{table}[ht]
	\caption{Simulation Results: $d=160$} 
	\label{tb_d_160}
	\centering
	\begin{tabular}{lS[table-format=5.2]S[table-format=4.2]S[table-format=1.3]S[table-format=1.4]}
		\hline
		{Method} & {Time (sec.)} & {Relative Time} & {Coverage Rate} & {CI Length} \\  
		\hline
		\underline{$n=10^5$}\\
		S-subGD & 0.71 & 1.00 & 0.954 & 0.0299 \\ 
		S-subGD-all & 8.72 & 12.28 & 0.954 & 0.0299 \\ 
		QR & 40.44 & 56.96 & 0.930 & 0.0149 \\ 
		CONQUER-plugin & 1955.91 & 2754.80 & 0.952 & 0.0151 \\ 
		CONQUER-bootstrap & 276.53 & 389.48 & 0.943 & 0.0151 \\ 
		SGD-bootstrap & 82.20 & 115.77 & 0.929 & 0.1387 \\ 
		\\
		\underline{$n=10^6$}\\
		S-subGD & 8.37 & 1.00 & 0.962 & 0.0072 \\ 
		S-subGD-all & 88.54 & 10.58 & 0.962 & 0.0072 \\ 
		QR & 555.06 & 66.32 & 0.951 & 0.0049 \\ 
		CONQUER-plugin &{NA}  &{NA}  &    {NA} &       {NA} \\ 
		CONQUER-bootstrap & 3542.03 & 423.18 & 0.957 & 0.0048 \\ 
		SGD-bootstrap & 936.23 & 111.86 & 0.947 & 0.0151 \\ 
		\\
		\underline{$n=10^7$}\\
		S-subGD & 81.35 & 1.00 & 0.953 & 0.0021 \\ 
		S-subGD-all & 873.67 & 10.74 & 0.953 & 0.0021 \\ 
		QR & 1830.76 & 22.50 & 0.950 & 0.0016 \\ 
		CONQUER-plugin & NA &NA  &       NA &      NA \\ 
		CONQUER-bootstrap & 14692.08 & 180.60 & 0.952 & 0.0015 \\ 
		SGD-bootstrap & 9178.86 & 112.83 & 0.961 & 0.0022 \\ 
		\hline
	\end{tabular}
\end{table}
\begin{table}[ht]
	\caption{Simulation Results: $d=320$} 
	\label{tb_d_320}
	\centering
	\begin{tabular}{lS[table-format=5.2]S[table-format=4.2]S[table-format=1.3]S[table-format=1.4]}
		\hline
		{Method} & {Time (sec.)} & {Relative Time} & {Coverage Rate} & {CI Length} \\  
		\hline
		\underline{$n=10^5$}\\
		S-subGD & 1.56 & 1.00 & 0.986 & 0.0473 \\ 
		S-subGD-all & 37.35 & 23.94 & 0.986 & 0.0473 \\ 
		QR & 161.93 & 103.80 & 0.926 & 0.0141 \\ 
		CONQUER-plugin & 3813.97 & 2444.85 & 0.955 & 0.0151 \\ 
		CONQUER-bootstrap & 575.50 & 368.91 & 0.956 & 0.0150 \\ 
		SGD-bootstrap & 146.56 & 93.95 & 0.910 & 0.3674 \\ 
		\\
		\underline{$n=10^6$}\\
		S-subGD & 15.65 & 1.00 & 0.962 & 0.0083 \\ 
		S-subGD-all & 331.31 & 21.17 & 0.962 & 0.0083 \\ 
		QR & 1373.95 & 87.79 & 0.953 & 0.0048 \\ 
		CONQUER-plugin &{NA}  &{NA}  &    {NA} &       {NA} \\ 
		CONQUER-bootstrap & 4112.76 & 262.80 & 0.952 & 0.0048 \\ 
		SGD-bootstrap & 1642.90 & 104.98 & 0.922 & 0.0364 \\ 
		\\
		\underline{$n=10^7$}\\
		S-subGD & 166.40 & 1.00 & 0.963 & 0.0011 \\ 
		S-subGD-all & 3349.89 & 20.13 & 0.963 & 0.0011 \\ 
		QR &{NA}  &{NA}  &    {NA} &       {NA} \\ 
		CONQUER-plugin &{NA}  &{NA}  &    {NA} &       {NA} \\ 
		CONQUER-bootstrap &{NA}  &{NA}  &    {NA} &       {NA} \\ 
		SGD-bootstrap & 16322.79 & 98.09 & 0.954 & 0.0025 \\ 
		\hline
	\end{tabular}
\end{table}

\clearpage

\subsection{Empirical application: full estimation results}


We provide full estimation results.

\begin{table}[ht]
	\centering
	\caption{College Wage Premium: Combining 5-Year Data} 
\begin{tabular}{ccccc}
	\hline
	Year & Female & Male & Difference & Time (sec.) \\ 
	\hline
	\underline{$\tau=0.1$}\\
	1980 & 0.1689 & 0.1737 & -0.0048 & 375 \\ 
	& [0.1667,0.1711] & [0.1700,0.1774] & [-0.0087,-0.0010] &  \\ 
	1990 & 0.3091 & 0.2632 & 0.0458 & 356 \\ 
	& [0.3056,0.3125] & [0.2584,0.2680] & [0.0417,0.0500] &  \\ 
	2000 & 0.3340 & 0.2929 & 0.0411 & 395 \\ 
	& [0.3323,0.3357] & [0.2896,0.2961] & [0.0377,0.0446] &  \\ 
	2001-2005  & 0.3661 & 0.3363 & 0.0299 & 180 \\ 
	& [0.3595,0.3728] & [0.3297,0.3428] & [0.0175,0.0422] &  \\ 
	2006-2010 & 0.3911 & 0.3689 & 0.0221 & 448 \\ 
	& [0.3861,0.3961] & [0.3626,0.3753] & [0.0200,0.0242] &  \\ 
	2011-2015 & 0.3946 & 0.3737 & 0.0209 & 395 \\ 
	& [0.3921,0.3971] & [0.3677,0.3797] & [0.0134,0.0284] &  \\ 
	\underline{$\tau=0.2$}\\
	1980 & 0.2093 & 0.1974 & 0.0120 & 423 \\ 
	& [0.2072,0.2114] & [0.1944,0.2003] & [0.0078,0.0162] &  \\ 
	1990 & 0.3391 & 0.2821 & 0.0570 & 454 \\ 
	& [0.3352,0.3429] & [0.2796,0.2846] & [0.0532,0.0607] &  \\ 
	2000 & 0.3770 & 0.3084 & 0.0686 & 488 \\ 
	& [0.3737,0.3804] & [0.3061,0.3107] & [0.0650,0.0723] &  \\ 
	2001-2005 & 0.4076 & 0.3519 & 0.0557 & 195 \\ 
	& [0.4051,0.4101] & [0.3499,0.3540] & [0.0524,0.0589] &  \\ 
	2006-2010 & 0.4328 & 0.3897 & 0.0431 & 447 \\ 
	& [0.4299,0.4357] & [0.3861,0.3932] & [0.0411,0.0452] &  \\ 
	2011-2015 & 0.4385 & 0.4023 & 0.0362 & 463 \\ 
	& [0.4353,0.4416] & [0.4002,0.4043] & [0.0316,0.0407] &  \\ 
	\underline{$\tau=0.3$}\\
	1980 & 0.2272 & 0.2025 & 0.0247 & 417 \\ 
	& [0.2250,0.2295] & [0.1988,0.2063] & [0.0222,0.0271] &  \\ 
	1990 & 0.3538 & 0.2886 & 0.0652 & 421 \\ 
	& [0.3523,0.3554] & [0.2872,0.2900] & [0.0633,0.0672] &  \\ 
	2000 & 0.3902 & 0.3158 & 0.0745 & 472 \\ 
	& [0.3862,0.3943] & [0.3147,0.3169] & [0.0703,0.0786] &  \\ 
	2001-2005 & 0.4226 & 0.3582 & 0.0645 & 200 \\ 
	& [0.4193,0.4260] & [0.3560,0.3603] & [0.0603,0.0686] &  \\ 
	2006-2010 & 0.4508 & 0.3969 & 0.0539 & 490 \\ 
	& [0.4485,0.4532] & [0.3934,0.4005] & [0.0494,0.0585] &  \\ 
	2011-2015 & 0.4631 & 0.4143 & 0.0487 & 448 \\ 
	& [0.4610,0.4651] & [0.4112,0.4175] & [0.0443,0.0531] &  \\ 
	\hline
\end{tabular}
\end{table}

\begin{table}[ht]
\centering
\caption{College Wage Premium: Combining 5-Year Data (cont.)} 
\begin{tabular}{ccccc}
	\hline
	Year & Female & Male & Difference & Time (sec.) \\ 
	\hline
	\underline{$\tau=0.4$}\\
	1980 & 0.2393 & 0.2006 & 0.0386 & 426 \\ 
	& [0.2375,0.2410] & [0.1984,0.2029] & [0.0364,0.0409] &  \\ 
	1990 & 0.3628 & 0.2921 & 0.0707 & 402 \\ 
	& [0.3599,0.3657] & [0.2914,0.2929] & [0.0676,0.0738] &  \\ 
	2000 & 0.4009 & 0.3240 & 0.0769 & 412 \\ 
	& [0.3969,0.4048] & [0.3227,0.3252] & [0.0736,0.0802] &  \\ 
	2001-2005 & 0.4342 & 0.3641 & 0.0702 & 185 \\ 
	& [0.4312,0.4373] & [0.3611,0.3670] & [0.0651,0.0752] &  \\ 
	2006-2010 & 0.4661 & 0.4025 & 0.0636 & 492 \\ 
	& [0.4619,0.4703] & [0.3993,0.4057] & [0.0568,0.0705] &  \\ 
	2011-2015 & 0.4822 & 0.4243 & 0.0579 & 438 \\ 
	& [0.4807,0.4837] & [0.4197,0.4289] & [0.0527,0.0631] &  \\ 
	\underline{$\tau=0.5$}\\
	1980 & 0.2552 & 0.1993 & 0.0559 & 457 \\ 
	& [0.2521,0.2583] & [0.1979,0.2007] & [0.0529,0.0589] &  \\ 
	1990 & 0.3700 & 0.2940 & 0.0761 & 401 \\ 
	& [0.3681,0.3720] & [0.2934,0.2946] & [0.0740,0.0781] &  \\ 
	2000 & 0.4124 & 0.3331 & 0.0792 & 423 \\ 
	& [0.4101,0.4147] & [0.3322,0.3341] & [0.0773,0.0812] &  \\ 
	2001-2005 & 0.4459 & 0.3732 & 0.0727 & 185 \\ 
	& [0.4435,0.4483] & [0.3708,0.3756] & [0.0688,0.0766] &  \\ 
	2006-2010 & 0.4782 & 0.4135 & 0.0648 & 453 \\ 
	& [0.4765,0.4799] & [0.4112,0.4157] & [0.0614,0.0681] &  \\ 
	2011-2015 & 0.4962 & 0.4383 & 0.0579 & 438 \\ 
	& [0.4940,0.4984] & [0.4337,0.4429] & [0.0544,0.0614] &  \\ 
	\underline{$\tau=0.6$}\\
	1980 & 0.2728 & 0.2001 & 0.0727 & 461 \\ 
	& [0.2702,0.2755] & [0.1983,0.2019] & [0.0706,0.0748] &  \\ 
	1990 & 0.3804 & 0.2980 & 0.0824 & 439 \\ 
	& [0.3785,0.3823] & [0.2965,0.2996] & [0.0798,0.0849] &  \\ 
	2000 & 0.4244 & 0.3449 & 0.0795 & 411 \\ 
	& [0.4227,0.4261] & [0.3435,0.3463] & [0.0775,0.0815] &  \\ 
	2001-2005 & 0.4559 & 0.3879 & 0.0679 & 172 \\ 
	& [0.4532,0.4586] & [0.3857,0.3902] & [0.0635,0.0724] &  \\ 
	2006-2010 & 0.4899 & 0.4300 & 0.0599 & 501 \\ 
	& [0.4881,0.4918] & [0.4281,0.4319] & [0.0583,0.0615] &  \\ 
	2011-2015 & 0.5076 & 0.4536 & 0.0540 & 430 \\ 
	& [0.5040,0.5112] & [0.4491,0.4582] & [0.0514,0.0565] &  \\ 
	\hline
\end{tabular}
\end{table}

\begin{table}[ht]
\centering
\caption{College Wage Premium: Combining 5-Year Data (cont.)} 
\begin{tabular}{ccccc}
	\hline
	Year & Female & Male & Difference & Time (sec.) \\ 
	\hline
	\underline{$\tau=0.7$}\\
	1980 & 0.2902 & 0.2047 & 0.0855 & 462 \\ 
	& [0.2878,0.2925] & [0.2032,0.2062] & [0.0827,0.0883] &  \\ 
	1990 & 0.3909 & 0.3044 & 0.0865 & 397 \\ 
	& [0.3879,0.3939] & [0.3021,0.3066] & [0.0832,0.0898] &  \\ 
	2000 & 0.4318 & 0.3633 & 0.0685 & 487 \\ 
	& [0.4294,0.4343] & [0.3616,0.3650] & [0.0661,0.0709] &  \\ 
	2001-2005 & 0.4664 & 0.4083 & 0.0581 & 184 \\ 
	& [0.4642,0.4686] & [0.4047,0.4119] & [0.0533,0.0629] &  \\ 
	2006-2010 & 0.4965 & 0.4493 & 0.0473 & 483 \\ 
	& [0.4937,0.4993] & [0.4476,0.4509] & [0.0439,0.0507] &  \\ 
	2011-2015 & 0.5164 & 0.4718 & 0.0446 & 438 \\ 
	& [0.5131,0.5197] & [0.4681,0.4756] & [0.0429,0.0463] &  \\ 
	\underline{$\tau=0.8$}\\
	1980 & 0.3008 & 0.2182 & 0.0826 & 441 \\ 
	& [0.2991,0.3026] & [0.2165,0.2199] & [0.0801,0.0851] &  \\ 
	1990 & 0.3968 & 0.3180 & 0.0788 & 393 \\ 
	& [0.3943,0.3993] & [0.3162,0.3198] & [0.0750,0.0826] &  \\ 
	2000 & 0.4382 & 0.3937 & 0.0445 & 467 \\ 
	& [0.4356,0.4407] & [0.3907,0.3966] & [0.0409,0.0481] &  \\ 
	2001-2005 & 0.4725 & 0.4344 & 0.0380 & 164 \\ 
	& [0.4705,0.4744] & [0.4310,0.4379] & [0.0334,0.0427] &  \\ 
	2006-2010 & 0.5012 & 0.4707 & 0.0305 & 448 \\ 
	& [0.4986,0.5038] & [0.4682,0.4732] & [0.0276,0.0334] &  \\ 
	2011-2015 & 0.5238 & 0.4930 & 0.0307 & 456 \\ 
	& [0.5168,0.5307] & [0.4909,0.4952] & [0.0248,0.0367] &  \\ 
	\underline{$\tau=0.9$}\\
	1980 & 0.2889 & 0.2534 & 0.0355 & 453 \\ 
	& [0.2869,0.2910] & [0.2504,0.2565] & [0.0309,0.0401] &  \\ 
	1990 & 0.3901 & 0.3639 & 0.0262 & 412 \\ 
	& [0.3876,0.3926] & [0.3621,0.3657] & [0.0241,0.0283] &  \\ 
	2000 & 0.4456 & 0.4545 & -0.0090 & 495 \\ 
	& [0.4433,0.4479] & [0.4491,0.4600] & [-0.0126,-0.0053] &  \\ 
	2001-2005 & 0.4773 & 0.5073 & -0.0300 & 168 \\ 
	& [0.4749,0.4797] & [0.5014,0.5132] & [-0.0357,-0.0244] &  \\ 
	2006-2010 & 0.5082 & 0.5191 & -0.0109 & 492 \\ 
	& [0.5037,0.5127] & [0.5165,0.5218] & [-0.0138,-0.0081] &  \\ 
	2011-2015 & 0.5347 & 0.5417 & -0.0070 & 427 \\ 
	& [0.5319,0.5376] & [0.5403,0.5431] & [-0.0101,-0.0038] &  \\ 
	\hline
\end{tabular}
\end{table}

\subsection*{Empirical application: a single year for 2005, 2010, and 2015.}
We use a dataset composed of a single year sample for 2005, 2010, and 2015 without combining 5-year periods. As a result, the sample sizes for those years are around 25\% of the previous years. The computation times for those years are shorter but the confidence intervals are wider. The point estimates are similar to those in the baseline model. For 1980, 1990, and 2000, all results are the same with the baseline model.

\begin{table}[htb]
\centering
\caption{Summary Statistics: Single Years} 
\label{tb:summary_stat_no_merging}
\begin{tabular}{cccccc}
  \hline
Year & Sample Size & $\mathbb E(Female$) & $\mathbb E(Educ$) & $\mathbb E(Educ$|$Male$) & $\mathbb E(Educ$|$Female$) \\ 
  \hline
  1980 & 3,659,684 & 0.390 & 0.433 & 0.444 & 0.416 \\ 
  1990 & 4,192,119 & 0.425 & 0.543 & 0.537 & 0.550 \\ 
  2000 & 4,479,724 & 0.439 & 0.600 & 0.578 & 0.629 \\ 
  2005 & 931,533 & 0.447 & 0.646 & 0.620 & 0.678 \\ 
  2010 & 914,763 & 0.450 & 0.677 & 0.643 & 0.720 \\ 
  2015 & 919,702 & 0.446 & 0.695 & 0.652 & 0.747 \\ 
  \hline
\end{tabular}
\end{table}

\begin{figure}[htb]
    \centering
    \caption{College Wage Premium: Single Years}
    \label{fig:premium_design_03}
    \hskip15pt
    \begin{tabular}{c c c}
        \includegraphics[scale=0.23]{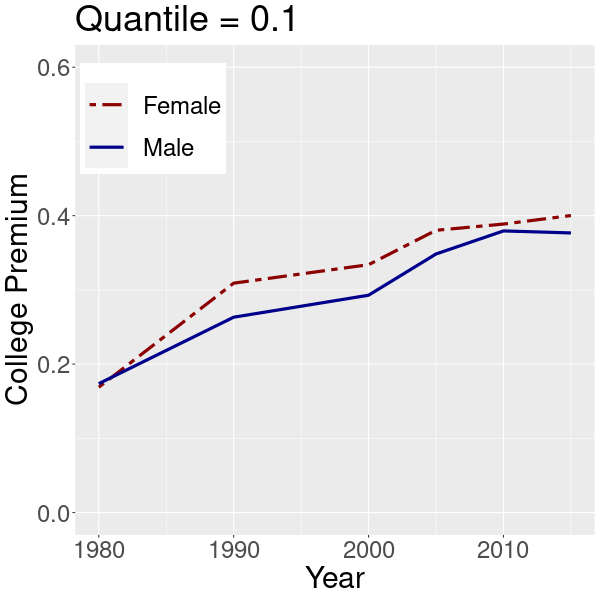} & \includegraphics[scale=0.23]{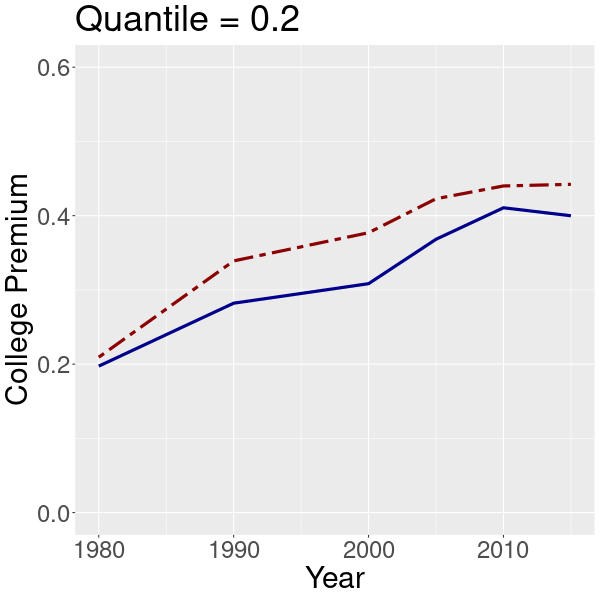} & \includegraphics[scale=0.23]{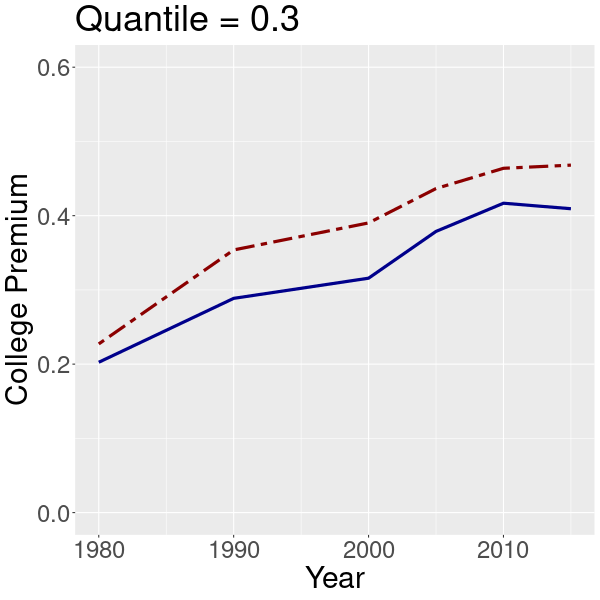} \\
        \includegraphics[scale=0.23]{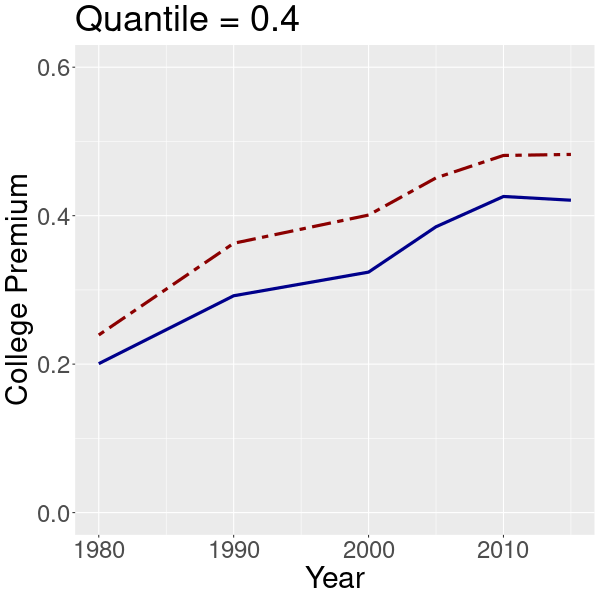} & \includegraphics[scale=0.23]{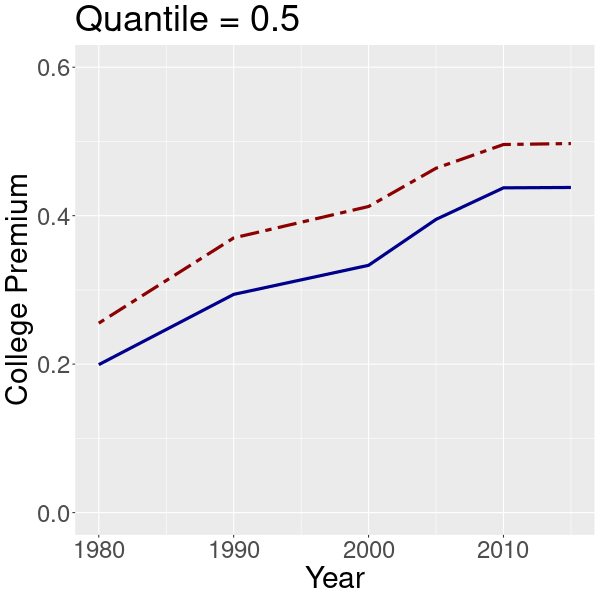} & \includegraphics[scale=0.23]{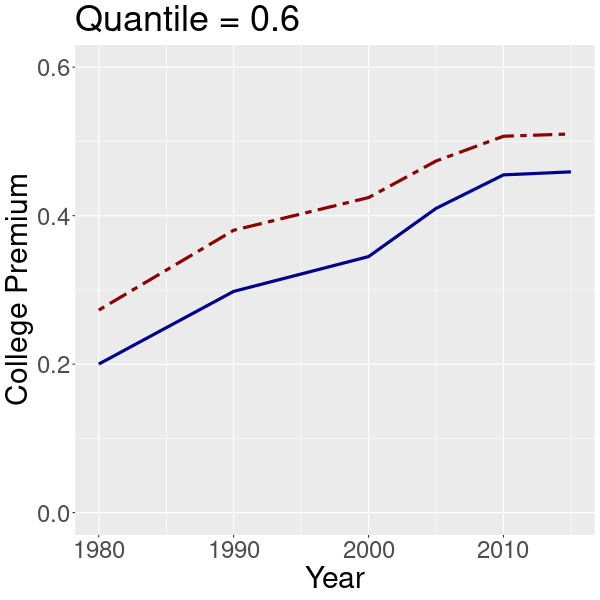} \\
        \includegraphics[scale=0.23]{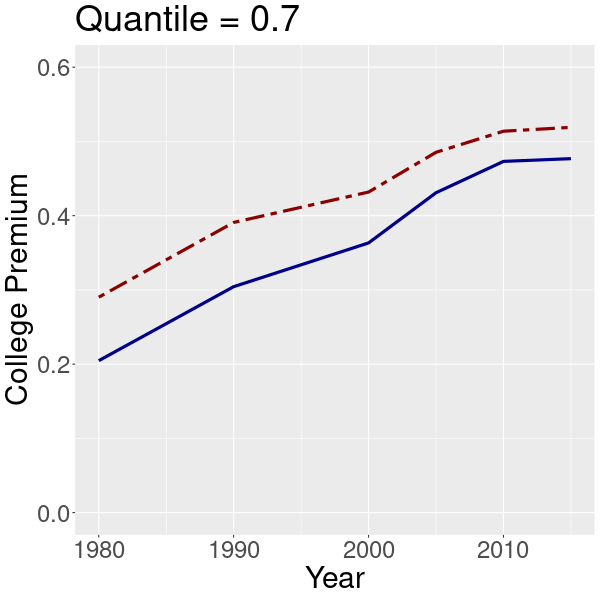} & \includegraphics[scale=0.23]{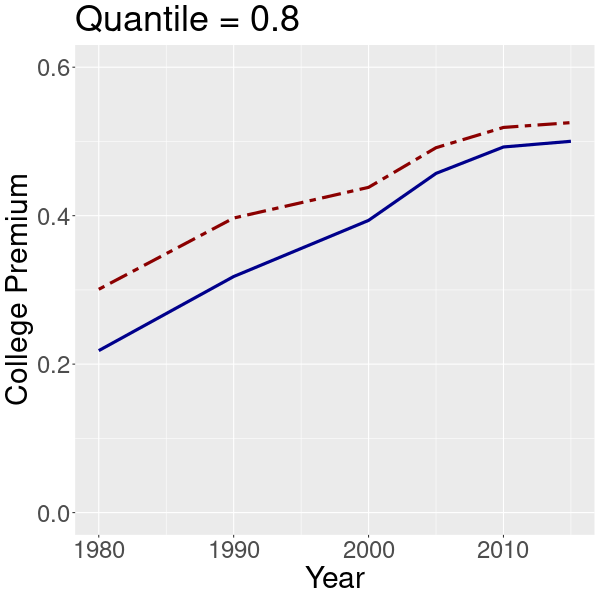} & \includegraphics[scale=0.23]{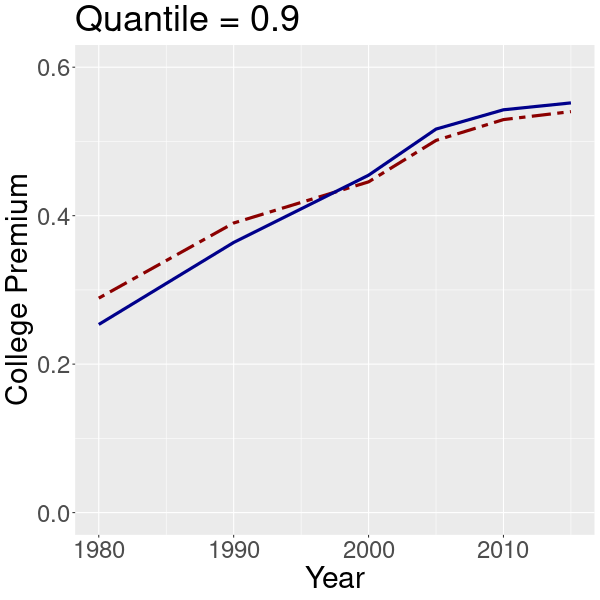} 
    \end{tabular}
    \flushleft{\footnotesize Notes. 
    The blue solid line (Male) comes from $\hat{\beta}_2$ each year and the red dot-dashed line (Female) does from the estimate $\hat{\beta}_2+\hat{\beta}_3$.}
\end{figure}

\begin{figure}[htb]
    \centering
    \caption{Female-Male College Premium Difference: Single Years}
    \label{fig:diff_design_03}
    \hskip15pt
    \begin{tabular}{c c c}
        \includegraphics[scale=0.23]{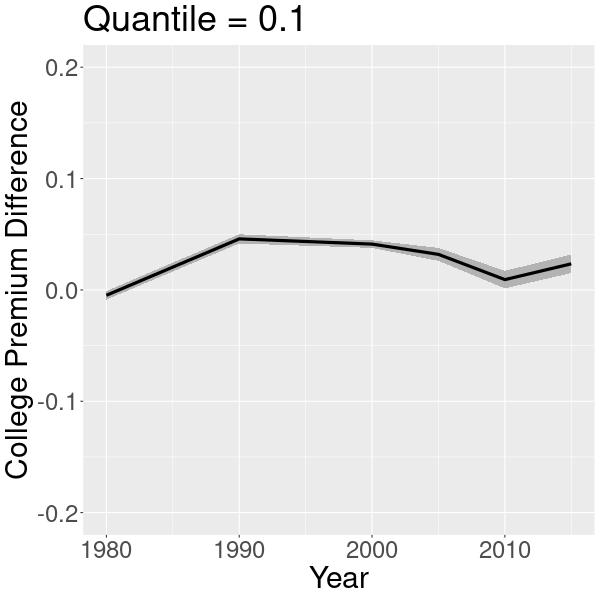} & \includegraphics[scale=0.23]{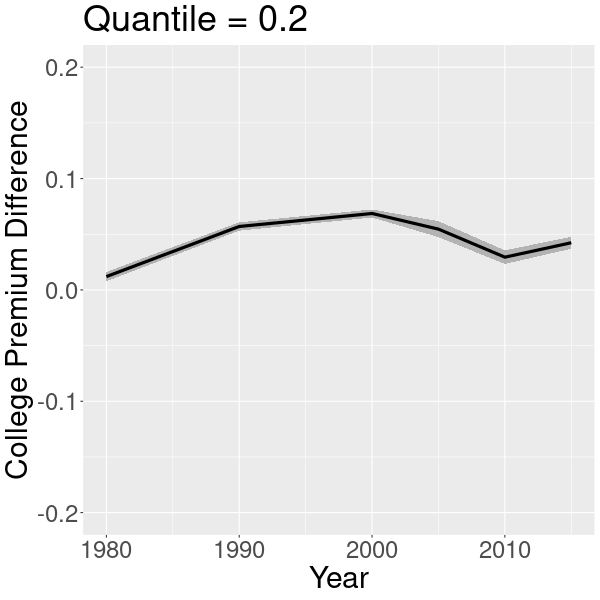} & \includegraphics[scale=0.23]{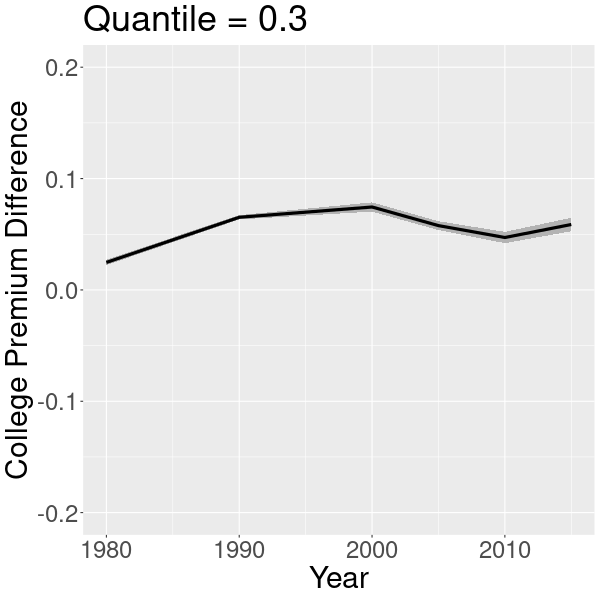} \\
        \includegraphics[scale=0.23]{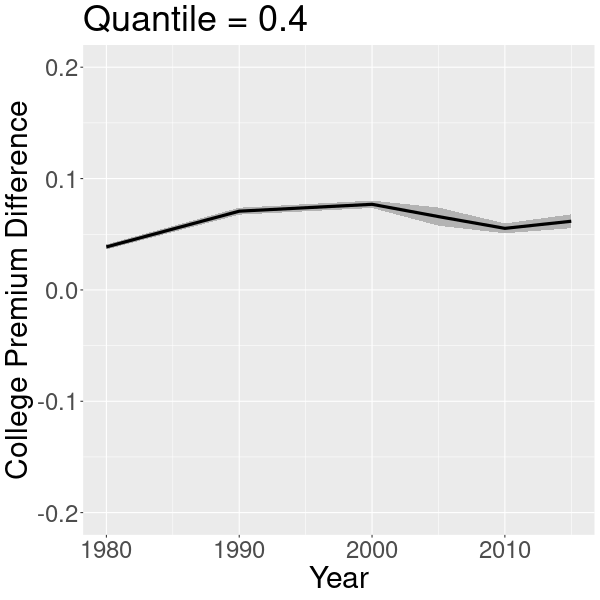} & \includegraphics[scale=0.23]{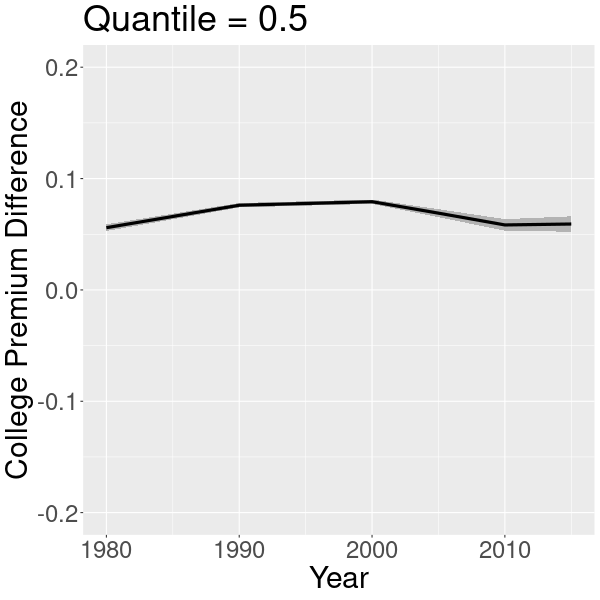} & \includegraphics[scale=0.23]{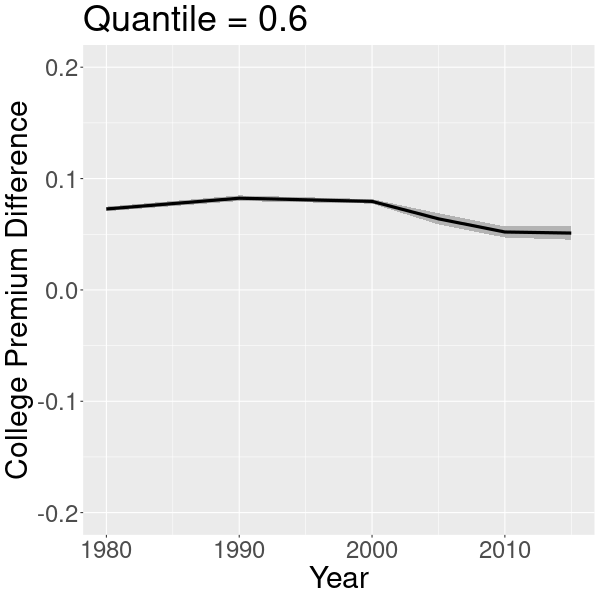} \\
        \includegraphics[scale=0.23]{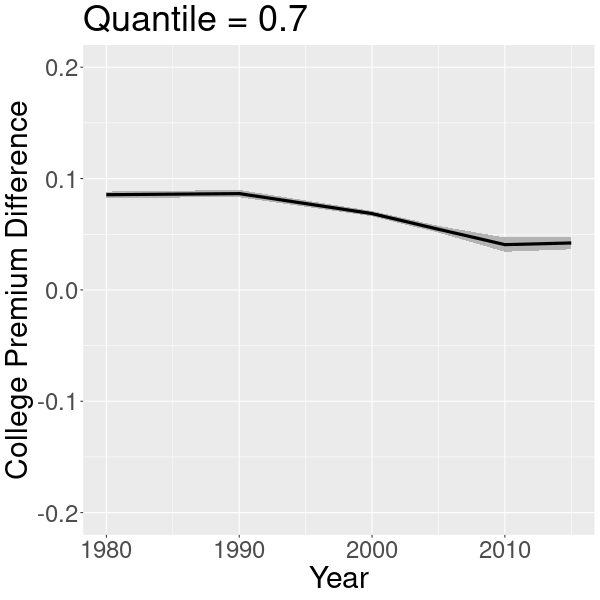} & \includegraphics[scale=0.23]{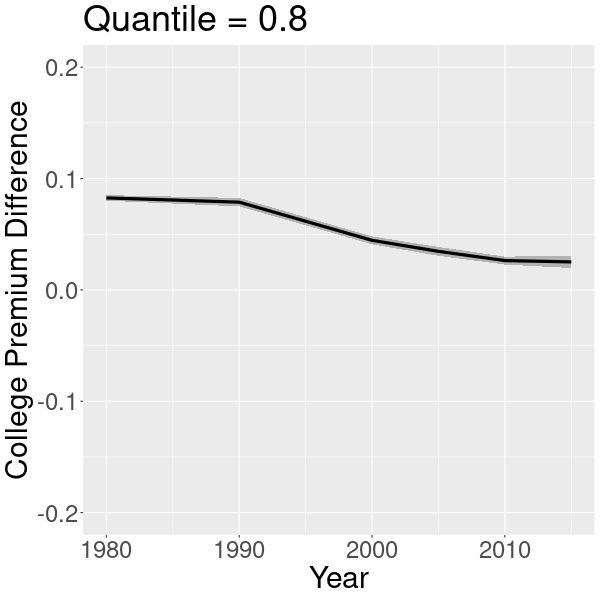} & \includegraphics[scale=0.23]{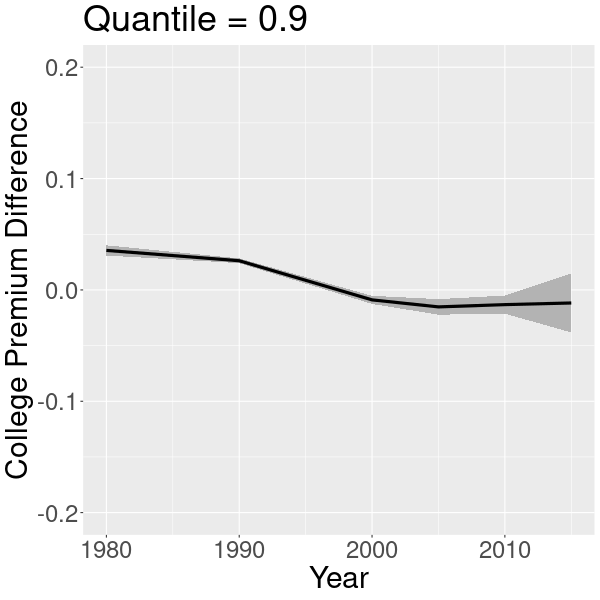} 
    \end{tabular}
    \flushleft{\footnotesize Notes. The graphs show the difference between female college premium and male college premium. The positive number denotes that female college premium is higher than male one. The solid line comes from $\hat{\beta}_3$ each year and the grey area denotes the pointwise 95\% confidence interval.}
\end{figure}\textbf{}

\begin{table}[ht]
\centering
\caption{College Wage Premium: Single Years} 
\begin{tabular}{ccccc}
  \hline
Year & Female & Male & Difference & Time (sec.) \\ 
  \hline
  \underline{$\tau=0.1$}\\
1980 & 0.1689 & 0.1737 & -0.0048 & 375 \\ 
   & [0.1667,0.1711] & [0.1700,0.1774] & [-0.0087,-0.0010] &  \\ 
  1990 & 0.3091 & 0.2632 & 0.0458 & 356 \\ 
   & [0.3056,0.3125] & [0.2584,0.2680] & [0.0417,0.0500] &  \\ 
  2000 & 0.3340 & 0.2929 & 0.0411 & 395 \\ 
   & [0.3323,0.3357] & [0.2896,0.2961] & [0.0377,0.0446] &  \\ 
  2005 & 0.3803 & 0.3484 & 0.0319 & 76.2 \\ 
   & [0.3763,0.3844] & [0.3411,0.3557] & [0.0261,0.0377] &  \\ 
  2010 & 0.3887 & 0.3795 & 0.0092 & 84.9 \\ 
   & [0.3763,0.4011] & [0.3709,0.3880] & [0.0014,0.0171] &  \\ 
  2015 & 0.4002 & 0.3767 & 0.0234 & 86.6 \\ 
   & [0.3951,0.4052] & [0.3701,0.3834] & [0.0151,0.0318] &  \\ 
  \underline{$\tau=0.2$}\\
1980 & 0.2093 & 0.1974 & 0.0120 & 423 \\ 
   & [0.2072,0.2114] & [0.1944,0.2003] & [0.0078,0.0162] &  \\ 
  1990 & 0.3391 & 0.2821 & 0.0570 & 454 \\ 
   & [0.3352,0.3429] & [0.2796,0.2846] & [0.0532,0.0607] &  \\ 
  2000 & 0.3770 & 0.3084 & 0.0686 & 488 \\ 
   & [0.3737,0.3804] & [0.3061,0.3107] & [0.0650,0.0723] &  \\ 
  2005 & 0.4228 & 0.3682 & 0.0546 & 76.1 \\ 
   & [0.4206,0.4250] & [0.3611,0.3754] & [0.0474,0.0617] &  \\ 
  2010 & 0.4400 & 0.4106 & 0.0294 & 89.5 \\ 
   & [0.4291,0.4510] & [0.3950,0.4263] & [0.0233,0.0355] &  \\ 
  2015 & 0.4422 & 0.3999 & 0.0424 & 90.5 \\ 
   & [0.4389,0.4456] & [0.3961,0.4036] & [0.0370,0.0477] &  \\ 
  \underline{$\tau=0.3$}\\
1980 & 0.2272 & 0.2025 & 0.0247 & 417 \\ 
   & [0.2250,0.2295] & [0.1988,0.2063] & [0.0222,0.0271] &  \\ 
  1990 & 0.3538 & 0.2886 & 0.0652 & 421 \\ 
   & [0.3523,0.3554] & [0.2872,0.2900] & [0.0633,0.0672] &  \\ 
  2000 & 0.3902 & 0.3158 & 0.0745 & 472 \\ 
   & [0.3862,0.3943] & [0.3147,0.3169] & [0.0703,0.0786] &  \\ 
  2005 & 0.4366 & 0.3788 & 0.0577 & 69.1 \\ 
   & [0.4306,0.4425] & [0.3728,0.3848] & [0.0538,0.0617] &  \\ 
  2010 & 0.4638 & 0.4168 & 0.0471 & 88.3 \\ 
   & [0.4593,0.4684] & [0.4096,0.4240] & [0.0420,0.0521] &  \\ 
  2015 & 0.4681 & 0.4094 & 0.0587 & 86.3 \\ 
   & [0.4644,0.4718] & [0.4025,0.4162] & [0.0527,0.0648] &  \\ 
   \hline
\end{tabular}
\end{table}

\begin{table}[ht]
\centering
\caption{College Wage Premium (cont.): Single Years} 
\begin{tabular}{ccccc}
  \hline
Year & Female & Male & Difference & Time (sec.) \\ 
  \hline
  \underline{$\tau=0.4$}\\
1980 & 0.2393 & 0.2006 & 0.0386 & 426 \\ 
   & [0.2375,0.2410] & [0.1984,0.2029] & [0.0364,0.0409] &  \\ 
  1990 & 0.3628 & 0.2921 & 0.0707 & 402 \\ 
   & [0.3599,0.3657] & [0.2914,0.2929] & [0.0676,0.0738] &  \\ 
  2000 & 0.4009 & 0.3240 & 0.0769 & 412 \\ 
   & [0.3969,0.4048] & [0.3227,0.3252] & [0.0736,0.0802] &  \\ 
  2005 & 0.4509 & 0.3851 & 0.0658 & 83.5 \\ 
   & [0.4410,0.4609] & [0.3824,0.3878] & [0.0577,0.0740] &  \\ 
  2010 & 0.4811 & 0.4259 & 0.0553 & 88.1 \\ 
   & [0.4780,0.4843] & [0.4206,0.4311] & [0.0508,0.0597] &  \\ 
  2015 & 0.4825 & 0.4208 & 0.0617 & 88.7 \\ 
   & [0.4770,0.4880] & [0.4128,0.4288] & [0.0555,0.0679] &  \\ 
  \underline{$\tau=0.5$}\\
1980 & 0.2552 & 0.1993 & 0.0559 & 457 \\ 
   & [0.2521,0.2583] & [0.1979,0.2007] & [0.0529,0.0589] &  \\ 
  1990 & 0.3700 & 0.2940 & 0.0761 & 401 \\ 
   & [0.3681,0.3720] & [0.2934,0.2946] & [0.0740,0.0781] &  \\ 
  2000 & 0.4124 & 0.3331 & 0.0792 & 423 \\ 
   & [0.4101,0.4147] & [0.3322,0.3341] & [0.0773,0.0812] &  \\ 
  2005 & 0.4639 & 0.3950 & 0.0689 & 75.8 \\ 
   & [0.4592,0.4686] & [0.3924,0.3976] & [0.0653,0.0725] &  \\ 
  2010 & 0.4959 & 0.4375 & 0.0583 & 84.9 \\ 
   & [0.4923,0.4994] & [0.4344,0.4407] & [0.0531,0.0636] &  \\ 
  2015 & 0.4972 & 0.4380 & 0.0592 & 83 \\ 
   & [0.4928,0.5015] & [0.4341,0.4419] & [0.0523,0.0661] &  \\ 
  \underline{$\tau=0.6$}\\
1980 & 0.2728 & 0.2001 & 0.0727 & 461 \\ 
   & [0.2702,0.2755] & [0.1983,0.2019] & [0.0706,0.0748] &  \\ 
  1990 & 0.3804 & 0.2980 & 0.0824 & 439 \\ 
   & [0.3785,0.3823] & [0.2965,0.2996] & [0.0798,0.0849] &  \\ 
  2000 & 0.4244 & 0.3449 & 0.0795 & 411 \\ 
   & [0.4227,0.4261] & [0.3435,0.3463] & [0.0775,0.0815] &  \\ 
  2005 & 0.4736 & 0.4098 & 0.0638 & 73.8 \\ 
   & [0.4692,0.4780] & [0.4065,0.4131] & [0.0588,0.0689] &  \\ 
  2010 & 0.5070 & 0.4550 & 0.0520 & 86.8 \\ 
   & [0.5006,0.5135] & [0.4492,0.4608] & [0.0470,0.0570] &  \\ 
  2015 & 0.5101 & 0.4590 & 0.0511 & 82.1 \\ 
   & [0.5060,0.5142] & [0.4546,0.4634] & [0.0451,0.0570] &  \\ 
   \hline
\end{tabular}
\end{table}

\begin{table}[ht]
\centering
\caption{College Wage Premium (cont.): Single Years} 
\begin{tabular}{ccccc}
  \hline
Year & Female & Male & Difference & Time (sec.) \\ 
  \hline
  \underline{$\tau=0.7$}\\
1980 & 0.2902 & 0.2047 & 0.0855 & 462 \\ 
   & [0.2878,0.2925] & [0.2032,0.2062] & [0.0827,0.0883] &  \\ 
  1990 & 0.3909 & 0.3044 & 0.0865 & 397 \\ 
   & [0.3879,0.3939] & [0.3021,0.3066] & [0.0832,0.0898] &  \\ 
  2000 & 0.4318 & 0.3633 & 0.0685 & 487 \\ 
   & [0.4294,0.4343] & [0.3616,0.3650] & [0.0661,0.0709] &  \\ 
  2005 & 0.4854 & 0.4309 & 0.0545 & 90.3 \\ 
   & [0.4816,0.4892] & [0.4274,0.4345] & [0.0514,0.0575] &  \\ 
  2010 & 0.5138 & 0.4731 & 0.0406 & 96.4 \\ 
   & [0.5080,0.5196] & [0.4639,0.4823] & [0.0342,0.0471] &  \\ 
  2015 & 0.5190 & 0.4768 & 0.0422 & 80.4 \\ 
   & [0.5109,0.5271] & [0.4727,0.4809] & [0.0365,0.0478] &  \\ 
  \underline{$\tau=0.8$}\\
1980 & 0.3008 & 0.2182 & 0.0826 & 441 \\ 
   & [0.2991,0.3026] & [0.2165,0.2199] & [0.0801,0.0851] &  \\ 
  1990 & 0.3968 & 0.3180 & 0.0788 & 393 \\ 
   & [0.3943,0.3993] & [0.3162,0.3198] & [0.0750,0.0826] &  \\ 
  2000 & 0.4382 & 0.3937 & 0.0445 & 467 \\ 
   & [0.4356,0.4407] & [0.3907,0.3966] & [0.0409,0.0481] &  \\ 
  2005 & 0.4916 & 0.4570 & 0.0346 & 71.5 \\ 
   & [0.4865,0.4966] & [0.4513,0.4626] & [0.0306,0.0386] &  \\ 
  2010 & 0.5189 & 0.4926 & 0.0263 & 86.2 \\ 
   & [0.5109,0.5269] & [0.4863,0.4988] & [0.0227,0.0299] &  \\ 
  2015 & 0.5254 & 0.5002 & 0.0251 & 84.7 \\ 
   & [0.5173,0.5335] & [0.4953,0.5052] & [0.0196,0.0306] &  \\ 
  \underline{$\tau=0.9$}\\
1980 & 0.2889 & 0.2534 & 0.0355 & 453 \\ 
   & [0.2869,0.2910] & [0.2504,0.2565] & [0.0309,0.0401] &  \\ 
  1990 & 0.3901 & 0.3639 & 0.0262 & 412 \\ 
   & [0.3876,0.3926] & [0.3621,0.3657] & [0.0241,0.0283] &  \\ 
  2000 & 0.4456 & 0.4545 & -0.0090 & 495 \\ 
   & [0.4433,0.4479] & [0.4491,0.4600] & [-0.0126,-0.0053] &  \\ 
  2005 & 0.5014 & 0.5167 & -0.0153 & 67.5 \\ 
   & [0.4913,0.5115] & [0.5092,0.5241] & [-0.0223,-0.0083] &  \\ 
  2010 & 0.5295 & 0.5427 & -0.0132 & 97.2 \\ 
   & [0.5256,0.5333] & [0.5346,0.5507] & [-0.0213,-0.0051] &  \\ 
  2015 & 0.5402 & 0.5520 & -0.0117 & 94.4 \\ 
   & [0.5173,0.5631] & [0.5475,0.5564] & [-0.0382,0.0147] &  \\ 
   \hline
\end{tabular}
\end{table}

\clearpage

\subsection*{Empirical application: robustness checks}
\begin{figure}[htb]
    \centering
    \caption{Sensitivity Analysis: Female-Male College Premium}
    \label{fig:sensitivity01}
    \hskip15pt
    \begin{tabular}{c c}
        \includegraphics[scale=0.30]{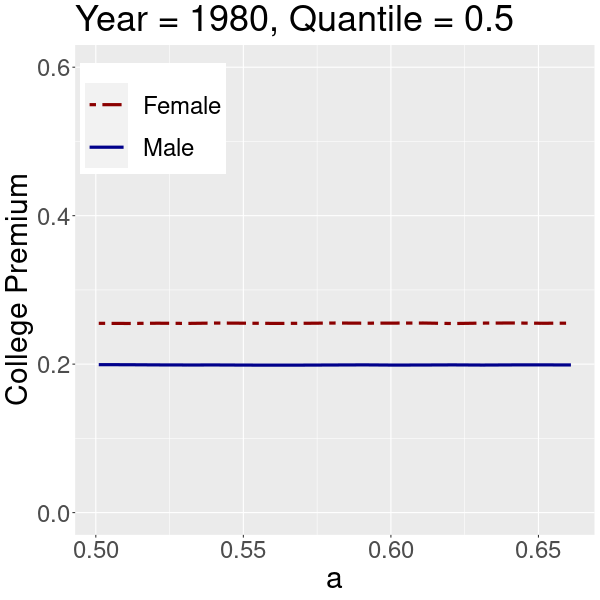} & \includegraphics[scale=0.30]{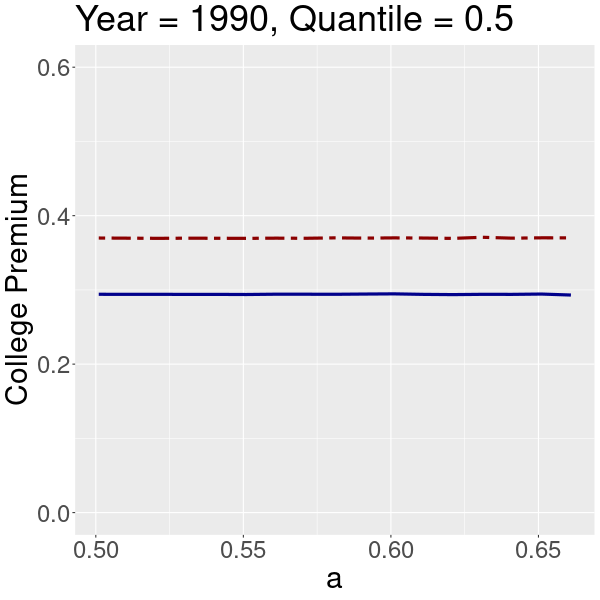} \\ 
        \includegraphics[scale=0.30]{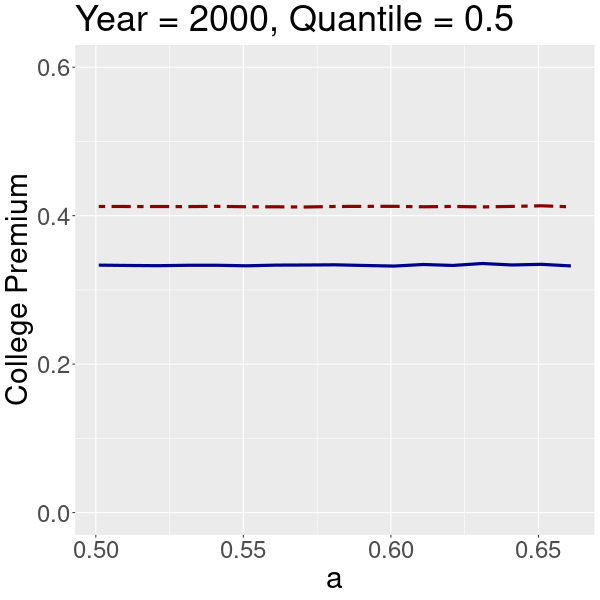} & \includegraphics[scale=0.30]{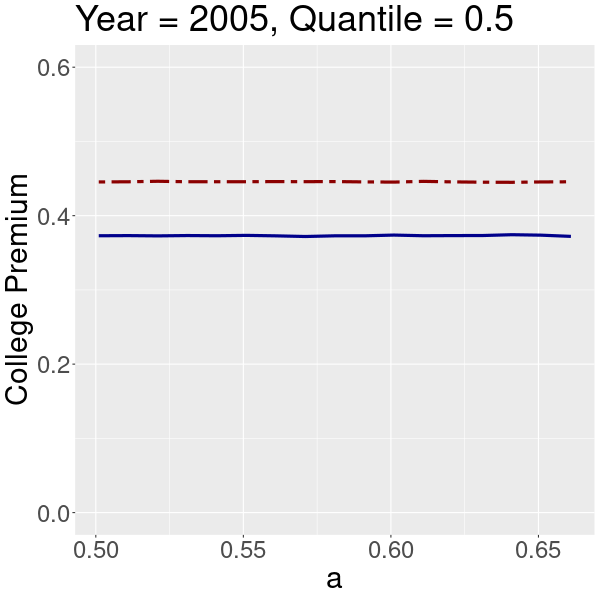} \\ 
        \includegraphics[scale=0.30]{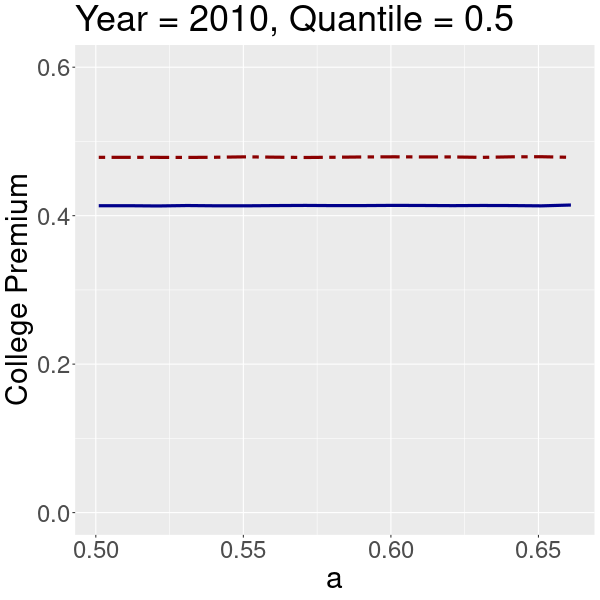} & \includegraphics[scale=0.30]{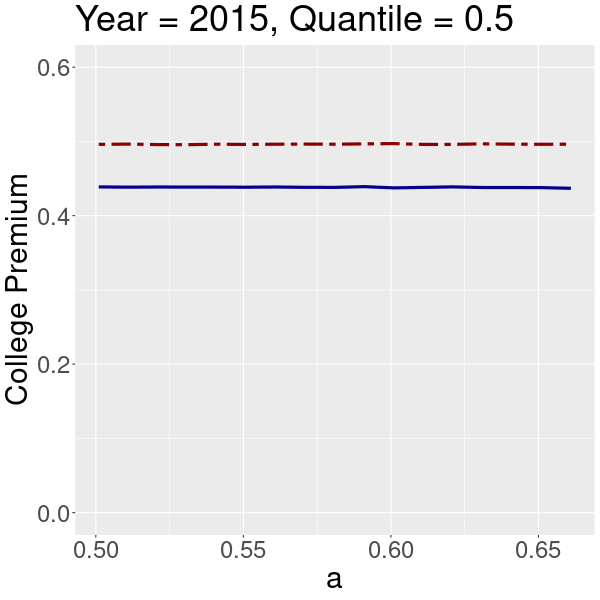} 
    \end{tabular}
    \flushleft{\footnotesize Notes.  The blue solid line (Male) comes from $\hat{\beta}_2$ each year and the red dot-dashed line (Female) does from the estimate $\hat{\beta}_2+\hat{\beta}_3$. We estimate the model over 17 equi-spaced points of $a$ in $[0.501, 0.661]$}.
\end{figure}

\begin{figure}[htb]
    \centering
    \caption{Sensitivity Analysis: Female-Male College Premium Differences}
    \label{fig:sensitivity02}
    \hskip15pt
    \begin{tabular}{c c}
        \includegraphics[scale=0.30]{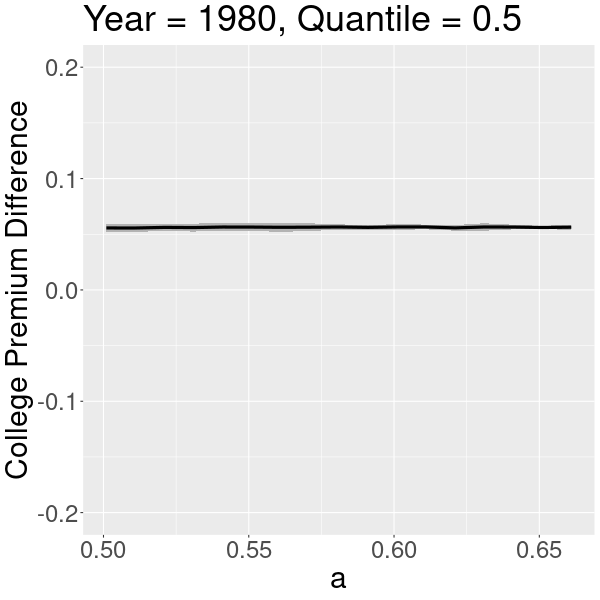} & \includegraphics[scale=0.30]{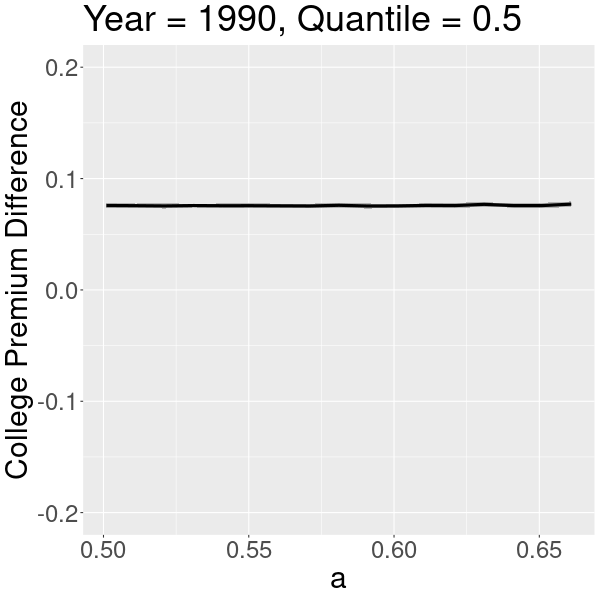} \\ 
        \includegraphics[scale=0.30]{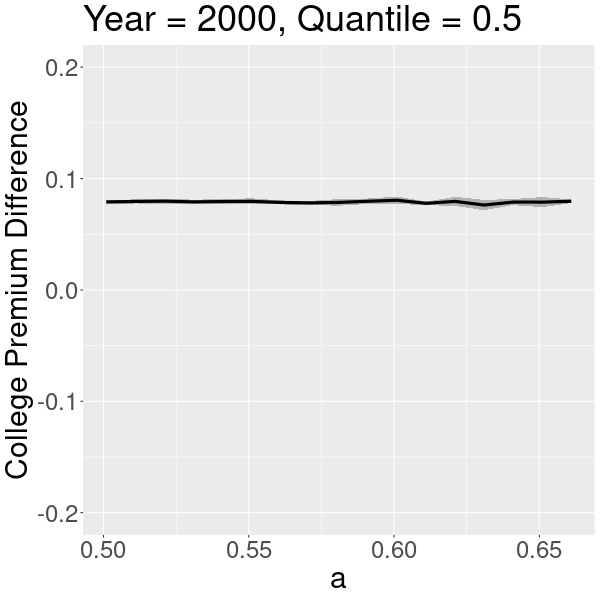} & \includegraphics[scale=0.30]{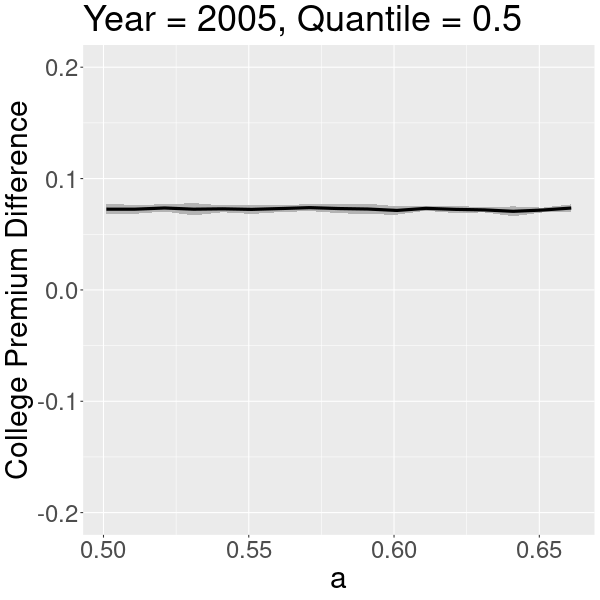} \\ 
        \includegraphics[scale=0.30]{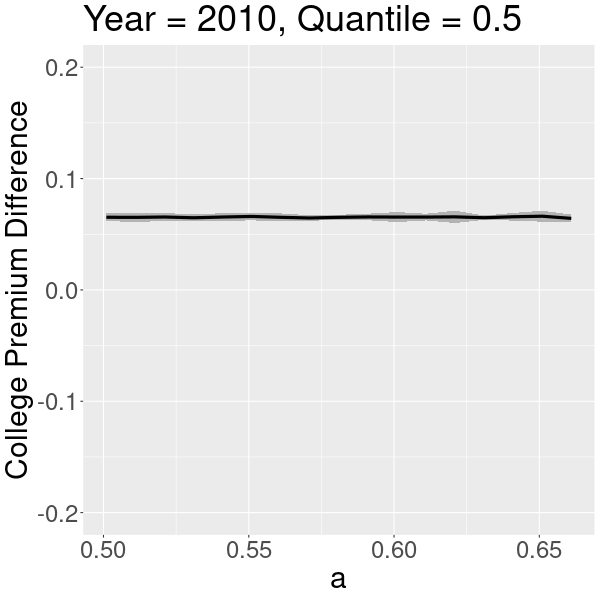} & \includegraphics[scale=0.30]{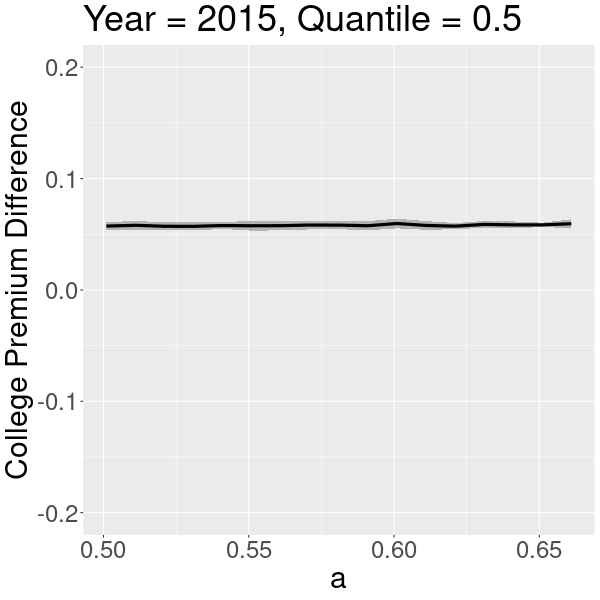} 
    \end{tabular}
    \flushleft{\footnotesize Notes.  The graphs show the difference between female college premium and male college premium. The positive number denotes that female college premium is higher than male one. The solid line comes from $\hat{\beta}_3$ each year and the grey area denotes the pointwise 95\% confidence interval. We estimate the model over 17 equi-spaced points of $a$ in $[0.501, 0.661]$}.
\end{figure}

To check the robustness of the results, we conduct some sensitivity analysis over different learning rates $\gamma_0 a^{-a}$. We keep using the rule-of-thumb method in Section \ref{sec:rule-of-thumb} for $\gamma_0$. Instead, We vary the value of $a$ over 17 equi-spaced points in $[0.501, 0.661]$. We estimate the model at quantile $\tau=0.5$ over all years. We use the samples that are bunching over 5 years after 2001 as in Section \ref{sec:empirical}. Figures \ref{fig:sensitivity01}-\ref{fig:sensitivity02} report the estimation results over different $a$ values. The graphs are flat over all areas and confirm that the estimation results are robust to the tuning parameter choice.

\textbf{}

\end{document}